\documentclass{jfm}

\usepackage{graphicx}
\usepackage{natbib}
\usepackage{math_pack}
\usepackage{color}
\usepackage{subfigure}
\usepackage{amsmath}
\usepackage{amssymb}
\usepackage{math_pack}

\ifCUPmtlplainloaded \else
  \checkfont{eurm10}
  \iffontfound
    \IfFileExists{upmath.sty}
      {\typeout{^^JFound AMS Euler Roman fonts on the system,
                   using the 'upmath' package.^^J}%
       \usepackage{upmath}}
      {\typeout{^^JFound AMS Euler Roman fonts on the system, but you
                   dont seem to have the}%
       \typeout{'upmath' package installed. JFM.cls can take advantage
                 of these fonts,^^Jif you use 'upmath' package.^^J}%
       \providecommand\upi{\pi}%
      }
  \else
    \providecommand\upi{\pi}%
  \fi
\fi

\ifCUPmtlplainloaded \else
  \checkfont{msam10}
  \iffontfound
    \IfFileExists{amssymb.sty}
      {\typeout{^^JFound AMS Symbol fonts on the system, using the
                'amssymb' package.^^J}%
       \usepackage{amssymb}%
       \let\le=\leqslant  
         
      }{}
  \fi
\fi

\ifCUPmtlplainloaded \else
  \IfFileExists{amsbsy.sty}
    {\typeout{^^JFound the 'amsbsy' package on the system, using it.^^J}%
     \usepackage{amsbsy}}
    {\providecommand\boldsymbol[1]{\mbox{\boldmath $##1$}}}
\fi

\newcommand\omegab{\boldsymbol{\omega}}
\newcommand\taub{\boldsymbol{\tau}}

\newsavebox{\astrutbox}
\sbox{\astrutbox}{\rule[-5pt]{0pt}{20pt}}

\definecolor{cinnamon}{rgb}{0.82, 0.41, 0.12}

\title[Suspensions of finite-size spheres in turbulent duct flow]{Suspensions of finite-size neutrally-buoyant spheres in turbulent duct flow}

\author[W. Fornari, H. Tabaei Kazerooni, J. Hussong and L. Brandt]%
{Walter Fornari$^1$%
  \thanks{Email address for correspondence: fornari@mech.kth.se},\ns
Hamid Tabaei Kazerooni$^{1,2}$, Jeanette Hussong$^2$ \\
and Luca Brandt$^1$}

\affiliation{$^1$Linn\'e Flow Centre and Swedish e-Science Research Centre (SeRC), KTH Mechanics,
SE-10044 Stockholm, Sweden\\[\affilskip]
$^2$Ruhr-Universit\"at Bochum, Chair of Hydraulic Fluid Machinery, Universit\"atsstra\ss e 150, 44801 Bochum, Germany}

\pubyear{2010}
\volume{650}
\pagerange{119--126}
\date{?; revised ?; accepted ?. - To be entered by editorial office}
\begin{document}

\maketitle

\begin{abstract}
We study the turbulent square duct flow of dense suspensions of neutrally-buoyant spherical particles. 
Direct numerical simulations (DNS) are performed in the range of volume fractions $\phi=0-0.2$, using the 
immersed boundary method (IBM) to account for the dispersed phase. 
Based on the hydraulic diameter a Reynolds number of $5600$ is considered.
We report  
flow features and particle statistics specific to this geometry, and compare the results to the case of two-dimensional channel flows. 
In particular, we observe that for $\phi=0.05$ and $0.1$, particles preferentially accumulate 
on the corner bisectors, close to the duct corners as also observed for laminar square duct flows of same duct-to-particle size ratios. At 
the highest volume fraction, particles preferentially accumulate in the core region. For channel flows, in the 
absence of lateral confinement particles are found instead to be uniformily distributed across the channel. We also observe 
that the intensity of the cross-stream secondary flows increases (with respect to the unladen case) with the volume 
fraction up to $\phi=0.1$, as a consequence of the high concentration of particles along the corner bisector. 
For $\phi=0.2$ the turbulence activity is strongly reduced and the intensity 
of the secondary flows reduces below that of the unladen case. The friction Reynolds number increases with $\phi$ in dilute 
conditions, as observed for channel flows. However, for $\phi=0.2$ the mean friction Reynolds number decreases below the value 
for $\phi=0.1$.

\end{abstract}


\section{Introduction}

Particle-laden turbulent flows are commonly encountered in many engineering and environmental processes. 
Examples include sediment transport in rivers, avalanches, slurries and chemical reactions involving 
particulate catalysts. Understanding the behavior of these suspensions is generally a difficult task due 
to the large number of parameters involved. Indeed, particles may vary in density, shape, size and 
stiffness, and when non-dilute particle concentrations are considered the collective suspension dynamics depends 
strongly on the mass and solid fractions. Even in Stokesian and laminar flows, different combinations of 
these parameters lead to interesting peculiar phenomena. In turbulence, the situation is further 
complicated due to the interaction between particles and vortical structures of different sizes. Hence, 
the particle behavior does not depend only on its dimensions and characteristic response time, but also on 
the ratio among these and the characteristic turbulent length and time scales. The turbulence features 
are also altered due to the presence of the dispersed phase, especially at high volume fractions. 
Because of the difficulty of treating the problem analytically, particle-laden flows are often studied 
either experimentally or numerically. In the context of wall-bounded flows, the suspension dynamics 
has often been studied in canonical flows such as channels and boundary layers. However, internal flows 
relevant to many industrial applications typically involve more complex, non-canonical geometries in which 
secondary flows, flow separation and other non-trivial phenomena are observed. It is hence important to 
understand the behavior of particulate suspensions in more complex and realistic geometries. We will here 
focus on turbulent square ducts, where gradients of the Reynolds stresses induce the generation of mean 
streamwise vortices. These are known as Prandtl's secondary motions of the second kind \citep{prandtl}. 
The suspension behavior subjected to these peculiar secondary flows will be investigated, as well as the 
influence of the solid phase on the turbulence features.

As said, interesting rheological behaviors can be observed already in the Stokesian regime. 
Among these we recall shear-thinning and thickening, jamming at high volume fractions and the 
generation of high effective viscosities and normal stress differences \citep{stickel2005,morris2009,
wagner2009}. Indeed, for these multiphase flows the response to the local deformation rate is altered 
and the effective viscosity $\mu_e$ changes with respect to that of the pure fluid $\mu$. 
Shear-thickening and normal stress differences are observed also in the laminar regime and are 
typically related to the formation of an anisotropic microstructure that arises due to the loss of 
symmetry in particle pair trajectories \citep{Morrispof08,picano2013,Morris2014}. In general, 
the effective viscosity of a suspension, $\mu_e$, has been shown to be a function of the 
particle Reynolds number $Re_p$, the P\'{e}clet number $Pe$ (quantifying thermal fluctuations), the volume fraction $\phi$ and, relevant to microfluidic application, of the system confinement \citep{fornariPRL,doyeux}.\\
Another important feature observed in wall-bounded flows is particle migration. Depending on the 
particle Reynolds number $Re_p$, different types of migrations are observed. In the viscous regime, particles irreversibly migrate towards 
the centerline in a pressure-driven Poiseuille flow. Hence, particles undergo a shear-induced migration as they move from high to low shear 
rate regions \citep{guazz2011,koh1994}. On the other hand, when inertial effects become important, 
particles are found to move radially away from both the centerline and the walls, towards an 
intermediate equilibrium position. \citet{segre1962} first observed this phenomenon in a tube and hence 
named it as the tubular pinch effect. 
This migration is mechanistically unrelated to the rheological properties of the flow and results from 
the fluid-particle interaction within the conduit. The exact particle focusing position has been shown 
to depend on the conduit-particle size ratio and on the bulk and particle Reynolds numbers \citep{matas,
morita}. In square ducts the situation is more complex. Depending on the same parameters, the focusing 
positions can occur at the wall bisectors, along heteroclinic orbits or only at the duct corners 
\citep{chun2006,abbas2014,nakagawa,kazerooni,lash2017}. 

Already in the laminar regime, the flow in conduits is altered by the presence of solid particles. Relevant to mixing, 
particle-induced secondary flows are generated in ducts, otherwise absent in the unladen 
reference cases as shown by \citet{amini,kazerooni}. Interesting results are found also in the transition 
regime from laminar to turbulent flow. It has been shown that the presence of particles can 
either increase or reduce the critical Reynolds number above which the transition occurs. In 
particular, transition depends upon the channel half-width to particle radius ratio $h/a$, the initial 
arrangement of particles and the solid volume fraction $\phi$ \citep{matas2003,Loisel2013,lashg2015}. 

In the fully turbulent regime, most studies have focused on dilute suspensions of heavy particles, 
smaller than the hydrodynamic scales, in channel flows. This is known as the one-way coupling regime 
\citep{balach-rev2010} as there is no back-influence of the solid phase on the fluid. These kind of 
particles are found to migrate from regions of high to low turbulence intensities (turbophoresis)
~\citep{reeks1983} and the effect is stronger when the turbulent near-wall characteristic time and the 
particle inertial time scale are similar \citep{soldati2009}. It was later shown by \citet{sardina2011,
sardina2012} that close to the walls particles also tend to form streaky particle patterns.\\
When the mass fraction is high, the fluid motion is altered by the presence of particles (two-way 
coupling regime) and it has been shown that turbulent near-wall fluctuations are reduced 
while their anisotropy is increased \citep{kulick1994}. The total drag is hence found to decrease 
\citep{zhao2010}.\\
Small heavy particles tend to accumulate in regions of high 
compressional strain and low swirling strength in turbulent duct flows, especially in the near-wall and vortex center 
regions \citep{Winkler04}. \citet{Sharma} showed that while passive tracers and low-inertia 
particles stay within the secondary swirling flows (circulating between the duct core and boundaries), 
high inertia particles accumulate close to the walls, mixing more efficiently in the streamwise 
direction. In particular, particles tend to deposit at the duct corners. More recently, 
\citet{noorani} studied the effect of varying the duct aspect ratio on the particle transport. 
These authors considered a higher bulk Reynolds number than \citet{Sharma} and found that in square 
ducts, particle concentration in the viscous sublayer is maximum at the centerplane. However, 
increasing the aspect ratio, the location of maximum concentration moves towards the corner as also 
the kinetic energy of the secondary flows increases closer to the corners.

In the four-way coupling regime, considering dense suspensions of finite-size particles in turbulent channel flows 
(with radius of about $10$ plus units), it was instead found that the large-scale streamwise vortices are 
mitigated and that fluid streamwise velocity fluctuations are reduced. As the solid volume fraction 
increases, fluid velocity fluctuation intensities and Reynolds shear stresses are found to decrease, 
however particle-induced stresses significantly increase and this results in an increase of the overall drag 
\citep{picano2015}. Indeed, \citet{lashgari2014} identified three regimes in particle-laden channel flow, depending on the different values of the 
solid volume fraction $\phi$ and the Reynolds number $Re$, each dominated by different 
components of the total stress. In particular, viscous, turbulent and particle-induced stresses 
dominate the laminar, turbulent and inertial shear-thickening regimes. The effects of solid-to-fluid 
density ratio $\rho_p/\rho_f$, mass fraction, polidispersity and shape have also been studied by 
\citet{forn2016,forn2018,lashgari2017,niazi}.\\
Recently, \citet{Lin} used a direct-forcing fictitious method to study turbulent duct flows laden 
with a dilute suspension of finite-size spheres heavier than the carrier fluid. Spheres with radius 
$a=h/10$ (with $h$ the duct half-width) were considered at a solid volume fraction $\phi = 2.36\%$. 
These authors show that particles sedimentation breaks the up-down symmetry of the mean 
secondary vortices. This results in a stronger circulation that transports the fluid downward in the 
bulk center region and upward along the side walls similarly to what observed for the duct 
flow over a porous wall by \citet{samanta2015}.
As the solid-to-fluid density ratio $\rho_p/\rho_f$ 
increases, the overall turbulence intensity is shown to decrease. However, mean secondary vortices at 
the bottom walls are enhanced and this leads to a preferential accumulation of particles at the face 
center of the bottom wall.

In the present work, we study the turbulence modulation and particle dynamics in turbulent square-duct 
flows laden with particles. In particular we consider neutrally-buoyant finite-size 
spheres with radius $a=h/18$ (where $h$ is the duct-half width), and increase the volume fraction up to 
$\phi=0.2$. We use data from direct numerical simulations (DNS) that fully describe the solid phase 
dynamics via an immersed boundary method (IBM). We show that up to $\phi=0.1$, particles preferentially 
accumulate close to the duct corners as also observed for small inertial particles and for laminar duct flows 
laden with spheres of comparable $h/a$ and $\phi$. 
At the highest volume fraction, instead, we see a clear
particle migration towards the core region, a feature that is absent in turbulent channel flows with similar $\phi$. 
Concerning the fluid phase, the intensity of the secondary flows and the mean friction 
Reynolds number increase with the volume fraction up to $\phi=0.1$. However, for $\phi=0.2$ we find a 
strong reduction in the turbulence activity. The intensity of the secondary flows decreases below the value 
of the unladen reference case. In contrast to what observed for channel flow, the mean friction Reynolds 
number at $\phi=0.2$ is found to be smaller than for $\phi=0.1$. Hence, the contribution of 
particle-induced stresses to the overall drag is lower than what observed in a channel flow.

\section{Methodology}\label{sec:Methodology}
\vspace{10pt}
\subsection{Numerical Method}
During the last years, various methods have been proposed to perform interface-resolved direct numerical 
simulations (DNS) of particulate flows. The state of art and the different principles 
and applications have been recently documented in the comprehensive review article by \cite{maxey2017}. In the 
present study, the immersed boundary method (IBM) originally proposed by \cite{uhlmann2005} and modified 
by \cite{breugem2012} has been used to simulate suspensions of finite-size neutrally-buoyant spherical 
particles in turbulent square duct flow. The fluid phase is described in an Eulerian framework by the 
incompressible Navier-Stokes equations:
\begin{equation}
\label{div_f}
\div \vec u_f = 0
\end{equation}
\begin{equation}
\label{NS_f}
\pd{\vec u_f}{t} + \vec u_f \cdot \grad \vec u_f = -\frac{1}{\rho_f}\grad p + \nu \grad^2 \vec u_f + \vec f
\end{equation}
where $\vec u_f$ and $p$ are the velocity field and pressure, while $\rho_f$ and $\nu$ are the density 
and kinematic viscosity of the fluid phase. The last term on the right hand side of equation~(\ref{NS_f}) 
$\vec f$ is the localized IBM force imposed to the flow to model the boundary condition at the moving 
particle surface (i.e. $\vec u_f|_{\partial \mathcal{V}_p} = \vec u_p + \vec \omegab_p \times \vec r$). 
The dynamics of the rigid particles is determined by the Newton-Euler Lagrangian equations:
\begin{align}
\label{lin-vel}
\rho_p V_p \td{\vec u_p}{t} &= \oint_{\partial \mathcal{V}_p}^{} \vec \taub \cdot \vec n\, dS\\
\label{ang-vel}
I_p \td{\vec \omegab_p}{t} &= \oint_{\partial \mathcal{V}_p}^{} \vec r \times \vec \taub \cdot \vec n\, dS
\end{align}
where $\vec u_p$ and $\vec \omegab_p$ are the linear and angular velocities of the particle. 
In equations~(\ref{lin-vel}) and (\ref{ang-vel}), $V_p = 4\upi a^3/3$ and 
$I_p=2 \rho_p V_p a^2/5$ represent the particle volume and moment of inertia, 
$\vec \taub = -p \vec I + \nu\rho_f \left(\grad \vec u_f + \grad \vec u_f^T \right)$ 
is the fluid stress tensor, $\vec r$ indicates the distance from the center of the particles, and $\bf{n}$ 
is the unit vector normal to the particle surface $\partial \mathcal{V}_p$. 

In order to solve the governing equations, the fluid phase is discretized on a spatially uniform 
staggered Cartesian grid using a second-order finite-difference scheme. An explicit third order 
Runge-Kutta scheme is combined with a standard pressure-correction method to perform the time 
integration at each sub-step. The same time integration scheme has also been used for the 
evolution of eqs.~(\ref{lin-vel}) and (\ref{ang-vel}). For the solid phase, each particle surface 
is described by $N_L$ uniformly distributed Lagrangian points. The force exchanged by the fluid 
on the particles is imposed on each $l-th$ Lagrangian point. This force is related to the Eulerian 
force field $\vec f$ by the expression 
$\vec f_{ijk} = \sum_{l=1}^{N_L} \vec F_l \delta_d(x_{ijk} - \vec X_l) \Delta V_l$, where 
$\Delta V_l$ is the volume of the cell containing the $l-th$ Lagrangian point and 
$\delta_d$ is the regularized Dirac delta function $\delta_d$.
Here, $\vec F_l$ is the force (per unit mass) at each Lagrangian 
point, and it is computed as $\vec F_l=(\vec U_p(\vec X_l)-\vec U_{l}^*)/\Delta t$, where 
$\vec U_p=\vec u_p + \vec \omegab_p \times \vec r$ is the velocity at the lagrangian point 
$l$ at the previous time-step, while $\vec U_{l}^*$ is the interpolated first prediction 
velocity at the same point. 
An iterative algorithm with second order spatial accuracy is developed to calculate 
this force field. To maintain accuracy, eqs.~(\ref{lin-vel}) and (\ref{ang-vel}) are 
rearranged in terms of the IBM force field, 
\begin{align}
\label{lin-vel-ibm}
\rho_p V_p \td{\vec u_p}{t} &= -\rho_f \sum_{l=1}^{N_l} \vec F_l \Delta V_l + \rho_f \td{}{t} \int_{\mathcal{V}_p}^{} \vec u_f\, dV \\
\label{ang-vel-ibm}
I_p \td{\vec \omegab_p}{t} &= -\rho_f \sum_{l=1}^{N_l} \vec r_l \times \vec F_l \Delta V_l + \rho_f \td{}{t} \int_{\mathcal{V}_p}^{} \vec r \times \vec u_f\, dV 
\end{align}
where $\vec r_l$ is the distance between the center of a particle and the $l-th$ 
Lagrangian point on its surface. 
The second terms on the right-hand sides are corrections 
that account for the inertia of the fictitious fluid contained within the particle volume. 
Particle-particle and particle-wall interactions are also considered. Well-known models 
based on Brenner's asymptotic solution~\citep{brenner1961} are employed to correctly predict 
the lubrication force when the distance between particles as well as particles and walls is 
smaller than twice the mesh size. Collisions are modelled using a soft-sphere collision model, 
with a coefficient of restitution of $0.97$ to achieve an almost elastic rebound of particles. 
Friction forces are also taken into account~\citep{costa2015}. For more detailed discussions 
of the numerical method and of the mentioned models the reader is refereed to previous 
publications~\citep{breugem2012,picano2015,forn2016,fornari2016JFM,lashg2016}.\\
Periodic boundary conditions for both solid and liquid phases are imposed in the streamwise 
direction. The \emph{stress immersed boundary method} is used in the remaining directions to 
impose the no-slip/no-penetration conditions at the duct walls. The stress immersed boundary method has 
originally been developed to simulate the flow around rectangular-shaped obstacles in a fully 
Cartesian grid~\citep{breugem2014}. In this work, we use this method to enforce the fluid velocity 
to be zero at the duct walls. 
For more details on the method, the reader is referred to the works of \cite{breugem2005} and 
\cite{pourquie2009}. This approach has already been used in our group \citep{kazerooni} to study the 
laminar flow of large spheres in a squared duct.

\subsection{Flow geometry}
\vspace{8pt}
We investigate the turbulent flow of dense suspensions of neutrally-buoyant spherical particles 
in a square duct. The simulations are performed in a Cartesian computational domain of size 
$L_x=12h$, $L_z=2h$ and $L_y=2h$ where $h$ is the duct half-width and $x$, $y$ and $z$ are the 
streamwise and cross-stream directions. The domain is uniformly ($ \Delta x = \Delta z = 
\Delta y$) meshed by 2592$\times$432$\times$432 Eulerian grid points in the streamwise and 
cross-flow directions. The bulk velocity of the entire mixture $U_b$ is kept constant by 
adjusting the streamwise pressure gradient to achieve the constant bulk 
Reynolds number $Re_b=U_b2h/\nu=5600$. Based on the data provided by \cite{pinelli2010}, 
$Re_b=5600$ corresponds to a mean friction Reynolds number $Re_\tau = \bar U_*h/\nu =185$ for an unladen 
case, where $\bar U_* = \sqrt{\langle \tau_w \rangle/\rho_f}$ is the friction velocity calculated using the mean 
value of the shear stress $\tau_w$ along the duct walls.

We consider three different solid volume fractions of $\phi=5, 10$ and $20\%$ which correspond 
to 3340, 6680 and 13360 particles respectively. The reference unladen case is also considered 
for direct comparison. In all simulations, the duct-to-particle size ratio is fixed to $h/a=18$, 
and the particles are randomly initialized in the computational domain with zero translational 
and angular velocities. The number of Eulerian grid points per particle diameter is 24 
($\Delta x = 1/24$) whereas the Lagrangian mesh on the surface of the particles consists of 
1721 grid points. 

\begin{figure}
   \centering
   \includegraphics[width=0.7\textwidth]{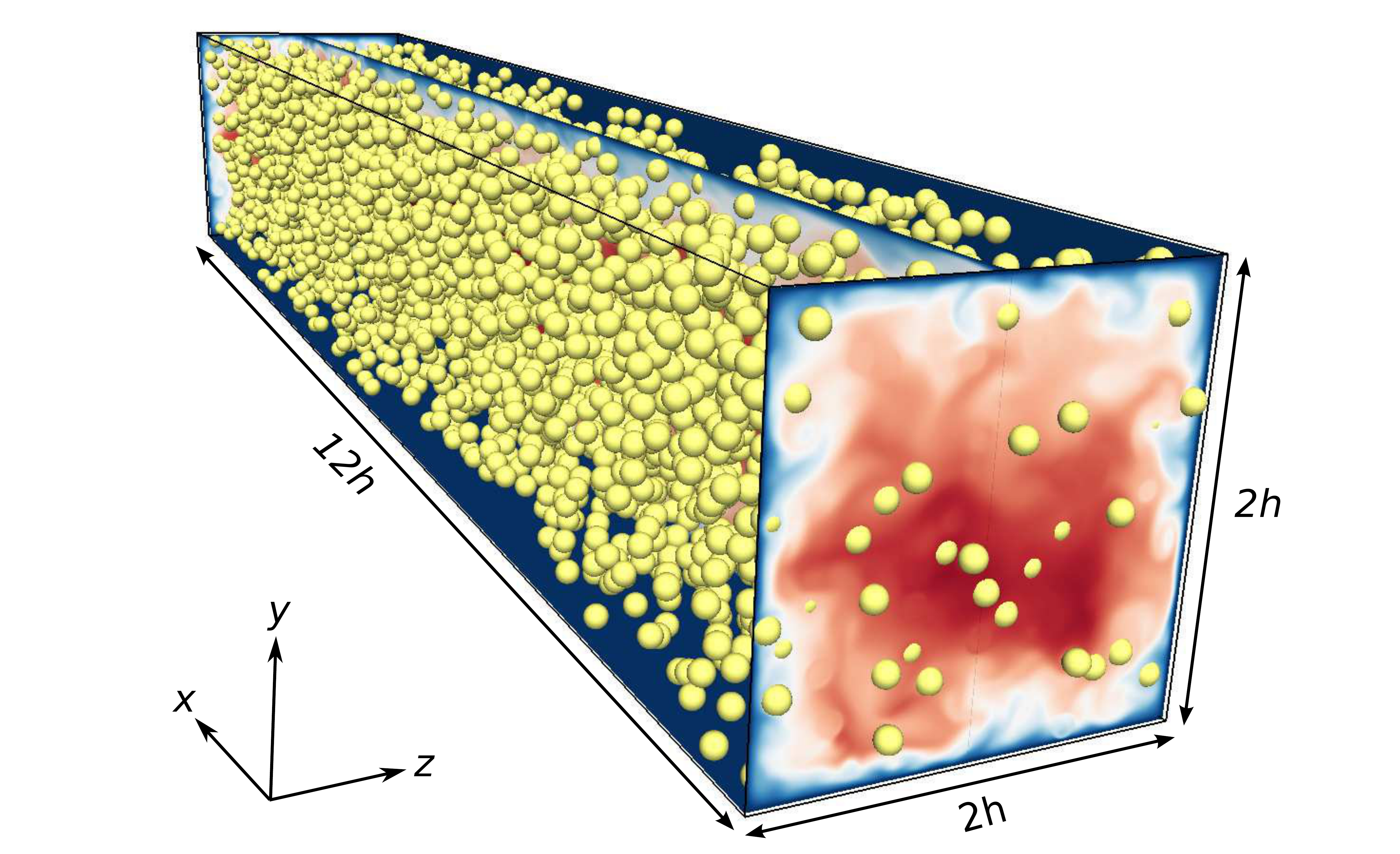}
   
\caption{Instantaneous snapshot of the magnitude of the velocity together with the solid particles; the solid  volume fraction 
$\phi=0.1$.}
\label{fig:snap}
\end{figure}

The simulations start from the laminar duct flow and the noise introduced by a high amplitude 
localised disturbance in the form of two counter-rotating streamwise vortices 
\citep{henningson1991}. Due to this disturbance and to the noise added by the particles, 
transition naturally occurs at the chosen Reynolds number. The statistics are collected after the 
initial transient phase of about $100 \, h/U_b$, using an averaging period of at least $600 \, h/U_b$ 
\citep{huser93,vinuesa14} (except for $\phi=20\%$ where $\sim 400 \, h/U_b$ are found to be enough to obtain 
converged statistics). A summary of the simulations is presented in table~\ref{tab:cases} 
while an instantaneous snapshot showing the magnitude of the streamwise velocity for $\phi=0.1$, 
together with the solid particles, is shown in figure~\ref{fig:snap}.

\begin{table}
  \begin{center}
\def~{\hphantom{0}}
  \begin{tabular}{cccccc}\\
\,\,\,\,\,\,\,\,\,  \,Case & $\phi=0.0 \,\,\,\,\,\,\,\,\,\,$  & $\phi=0.05$ & $\phi=0.1$  & $\phi=0.2 \,\,\,$ \\[3pt]
       $ \,\,\,\,\,\,\,\,\,\,\,\, \, \,  \,N_p$  & $0 \,\,\,\,\,\,\,\,\,\,$ & $3340$ & $6680$ & $13360$ \\[3pt]
      $ \,\,\,\,\,\,\,\,\,\,\,\, \, \,Re_b$  &  \,\,\,\,\,\,\, &  \,\,\,\,\,\,\, &  $5600$\\[3pt]
       $\, L_x \times L_y \times L_z$  &  \,\,\,\,\,\,\, &  \,\,\,\,\,\,\, &  $12h \times 2h \times 2h$\\[3pt]
       $N_x \times N_y \times N_z$  &  \,\,\,\,\,\,\, &  \,\,\,\,\,\,\, &  $2592 \times 432 \times 432$\\[3pt]
  \end{tabular}
  \caption{Summary of the different simulations cases. $N_p$ indicates the number of particles 
whereas $N_x$, $N_y$ and $N_z$ are the number of grid points in each direction.}
 \label{tab:cases}
 \end{center}
\end{table}

\section{Results}\label{sec:Results}
\vspace{10pt}
\subsection{Validation}
\begin{figure}
   \centering
   \includegraphics[width=0.6\textwidth]{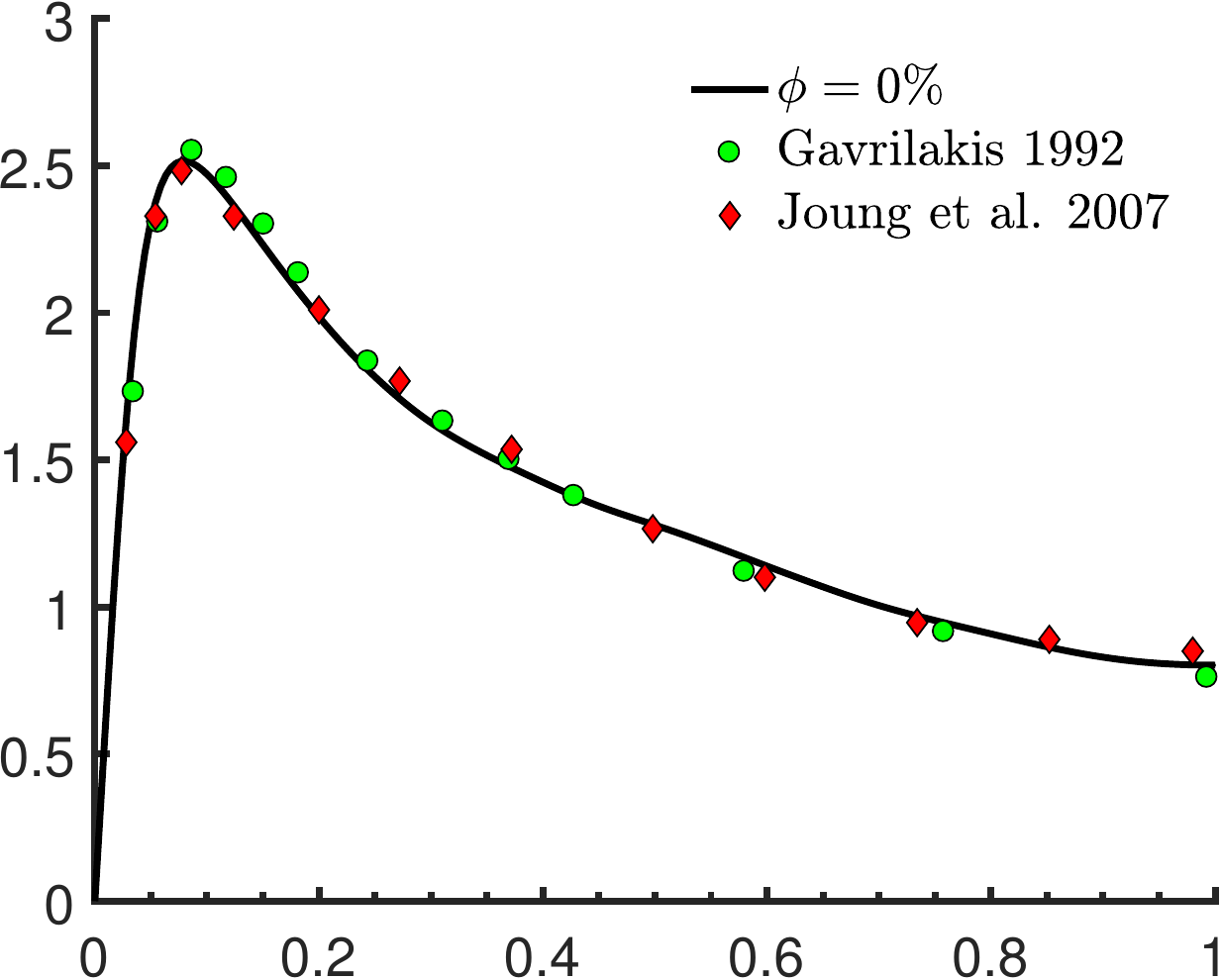}
    \put(-245,85){\rotatebox{90}{\large $u_{f,rms}'$}}
    \put(-115,-10){{\large $y/h$}}
   
\caption{Streamwise velocity fluctuation at the wall-bisector from the present simulation at $Re_b=4410$, and from 
the data by \citet{gavrilakis1992} at $Re_b=4410$ and \citet{joung2007} at $Re_b=4440$.}
\label{fig:valid}
\end{figure}

The code used in the present work has been already validated against several different cases in 
previous studies~\citep{breugem2012, picano2015, fornari2016JFM,kazerooni}. To further investigate 
the accuracy of the code, we calculate the friction factor $f=8\left(\bar U_*/U_b\right)^2$ for the reference unladen case with $Re_b=5600$, 
and compare it with the value obtained from the empirical correlation by 
\citet{jones1976}
\begin{equation}
 1/f^2=2log_{10}(1.125Re_{b}f^{1/2})-0.8
\end{equation}
The same value of $f=0.035$ is obtained from the simulation and the empirical formula. This corresponds to a 
mean $Re_{\tau}=185$.\\ 
We also performed a simulation at lower $Re_b=4410$ and $\phi=0$, see figure~\ref{fig:valid}, where we report the 
profile of the streamwise velocity fluctuation at the wall-bisector, normalized by the local 
friction velocity $U_*$. This is compared to the results by \citet{gavrilakis1992} and \citet{joung2007} 
at $Re_b=4410$ and $4440$. We see a good agreement with both works.

\subsection{Mean velocities, drag, and particle concentration}

In this section we report and discuss the results obtained for the different solid volume fractions 
$\phi$ considered. The phase-ensemble averages for the fluid (solid) phase have been calculated by 
considering only the points located outside (inside) of the volume occupied by the particles. The 
 statistics reported  are obtained by further averaging over the eight symmetric triangles that form the 
duct cross section.

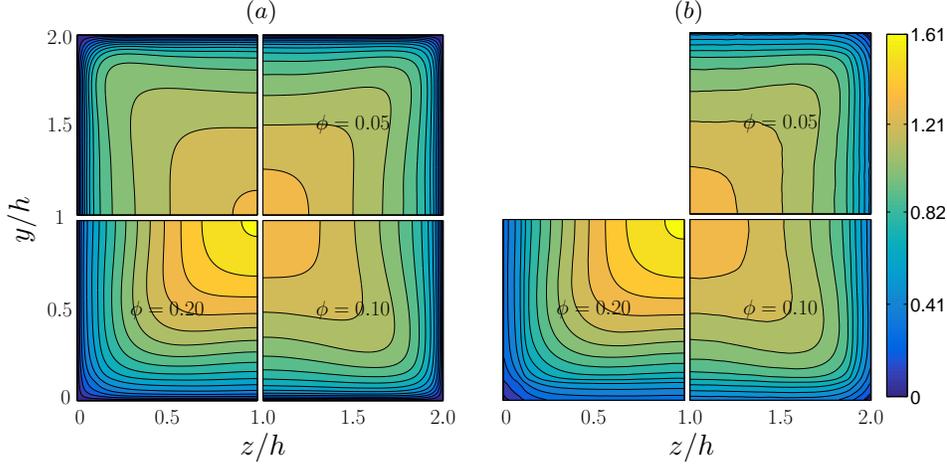
\begin{figure}
   \centering
   \includegraphics[width=0.41\textwidth]{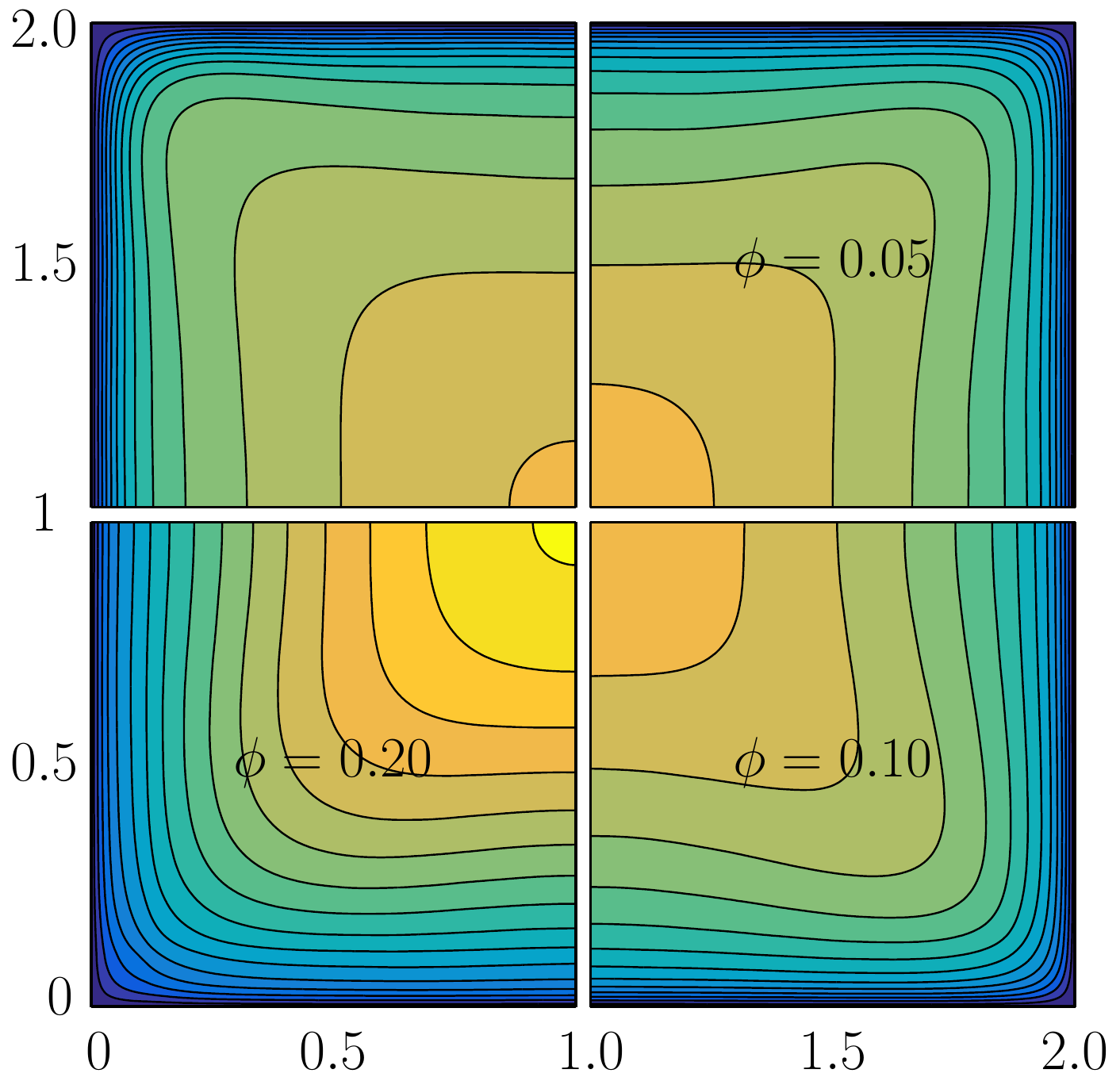}
   \put(-171,70){\rotatebox{90}{\large $y/h$}}
   \put(-83,-10){\large $z/h$}
   \put(-81,155){\footnotesize $(a)$}
   \hspace{0.5 cm}\includegraphics[width=0.445\textwidth]{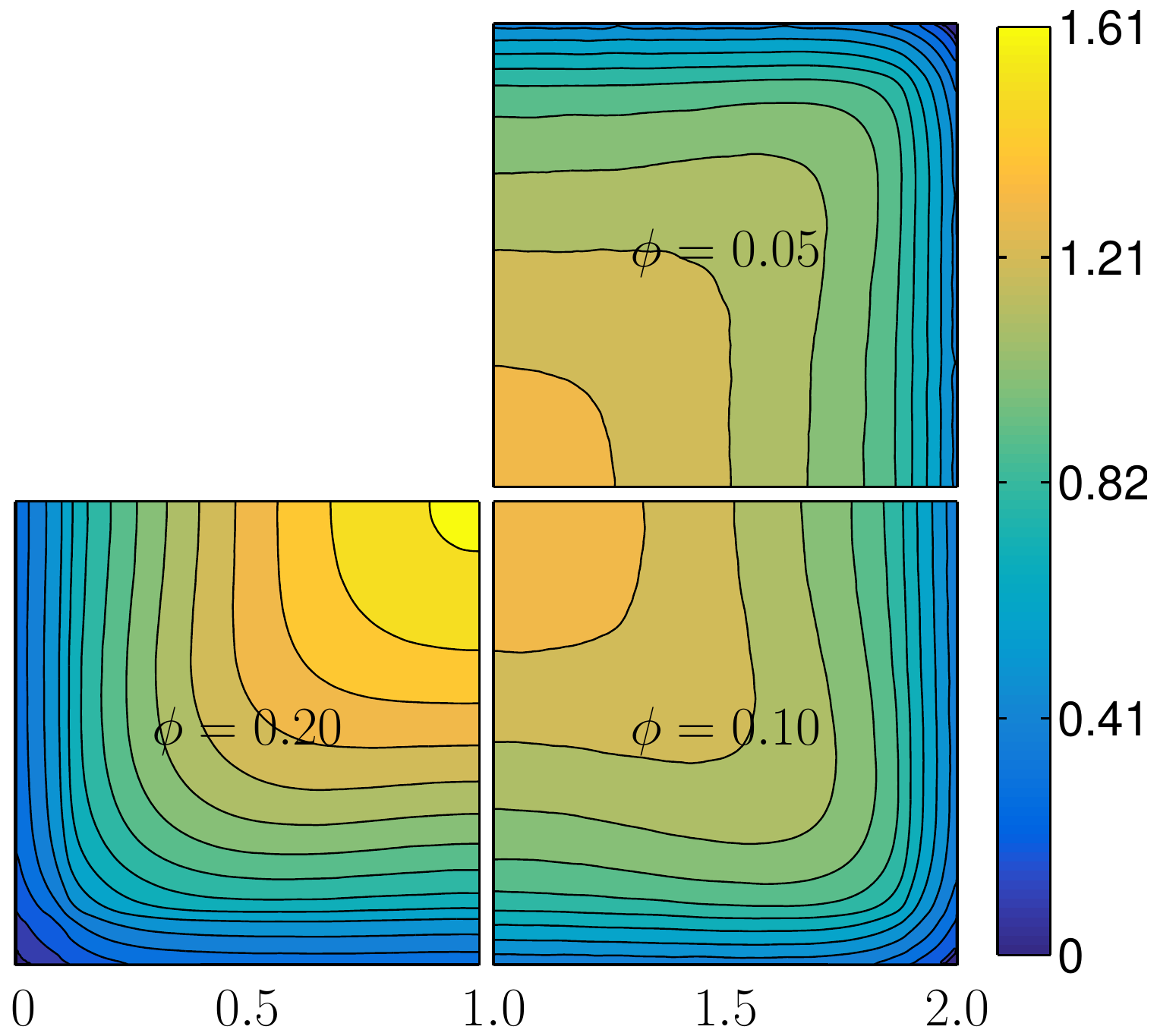}
   \put(-104,155){\footnotesize $(b)$}
   \put(-106,-10){\large $z/h$}
  \vspace{3pt}        
  \caption{Contours of streamwise mean fluid (a) and particle velocity (b) in outer units. In each 
figure, the top-left, top-right, bottom-left and bottom-right quadrants show the data for $\phi=0.0, 
0.05, 0.1$ and $0.2$.}
\label{fig:Umean}
\end{figure}

The streamwise mean fluid and particle velocities in outer units (i.e. normalized by the bulk velocity 
$U_b$), $U_{f/p}(y,z)$, are illustrated in figure~\ref{fig:Umean}(a,b) for all $\phi$. The contour 
plots are divided in four quadrants showing results for $\phi=0.0$ (top-left), $0.05$ (top-right), $0.1$ (bottom-right) and $0.2$ (bottom-left). The streamwise mean 
particle velocity contours closely resemble those of the fluid phase. In particular we observe that 
the maximum velocity at the core of the duct grows with $\phi$. The increase with $\phi$ is similar 
to that reported for turbulent channel flows \citep{picano2015}, except for $\phi=0.2$ where the 
increase of $U_{f/p}(y,z)$ in the duct core is substantially larger. 
We observe that the convexity of the mean velocity contours also increases 
with the volume fraction up to $\phi=0.1$. 
This is due to the increased intensity of secondary flows that convect mean 
velocity from regions of large shear along the walls towards regions of low shear along the corner 
bisectors \citep{prandtl,gessner,vinuesa14}. For $\phi=0.2$ the secondary flow intensity is 
substantially reduced and accordingly also the convexity of the contours of $U_{f/p}$ reduces.\\

Next, we show in figure~\ref{fig:Umean_bis}(a) and (b)  the streamwise mean fluid and particle velocity profiles 
along the wall-bisector in outer and inner units ($U_{f/p}(y)$ and $U_{f/p}^+$). The local value of 
the friction velocity at the wall-bisector, $U_*=\sqrt{\tau_{w,bis}/\rho_f}$, is used to normalize 
$U_{f/p}(y)$. Solid lines are used for $U_{f}(y)$ while symbols are used for $U_{p}(y)$. We observe 
that the mean velocity profiles of the two phase are almost perfectly overlapping at equal $\phi$, except 
very close to the walls ($y^+ \le 30$) where particles have a relative tangential motion (slip velocity). 
Note also that the mean particle velocity decreases with $\phi$ very close to the walls. 
We also observe that by increasing $\phi$, the profiles of $U_{f/p}(y)$ tend towards the laminar 
parabolic profile with lower velocity near the wall and larger velocity at the centerline, $y/h=1$. 
Concerning $U_{f/p}^+(y^+)$, we observe a progressive downward shift of the profiles with the volume 
fraction $\phi$ denoting a drag increase, at least up to $\phi=0.1$. 

The mean-velocity profiles still follow the log-law \citep{pope2000}
\begin{equation}
U_{f/p}^+(y^+) = \frac{1}{k} log(y^+) + B
\end{equation}
where $k$ is the von K\'arm\'an constant and $B$ is an additive coefficient. For the unladen case 
with $Re_b=4410$, \citet{gavrilakis1992} fitted the data between $y^+=30$ and $100$ to find $k=0.31$ 
and $B=3.9$. In the present simulations, the extent of the log-region is larger due to the higher 
bulk Reynolds number and we hence fit our data between $y^+=30$ and $140$. The results are reported 
in table \ref{tab:k_b}. For the unladen case we obtain results in agreement with those of 
\citet{gavrilakis1992} although our constant $B$ is slightly smaller. Increasing $\phi$, $k$ decreases and 
the additive constant $B$ decreases becoming negative at $\phi=0.1$. As 
shown by \citet{virk1975}, the reduction in $k$ should lead to drag reduction while the opposite 
effect is achieved by decreasing $B$. 

\begin{figure}
   \centering
   \includegraphics[width=0.42\textwidth]{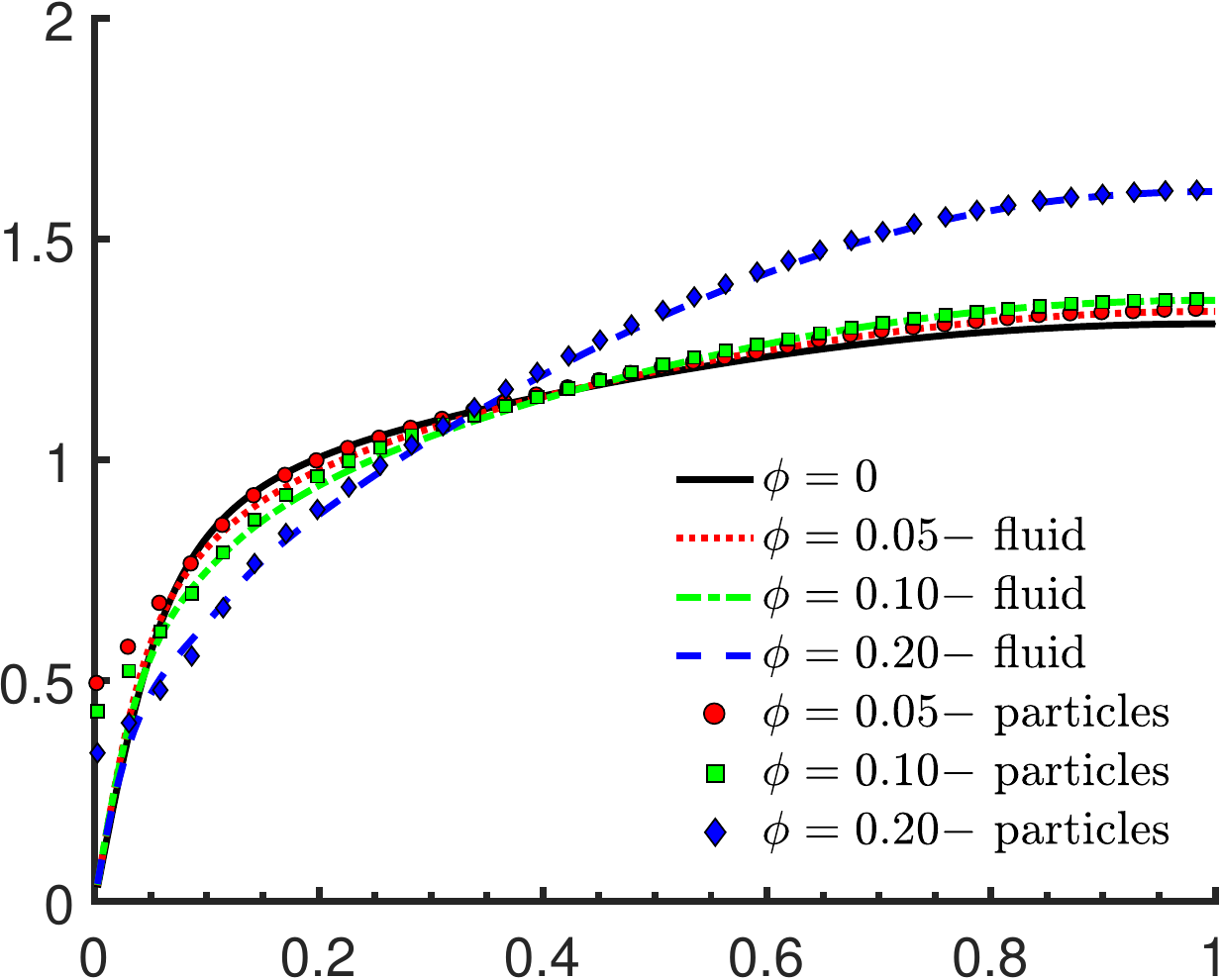}\hspace{1 cm}
   \includegraphics[width=0.42\textwidth]{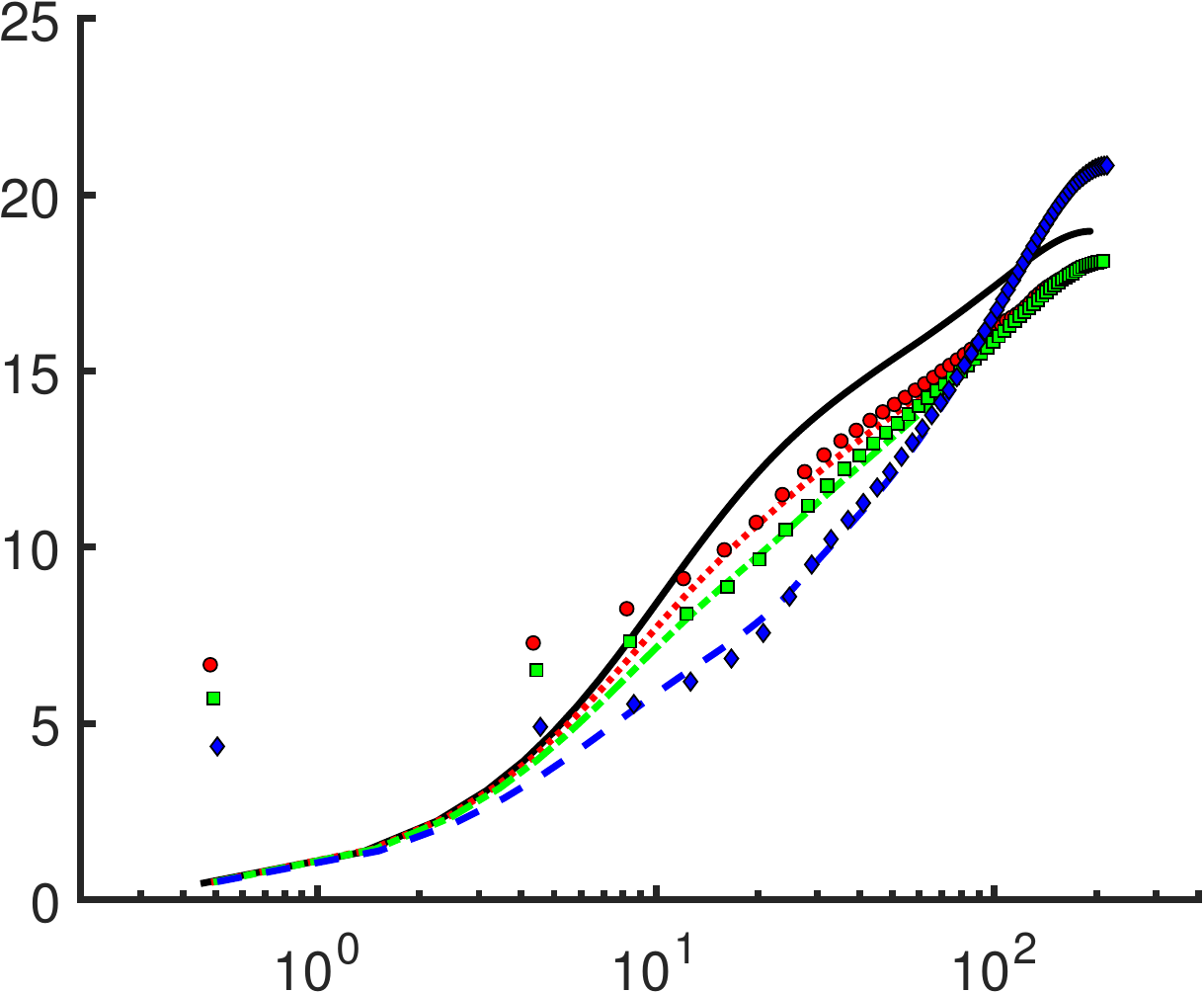}
   \put(-277,135){\footnotesize $(a)$}
   \put(-367,70){\rotatebox{90}{\large $U_{f/p}$}}
   \put(-182,70){\rotatebox{90}{\large $U_{f/p}^+$}}
   \put(-78,-10){{\large $y^+$}}
   \put(-278,-10){{\large $y/h$}}
   \put(-75,134){\footnotesize $(b)$}
  \vspace{3pt}        
  \caption{Streamwise mean fluid and particle velocity along the wall-bisector at $z/h=1$ in outer 
(a) and inner units (b). Lines are used for fluid velocity profiles while symbols are used for 
particle velocity profiles.}
\label{fig:Umean_bis}
\end{figure}
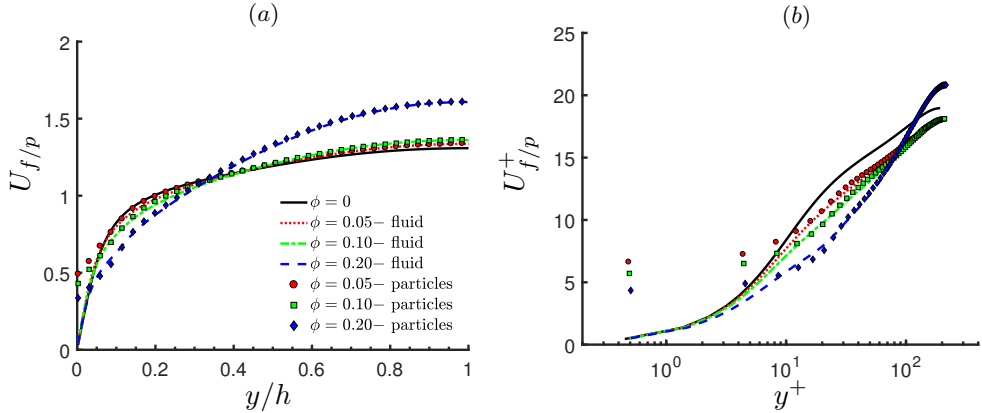
\begin{table}
  \begin{center}
\def~{\hphantom{0}}
  \begin{tabular}{lccccc}\\
\,\,\,\,\,\,\,\,\,  \,Case & $\phi=0.0 $  & $\phi=0.05$ & $\phi=0.1$  & $\phi=0.2 \,\,\,$ \\[3pt]
       $ \,\,\,\,\,\,\,\,\,\,\,\,\,\, \, \,  \,k$  & $0.31$ & $0.30$ & $0.26 $ & $0.16$ \\[3pt]
       $ \,\,\,\,\,\,\,\,\,\,\,\,\,\, \, \,  \,B$  & $3.1$ & $0.9$ & $-1.6$ & $-11.5$ \\[3pt]
       $ \,\,\,\,\,\,\,\,\,\,\,\,\,\, \, \,  \,Re_{\tau,bis}$  & $193$ & $208$ & $211$ & $217$ \\[3pt]
       $ \,\,\,\,\,\,\,\,\,\,\,\,\,\, \, \,  \,Re_{\tau,mean}$  & $185$ & $202$ & $210$ & $207$ \\[3pt]
       $ \,\,\,\,\,\,\,\,\,\,\,\,\,\, \, \,  \,Re_{\tau,2D}$  & $180$ & $195$ & $204$ & $216$ \\[3pt]
  \end{tabular}
  \caption{The von K\'arm\'an constant and additive constant $B$ of the log-law at the 
wall-bisector estimated from the present simulations for different volume fractions $\phi$. Here 
$k$ and $B$ have been fitted in the range $y^+ \in [30,140]$. The friction Reynolds number 
calculated at the wall-bisector $Re_{\tau,bis}$, the mean friction Reynolds number 
$Re_{\tau,mean}$ (based on $\langle \tau_w \rangle$, averaged over the walls), and the corresponding friction Reynolds number 
found by \citet{picano2015} for turbulent channel flow, $Re_{\tau,2D}$, are also reported.}
 \label{tab:k_b}
 \end{center}
\end{table}

\begin{figure}
   \centering
   \includegraphics[width=0.42\textwidth]{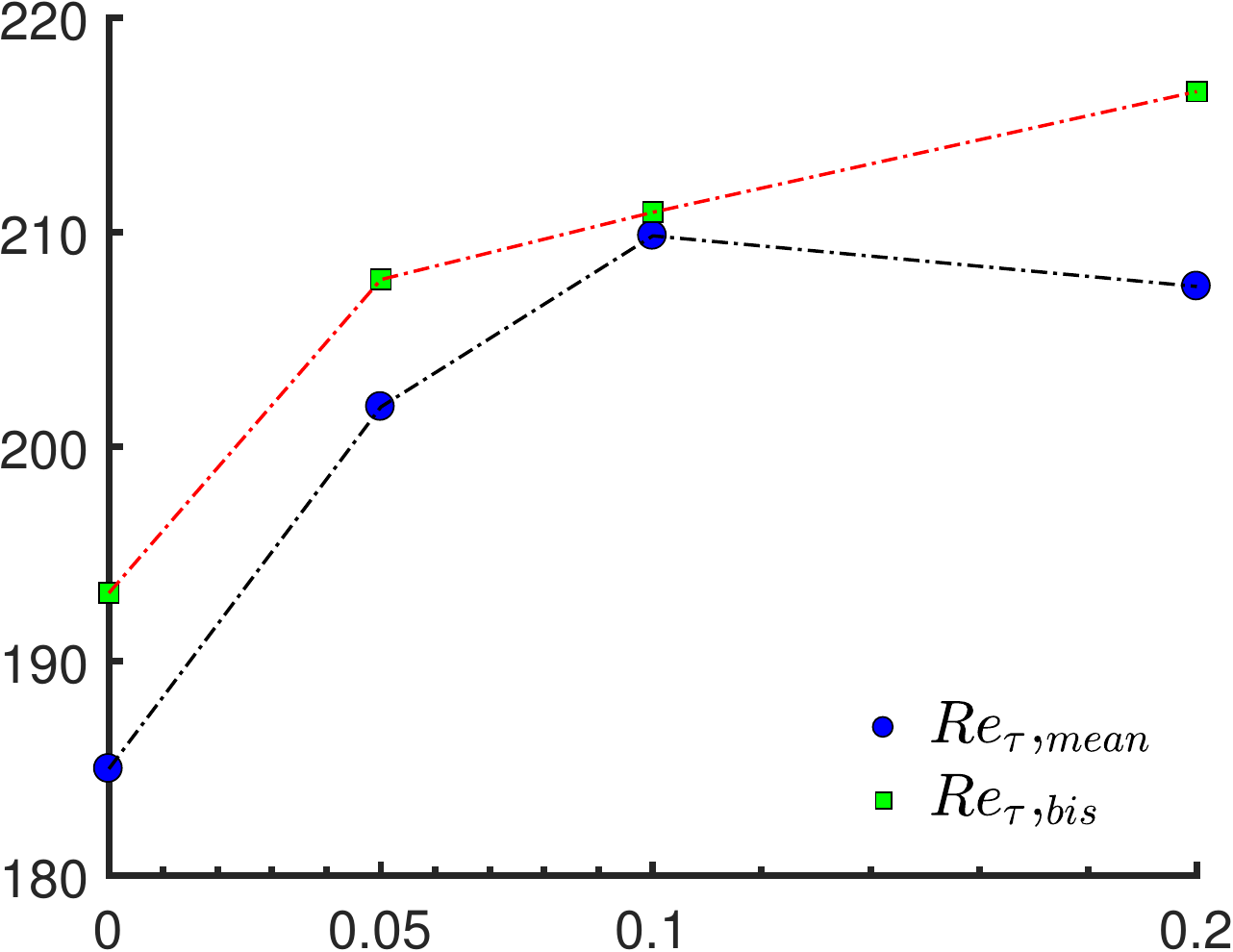}
   \put(-176,64){\rotatebox{90}{\large $Re_{\tau}$}}
 \hspace{0.6 cm}  \includegraphics[width=0.42\textwidth]{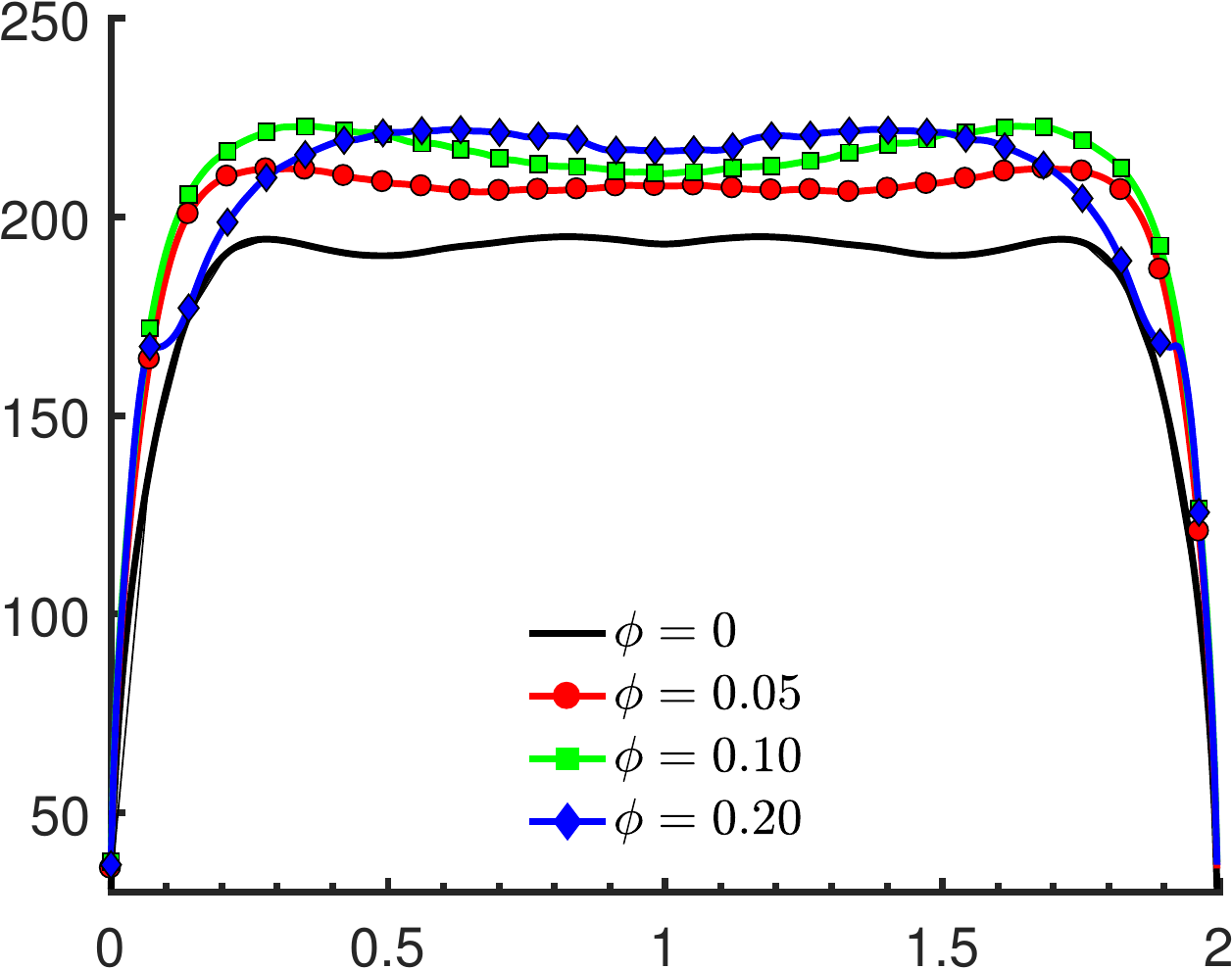}
   \put(-258,130){\footnotesize $(a)$}
   \put(-258,-10){\large $\phi$}
   \put(-85,130){\footnotesize $(b)$}   
   \put(-80,-10){\large $z/h$}   
   \put(-176,64){\rotatebox{90}{\large $Re_{\tau}$}}
  \vspace{3pt}        
  \caption{(a) Mean friction Reynolds number $Re_{\tau,mean}$ and friction Reynolds number at the 
wall bisector $Re_{\tau,bis}$ for different volume fractions $\phi$. (b) Profile of $Re_{\tau}$ 
along the duct wall.}
\label{fig:frRe}
\end{figure}

The combinations of $k$ and $B$ pertaining our simulations correspond to an overall drag increase at the 
wall-bisector for increasing $\phi$ \cite[as also shown for channel flows of comparable $h/a$ and 
$Re_b$ by][]{picano2015}. 
To show this, we report in figure~\ref{fig:frRe}(a) and in 
table \ref{tab:k_b} the friction Reynolds number calculated at the wall-bisector, $Re_{\tau,bis}$ as 
function of the volume fraction $\phi$. The results of \citet{picano2015} are also reported for 
comparison in table~\ref{tab:k_b}. We see that although the initial value of $Re_{\tau,bis}$ for 
$\phi=0.0$ is substantially larger than that of the corresponding channel flow at $Re_b=5600$, the 
increase with the volume fraction is smaller in the duct. Indeed for $\phi=0.2$ we find that $Re_{\tau,bis}\simeq Re_{\tau,2D}$. 
The increase in drag is hence less pronounced in square duct flow compared to the ideal channel flow case.\\
It is also interesting to observe the behavior of the mean friction Reynolds number 
$Re_{\tau,mean}$  as function of $\phi$, as this directly relates to the overall pressure drop along the duct. 
From figure~\ref{fig:frRe}(a) 
and table \ref{tab:k_b} we see that it strongly increases with the volume fraction up to 
$\phi=0.1$, while a reduction of $Re_{\tau,mean}$ takes place when $\phi$ is further increased. Observing the profiles of 
$Re_{\tau}$ along one wall (see figure~\ref{fig:frRe}b) we note that the friction Reynolds number 
increases with $\phi$, especially towards the corners. For $\phi=0.2$, however, the profile exhibits 
a sharp change at about $z/h = 0.1 \sim 2a$, and the maxima move towards the wall-bisector ($z/h \sim 
0.65$). This is probably due to the clear change in local particle concentration in the cross-section 
as we will explain in the following.

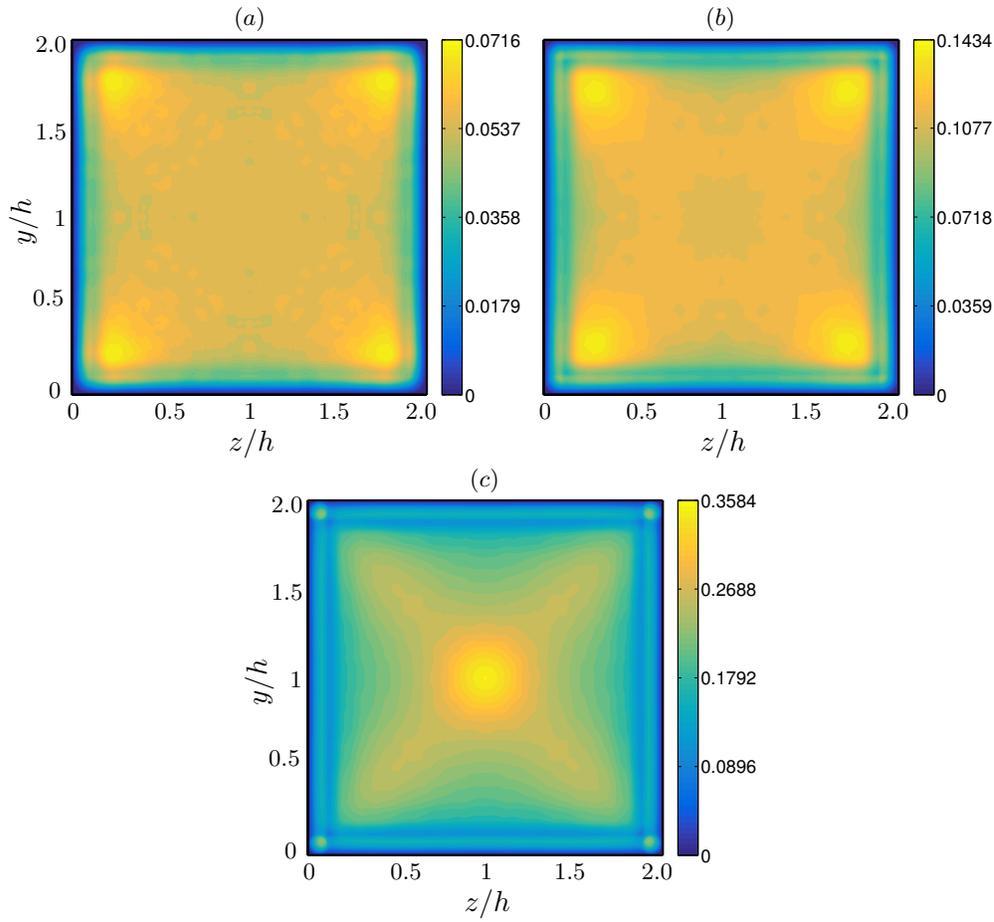
\begin{figure}
   \centering
   \includegraphics[width=.45\textwidth]{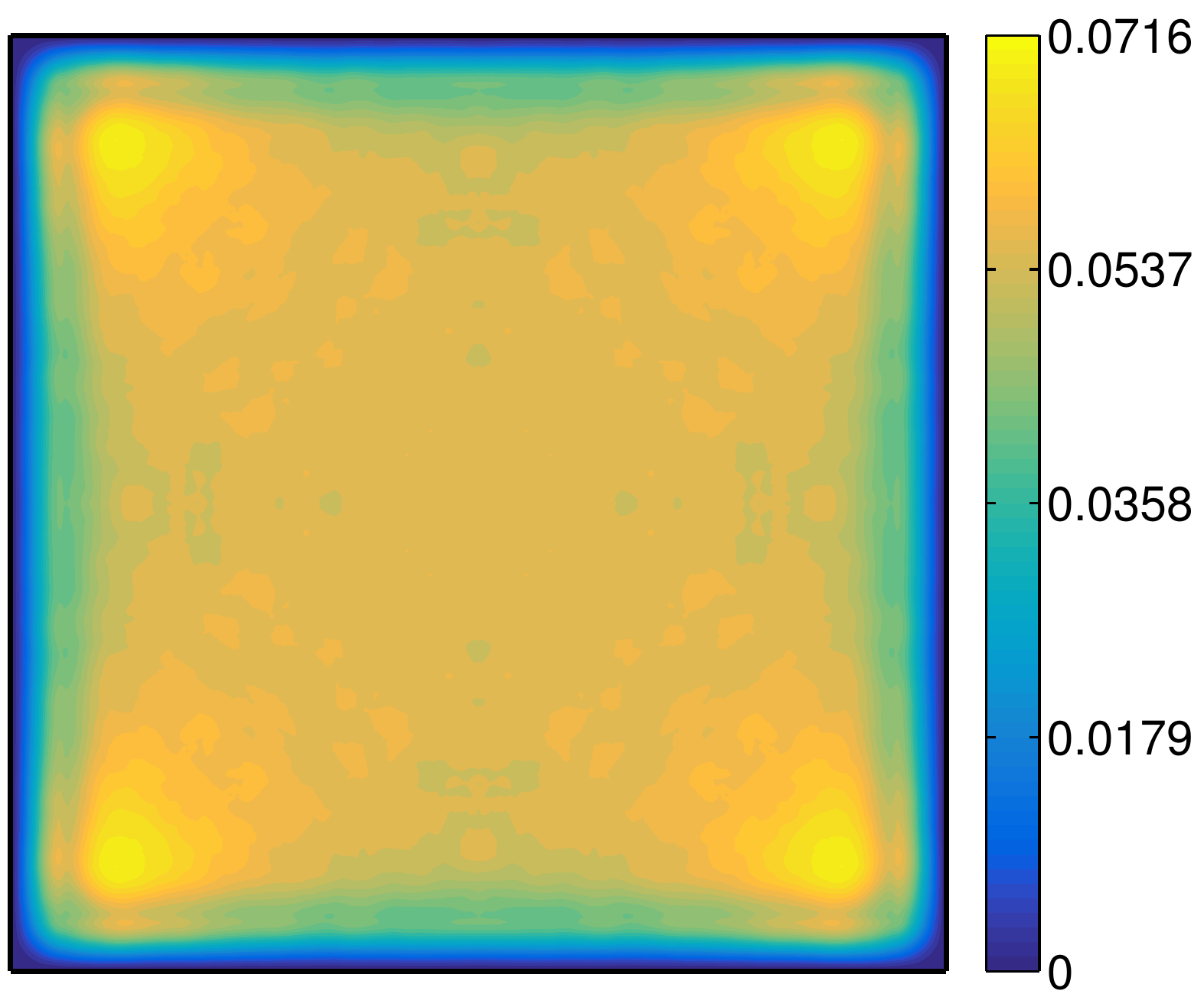}\hspace{3cm}
   \put(-110,143){\footnotesize $(a)$}
   \put(-180,2) {\footnotesize $0$}
   \put(-186,37) {\footnotesize $0.5$}
   \put(-178,67) {\footnotesize $1$}
   \put(-185,100) {\footnotesize $1.5$}
   \put(-185,133) {\footnotesize $2.0$}
   \put(-173,-6) {\footnotesize $0$}
   \put(-140,-6) {\footnotesize $0.5$}
   \put(-106,-6) {\footnotesize $1$}
   \put(-78,-6) {\footnotesize $1.5$}
   \put(-45,-6) {\footnotesize $2.0$}
   \put(-195,60){\rotatebox{90}{\large $y/h$}}
   \put(-112,-18) {\large $z/h$}
   \hspace{0.2 cm}\includegraphics[width=.45\textwidth]{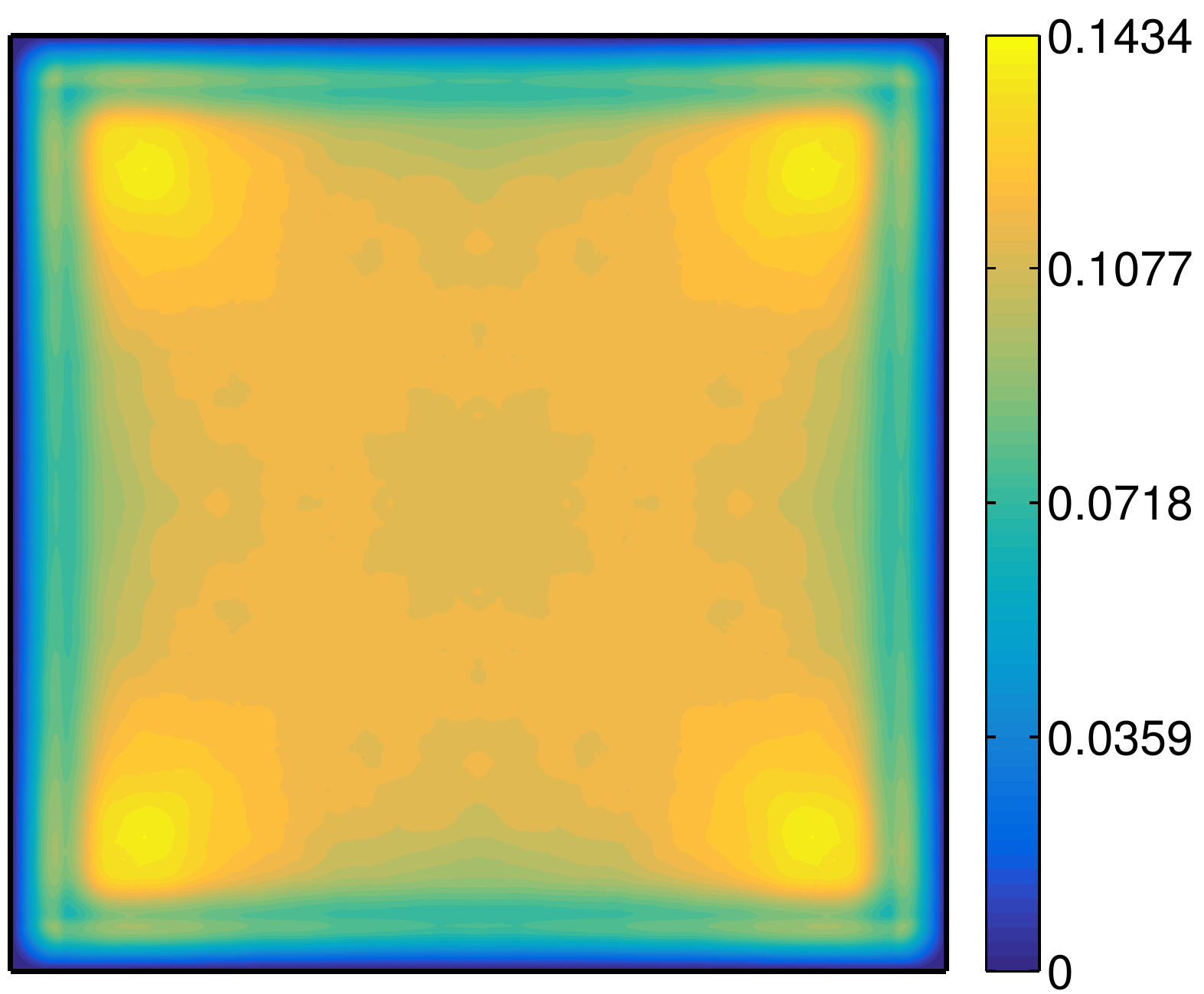}
   \put(-110,143){\footnotesize $(b)$}
   \put(-173,-6) {\footnotesize $0$}
   \put(-140,-6) {\footnotesize $0.5$}
   \put(-106,-6) {\footnotesize $1$}
   \put(-78,-6) {\footnotesize $1.5$}
   \put(-45,-6) {\footnotesize $2.0$}
   \put(-112,-18) {\large $z/h$}\\
   \vspace{0.1 cm}\includegraphics[width=.45\textwidth]{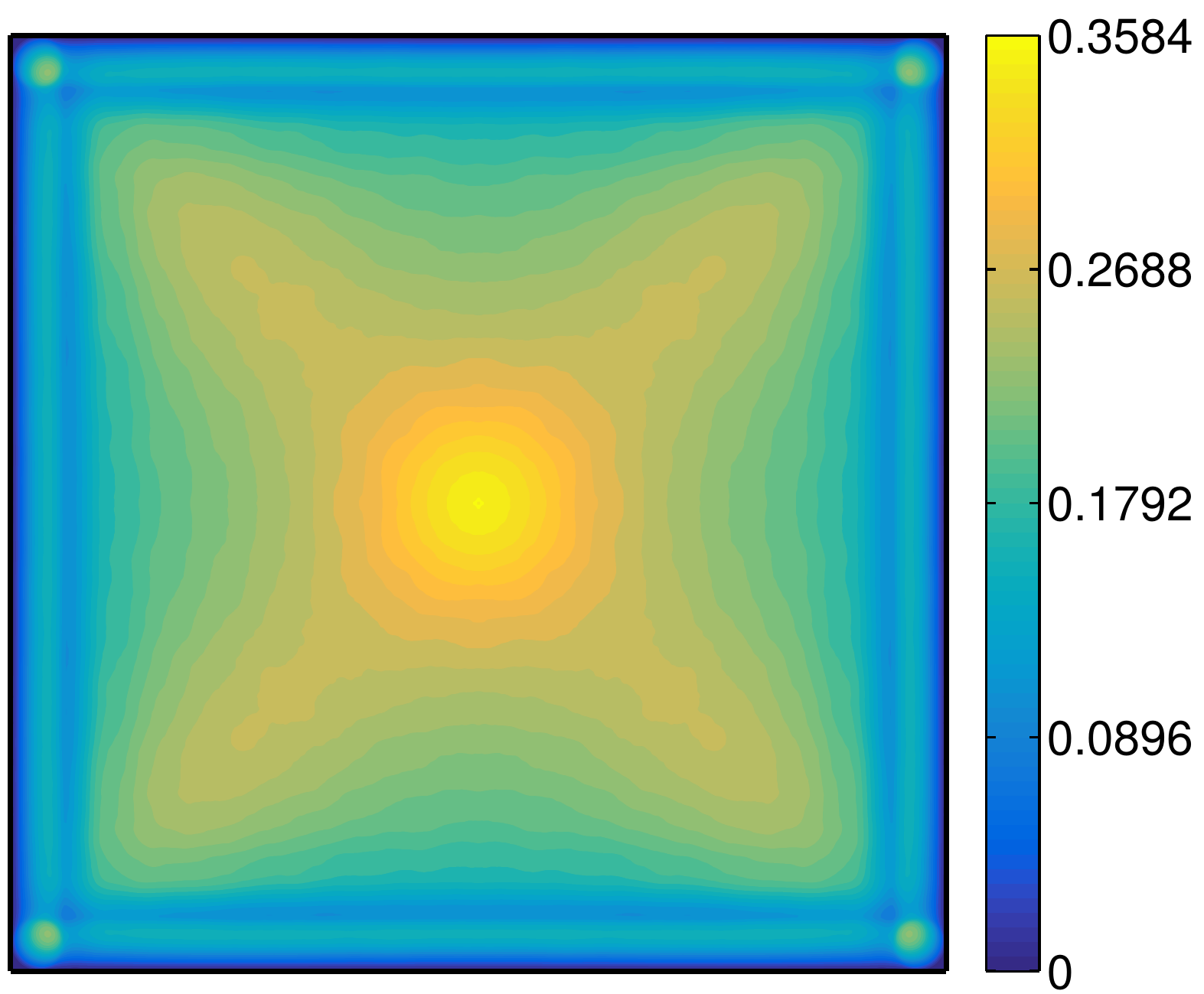}
   \put(-110,143){\footnotesize $(c)$}
   \put(-180,2) {\footnotesize $0$}
   \put(-186,37) {\footnotesize $0.5$}
   \put(-178,67) {\footnotesize $1$}
   \put(-185,100) {\footnotesize $1.5$}
   \put(-185,133) {\footnotesize $2.0$}
   \put(-173,-6) {\footnotesize $0$}
   \put(-140,-6) {\footnotesize $0.5$}
   \put(-106,-6) {\footnotesize $1$}
   \put(-78,-6) {\footnotesize $1.5$}
   \put(-45,-6) {\footnotesize $2.0$}
   \put(-195,60){\rotatebox{90}{\large $y/h$}}
   \put(-112,-18) {\large $z/h$}
  \vspace{3pt}
  \caption{Mean particle concentration $\Phi(y,z)$ in the duct cross-section for $\phi=0.05$ 
(a), $\phi=0.1$ (b) and $\phi=0.2$ (c).}
\label{fig:conc}
\end{figure}

\begin{figure}
   \centering
   \includegraphics[width=0.42\textwidth]{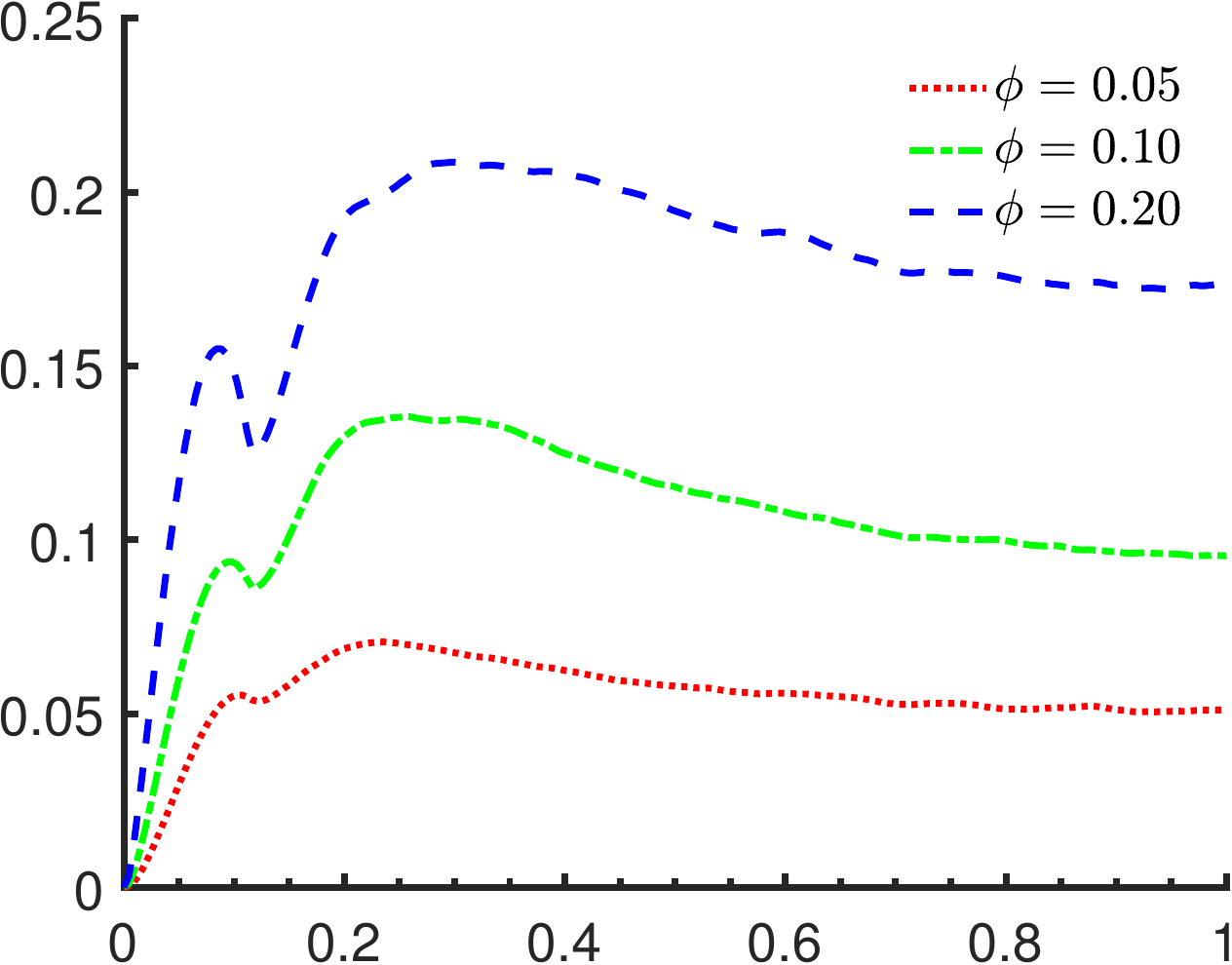}\hspace{1.1 cm}
   \includegraphics[width=0.42\textwidth]{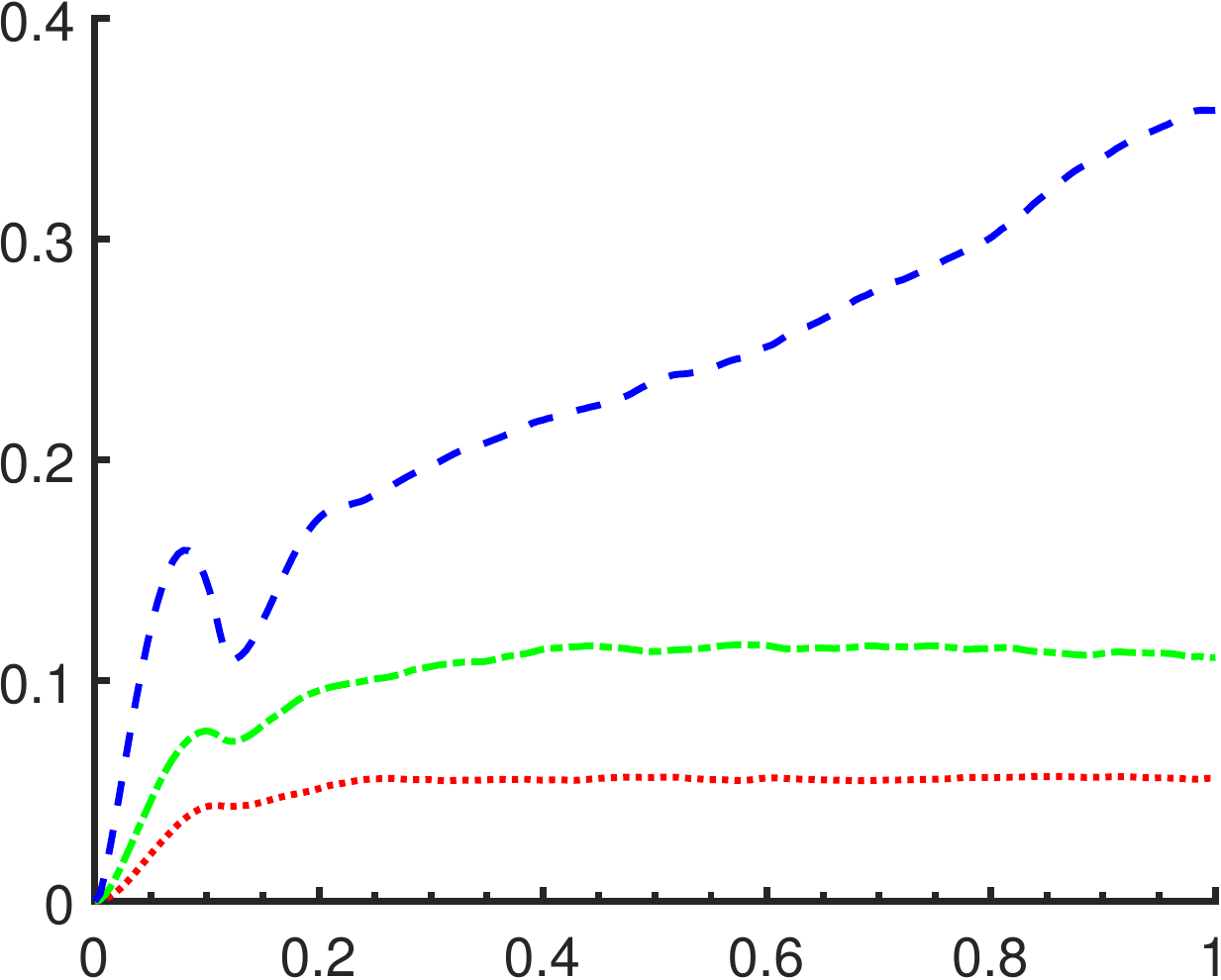}
   \put(-278,130){\footnotesize $(a)$}   
   \put(-85,130){\footnotesize $(b)$}   
   \put(-278,-10){\large $y/h$}   
   \put(-82,-10){\large $y/h$}   
   \put(-376,74){\rotatebox{90}{\large $\Phi$}}
   \put(-176,74){\rotatebox{90}{\large $\Phi$}}
  \vspace{3pt}        
  \caption{Mean particle concentration along a line at $z/h=0.2$ and at the wall-bisector, $z/h=1$.}
\label{fig:conc2}
\end{figure}

\begin{figure}
   \centering
   \includegraphics[width=0.47\textwidth]{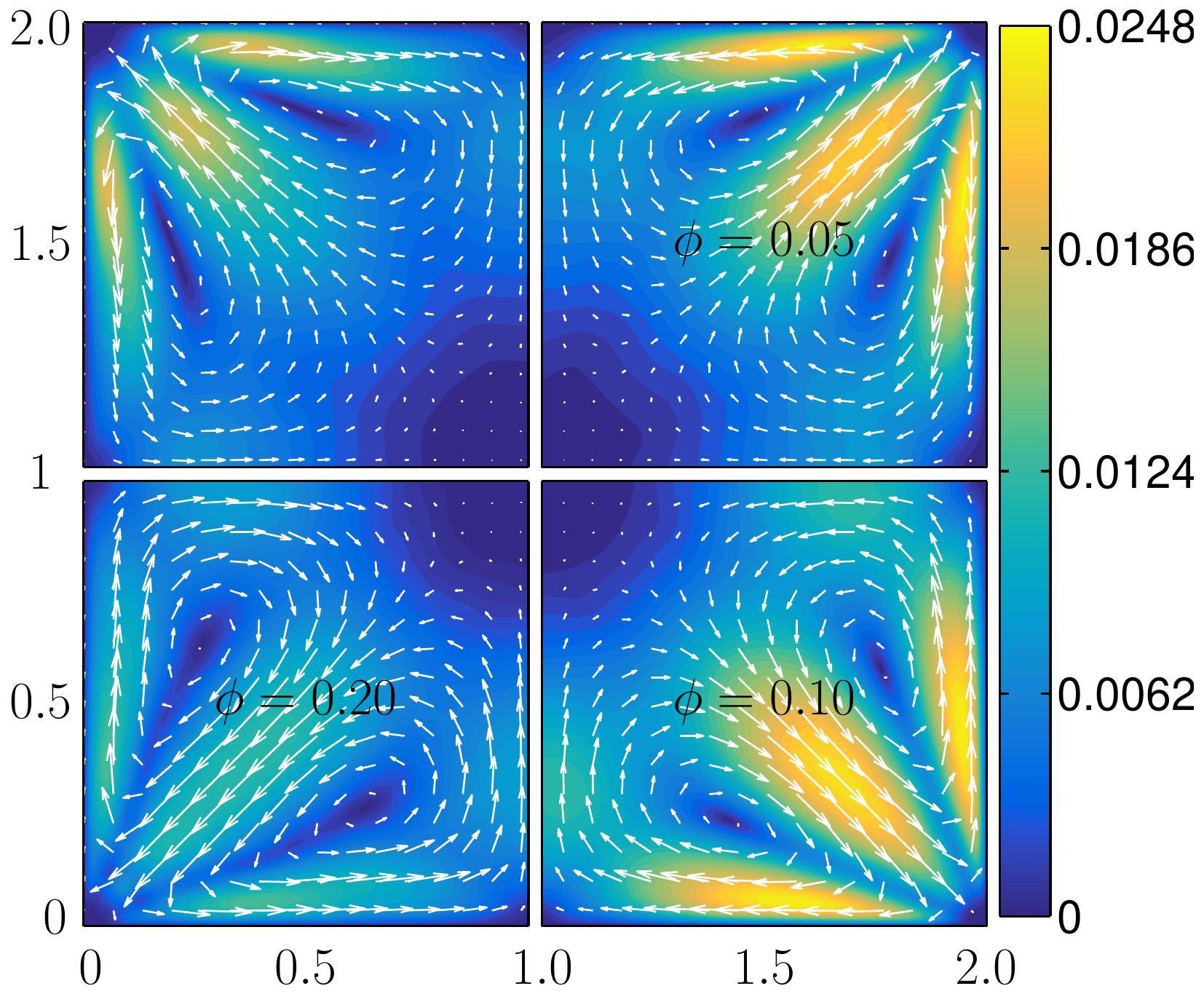}
   \put(-191,68){\rotatebox{90}{\large $y/h$}}
   \put(-107,-10){\large $z/h$}
   \put(-106,150){\footnotesize $(a)$}
   \includegraphics[width=0.445\textwidth]{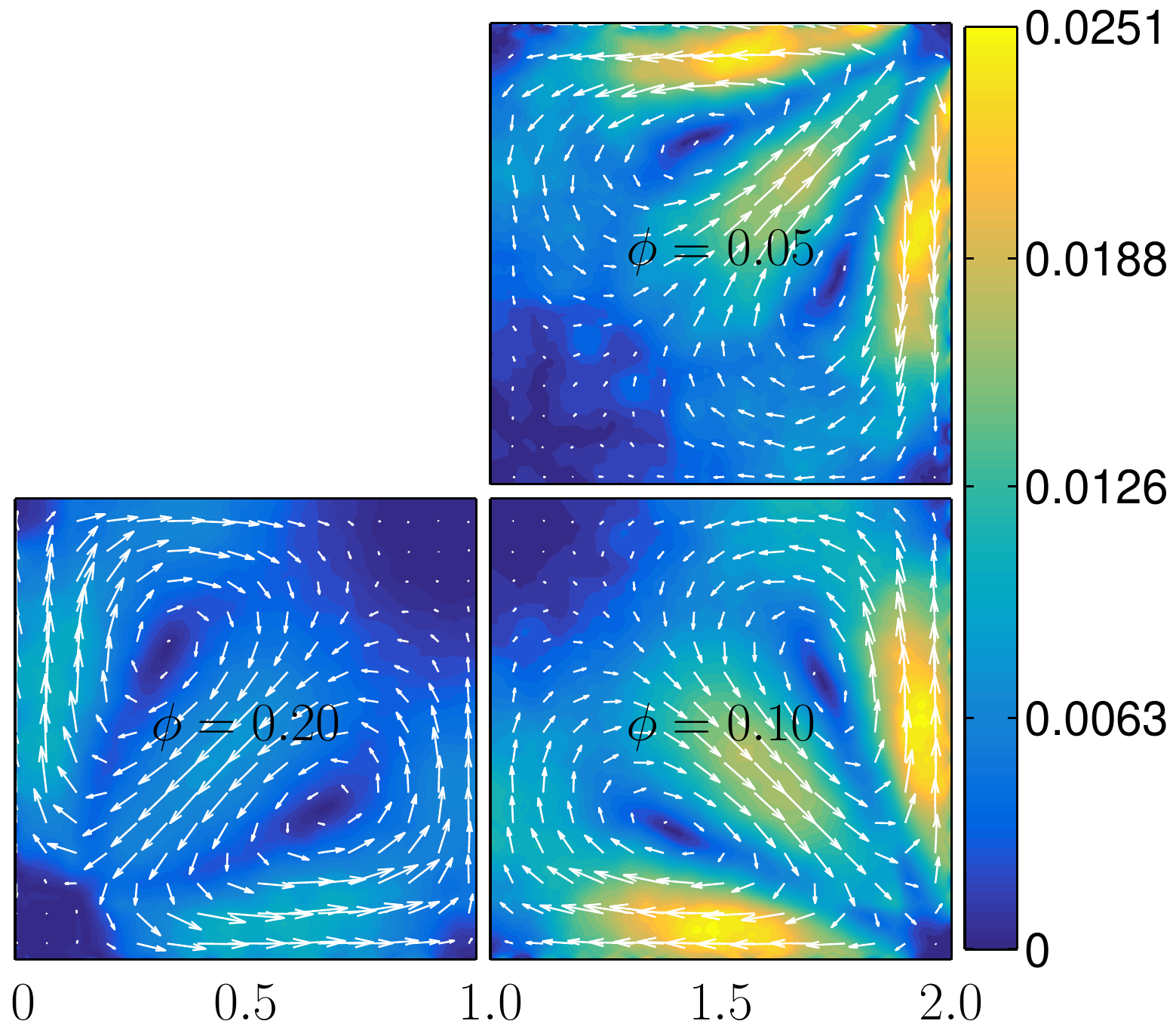}
   \put(-106,150){\footnotesize $(b)$}
   \put(-107,-10){\large $z/h$}
  \vspace{3pt}        
  \caption{Contours and vector fields of the secondary flow velocity $\sqrt{V_{f/p}^2+W_{f/p}^2}$ of the fluid (a) and solid phases (b).}
\label{fig:cs}
\end{figure}

The mean particle concentration over the duct cross-section $\Phi(y,z)$ is displayed 
in figure~\ref{fig:conc} for all $\phi$, whereas the particle concentration along the 
wall-bisector ($z/h=1$) and along a segment at $z/h=0.2$ is shown in figure~\ref{fig:conc2}. 
Finally, we report in figure~\ref{fig:cs} the secondary (cross-stream) velocities of both phases, defined as 
$\sqrt{V_{f/p}^2+W_{f/p}^2}$. We shall now discuss these 3 figures together.

\begin{figure}
   \centering
   \includegraphics[width=0.495\textwidth]{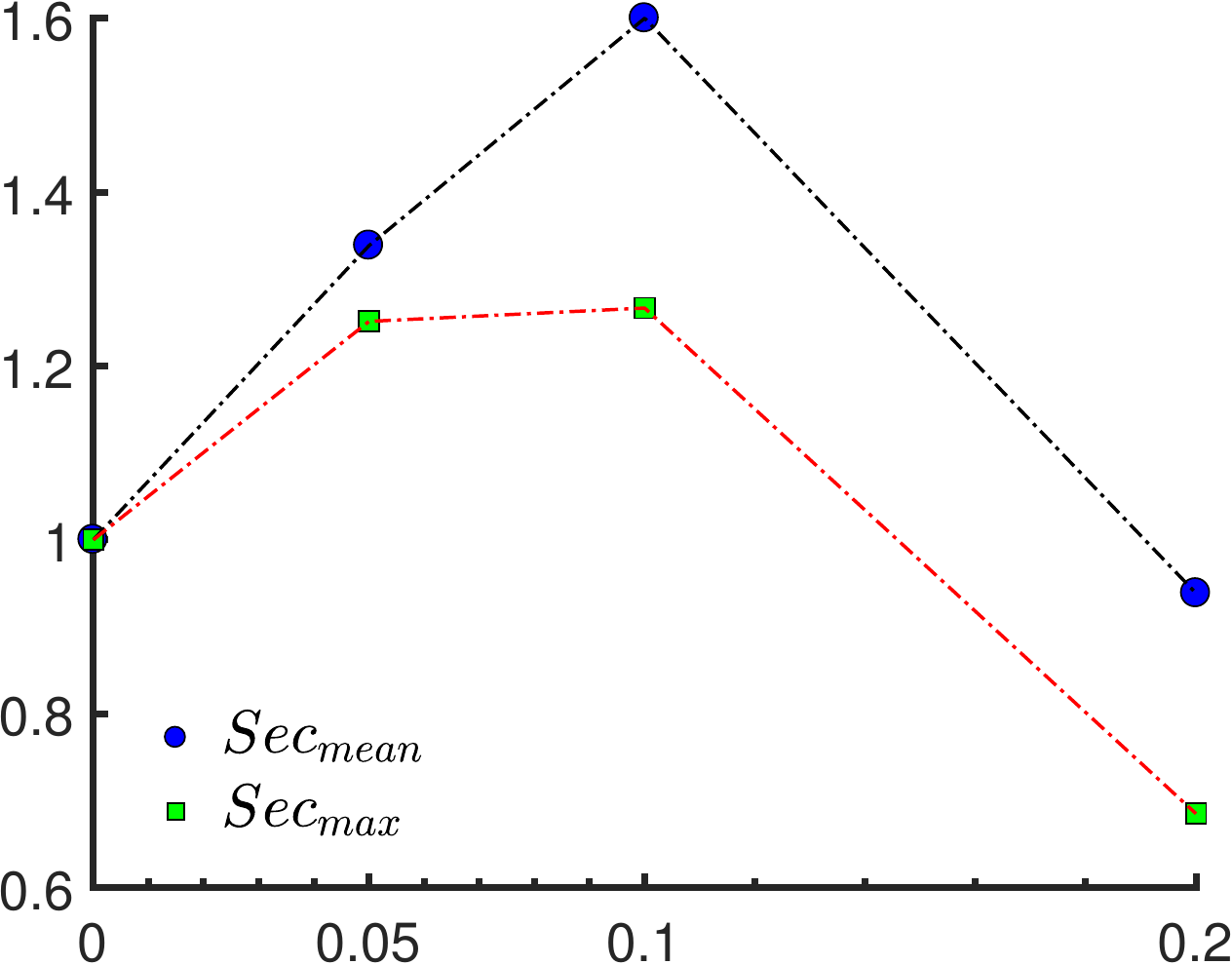}
   \put(-90,-10){\large $\phi$}
   \put(-210,30){\rotatebox{90}{\tiny $(\overline{\sqrt{V_f^2+W_f^2}})$/$(\overline{\sqrt{V_f^2+W_f^2}})_{0\%}$}}
\caption{Mean and maximum value of the secondary flow velocity of the fluid phase as function of the 
solid volume fraction $\phi$. Results are normalized by the values of the single-phase case.}
\label{fig:cs2}
\end{figure}

Two interesting observations are deduced from figure~\ref{fig:conc}: i) particle layers form close to the walls, 
and ii) for $\phi=0.05$ and $0.1$, the local particle concentration $\Phi(y,z)$ is higher close to the duct corners. 
We have recently reported a similar result for laminar duct flow at $Re_b=550$ 
and same duct-to-particle size ratio, $h/a=18$, and volume fractions, $\phi$ \citep{kazerooni}. At those $Re_b$ and 
$h/a$, particles undergo an inertial migration towards the walls and especially towards the corners, 
while the duct core is fully depleted of particles. Clearly, turbulence enhances mixing and the 
depletion of particles at the duct core disappears. We believe that the higher particle concentration 
close to the corners is here related to the interaction between particles and secondary motion. 
Indeed, $\Phi(y,z)$ is lower at the wall-bisector, where the secondary flow is directed away from the 
wall, and higher where the cross-stream fluid velocity is directed towards the corners (see 
figure~\ref{fig:cs}a). It is also interesting to observe that the presence of particles further 
enhances the fluid secondary flow around the corners for $\phi \le 0.1$. This can be easily 
seen from figure~\ref{fig:cs2}, where both the maximum and the mean value of the secondary fluid 
velocity are shown as function of the volume fraction $\phi$. The maximum value of the secondary cross-stream
velocity increases from about $2\%$ to $2.5\%$ of $U_b$. 
The relative increase of the mean $\sqrt{V_{f}^2+W_{f}^2}$ in comparison to the unladen case, is even larger 
than the increase in the maximum value at equal $\phi$. 
Similarly, in laminar ducts, as particles migrate towards the corners, secondary flows are generated. 


As shown in figure~\ref{fig:conc}, the particle concentration close to the corners increases with the volume fraction. However, 
the mean particle distribution in the cross-section changes at the highest volume fraction considered, $\phi=0.2$: the highest 
values of $\Phi(y,z)$ are now found at the center of the duct (see figure~\ref{fig:conc2}b). This is not the case in 
turbulent channel flows and hence it can be related to the additional confinement of the suspension given by the 
lateral walls. As previously mentioned, the streamwise mean fluid and particle velocities are also 
substantially higher at the duct center for $\phi=0.2$. 

To better quantify this effect, we analyse the numerical data by  
\citet{picano2015} and calculate the number of particles crossing the spanwise periodic boundaries 
per unit time $h/U_b$. For $\phi=20\%$, we find that in 1 unit of $h/U_b$ approximately $1\%$ of the total number of particles  
cross the lateral boundaries. Inhibiting this lateral migration with lateral walls has therefore important consequences on the flow structure. 
We also calculated the probability density function of the particle centroid position along the spanwise direction. It is found 
that while the mean position is always at the channel center, $z/h=1$, its variance grows in time as in a diffusive process. This is clearly 
not the case in a laterally confined geometry.\\
Concerning the secondary fluid velocity (figure~\ref{fig:cs}a), we note that for $\phi=0.2$ both the maximum and the mean 
values decrease below those of the unladen case. The presence of the solid phase leads to an 
increase of the secondary fluid velocity, which however saturates for volume fractions between $0.1$ and $0.2$. 
As shown by these mean velocity profiles, the turbulence activity is substantially reduced at $\phi=
0.2$. 

Vector and contour plots of the secondary motions of the solid phase are reported in figure~\ref{fig:cs}(b). 
These closely resemble those of the fluid phase. However, the velocities at the corner-bisectors are almost 
half of those pertaining the fluid phase (in agreement with the fact that the particle concentration is high at 
this locations). Conversely, the cross-stream particle velocity 
 is higher with respect to that of the fluid phase at the walls, close to the corners. 
For $\phi=0.2$ we have previously noticed that along the duct walls, the highest particle concentration $\Phi(y,z)$ is 
exactly at the corners. In agreement, we also observe from figure~\ref{fig:cs}(b) that the 
secondary particle velocity is negligible at these locations (i.e. particles tend to stay at these locations for long times).

\subsection{Velocity fluctuations}

\begin{figure}
   \centering
   \includegraphics[width=0.47\textwidth]{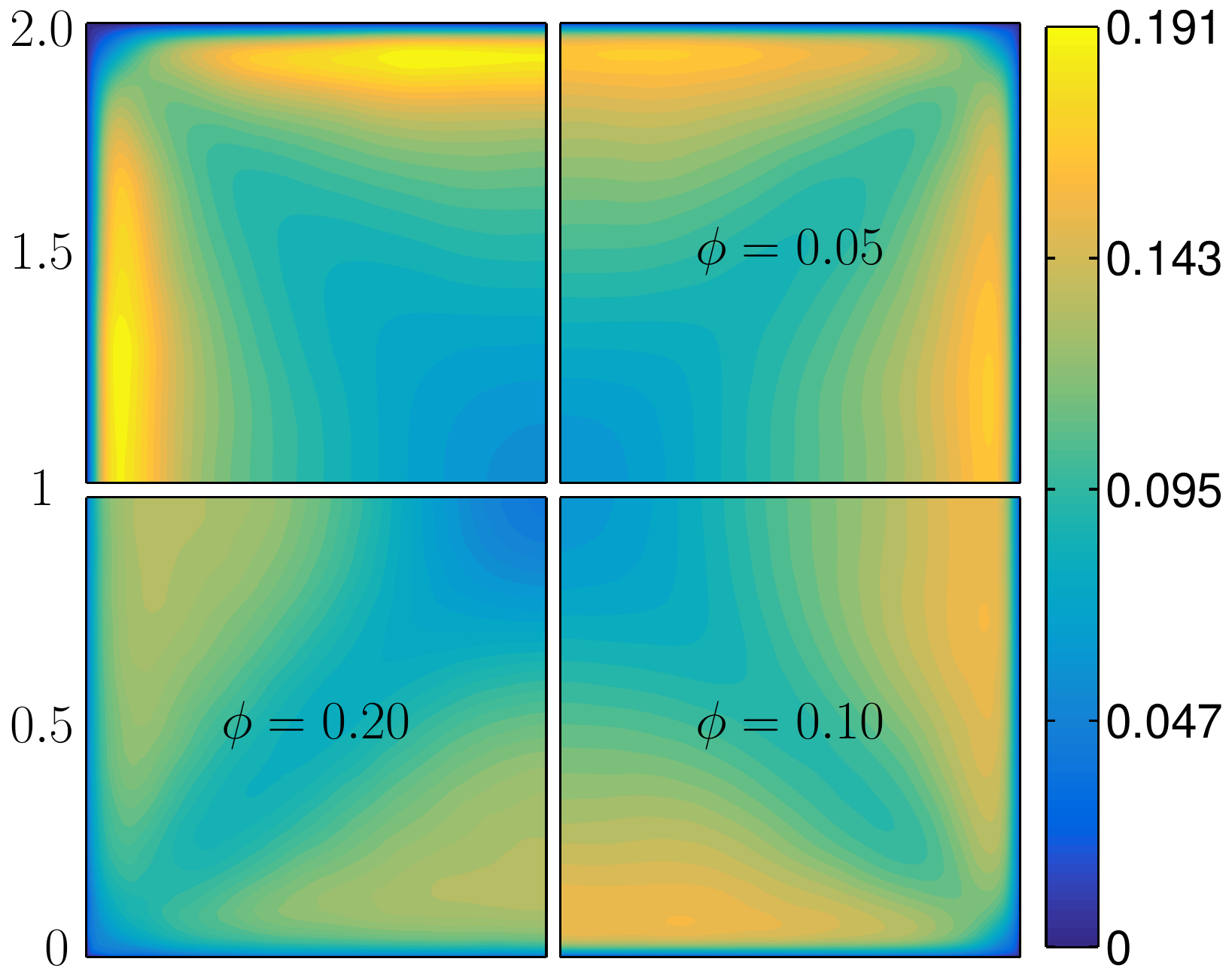}
   \put(-191,61){\rotatebox{90}{\large $y/h$}}
   \put(-105,145){\footnotesize $(a)$}
   \includegraphics[width=0.445\textwidth]{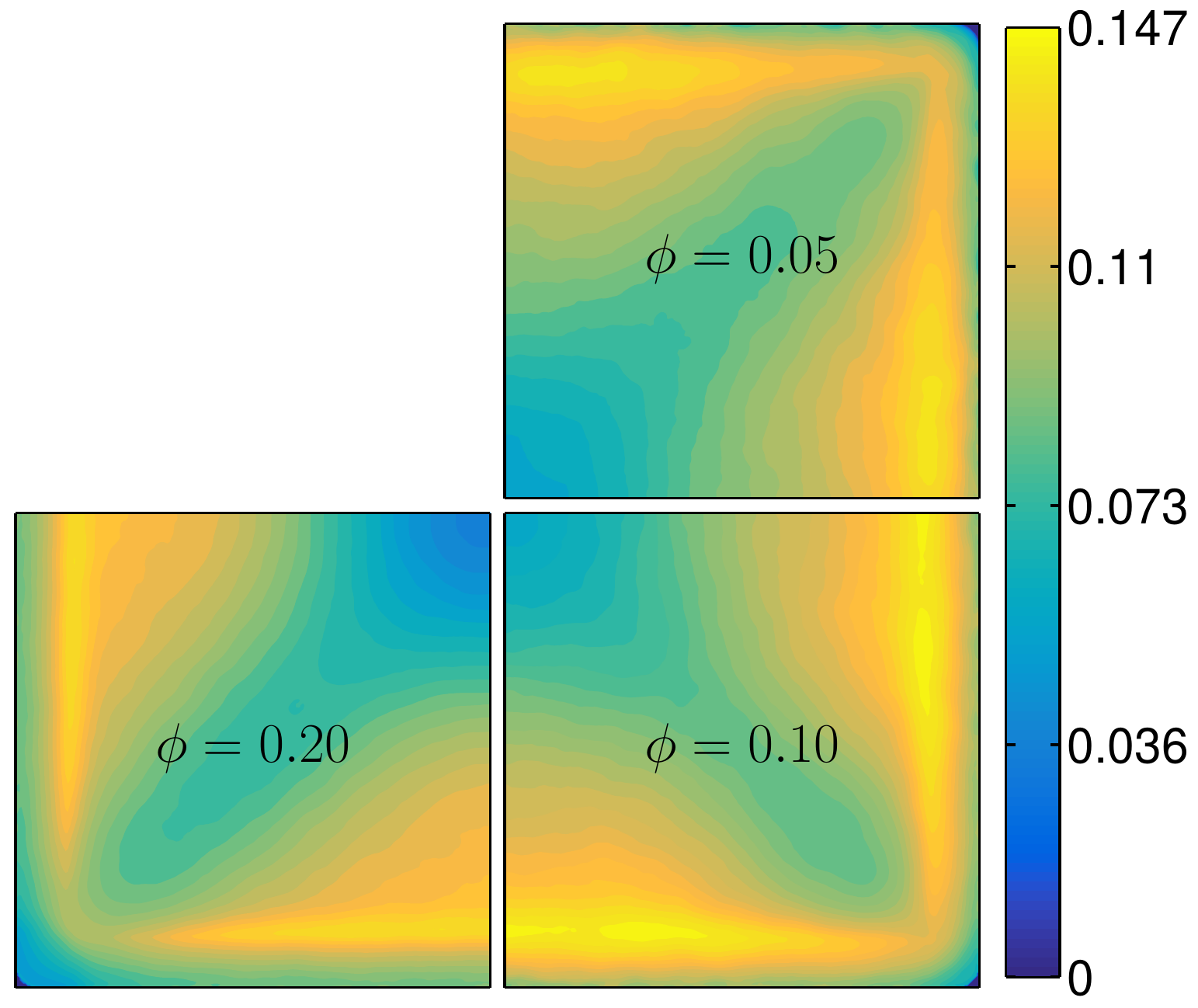}
   \put(-105,145){\footnotesize $(b)$}

   \vspace{0.5cm} \includegraphics[width=0.47\textwidth]{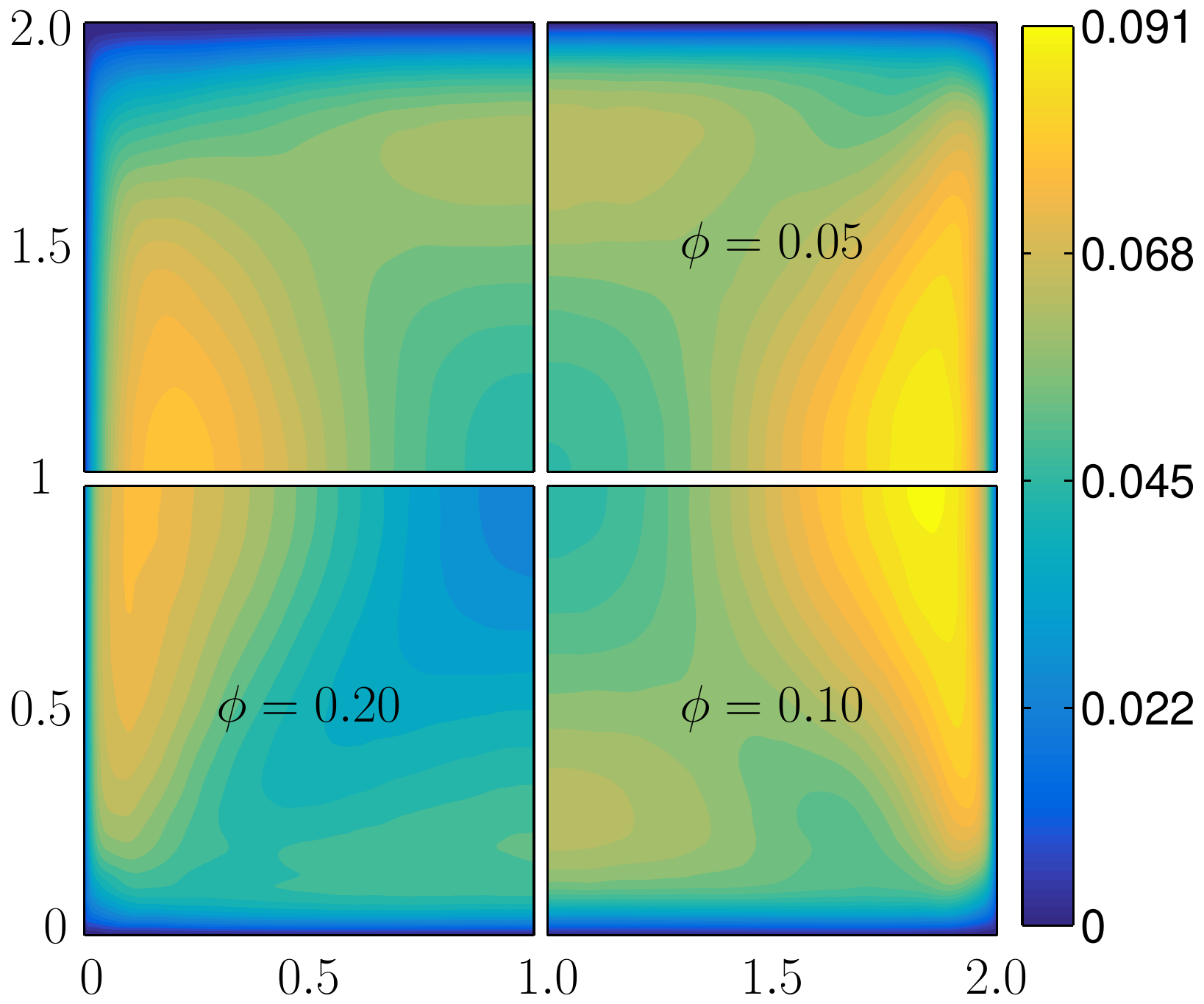}
   \put(-191,68){\rotatebox{90}{\large $y/h$}}
   \put(-105,152){\footnotesize $(c)$}
   \put(-105,-10){\large $z/h$}
   \includegraphics[width=0.445\textwidth]{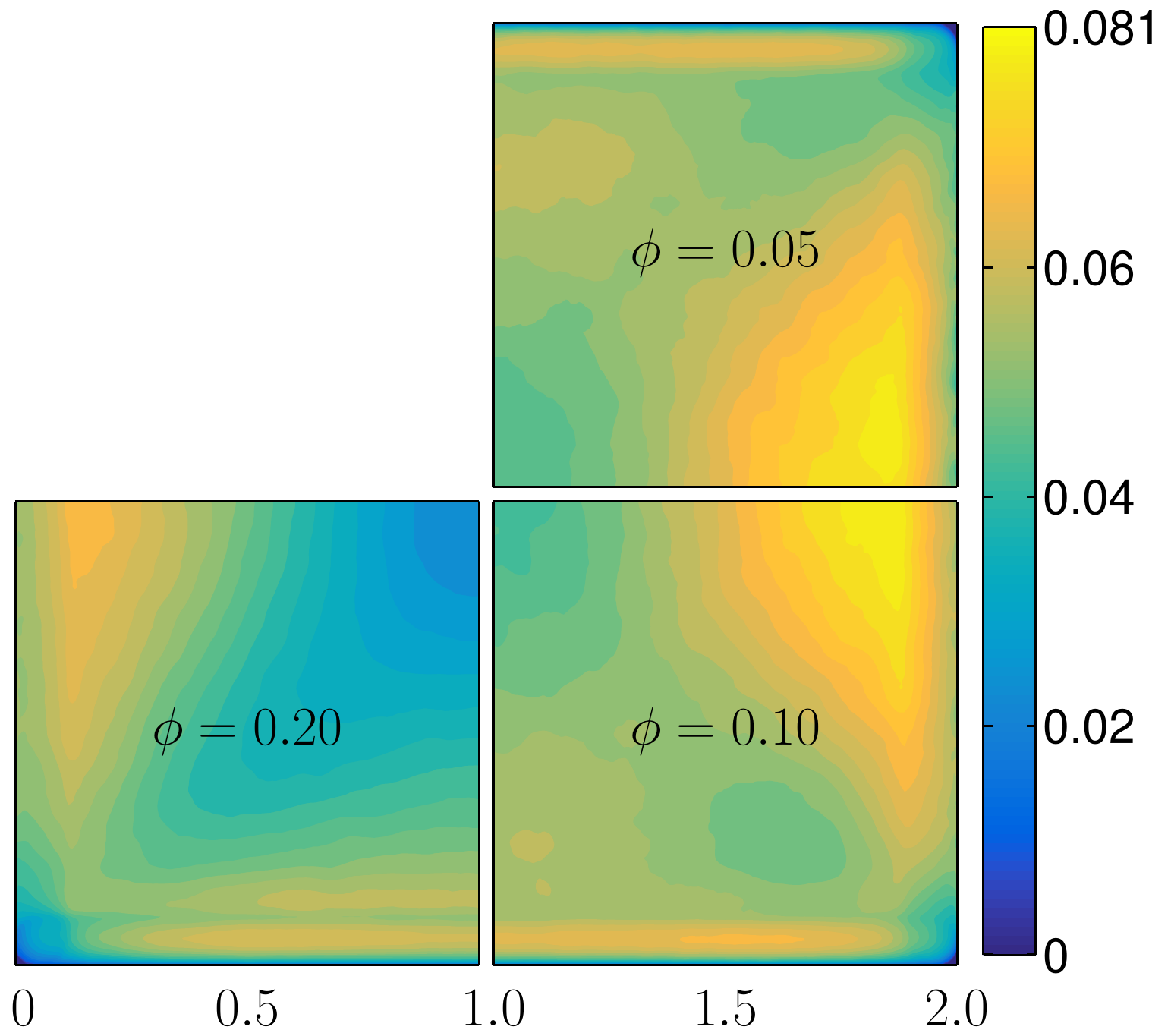}
   \put(-105,-10){\large $z/h$}
   \put(-105,152){\footnotesize $(d)$}\\
  \vspace{3pt}        
  \caption{(a) Root-mean-square of the streamwise fluid velocity fluctuations in outer units, $u_{f,rms}'$, for all $\phi$.
(b) Root-mean-square of the streamwise particle velocity fluctuations in outer units, $u_{p,rms}'$, for all $\phi$.
(c) Root-mean-square of the vertical fluid velocity fluctuations in outer units, $v_{f,rms}'$, for all $\phi$.
(d) Root-mean-square of the vertical particle velocity fluctuations in outer units $v_{p,rms}'$, for all $\phi$.}
\label{fig:frms1}
\end{figure}

\begin{figure}
   \centering
\includegraphics[width=0.41\textwidth]{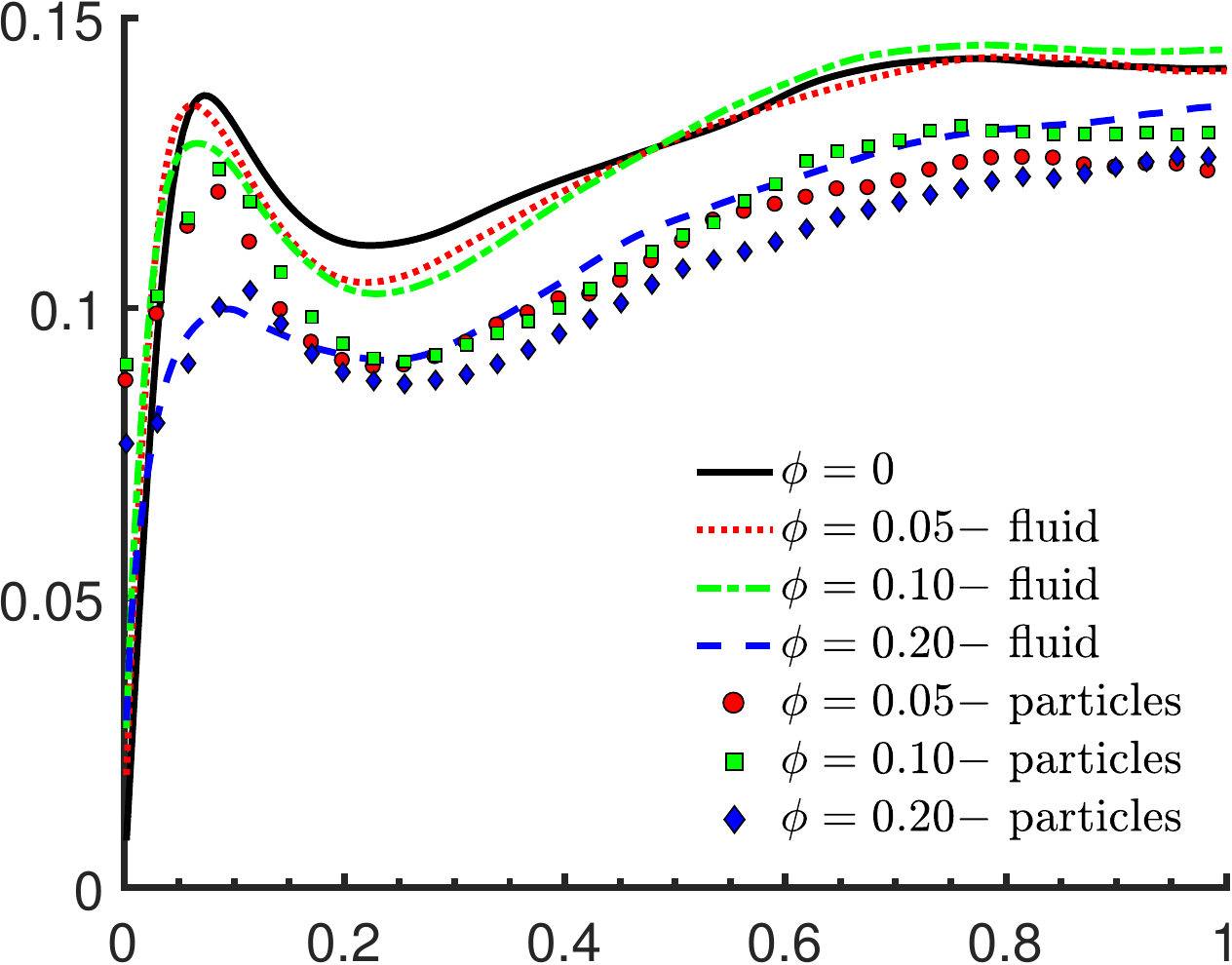}\hspace{1 cm}
   \includegraphics[width=0.41\textwidth]{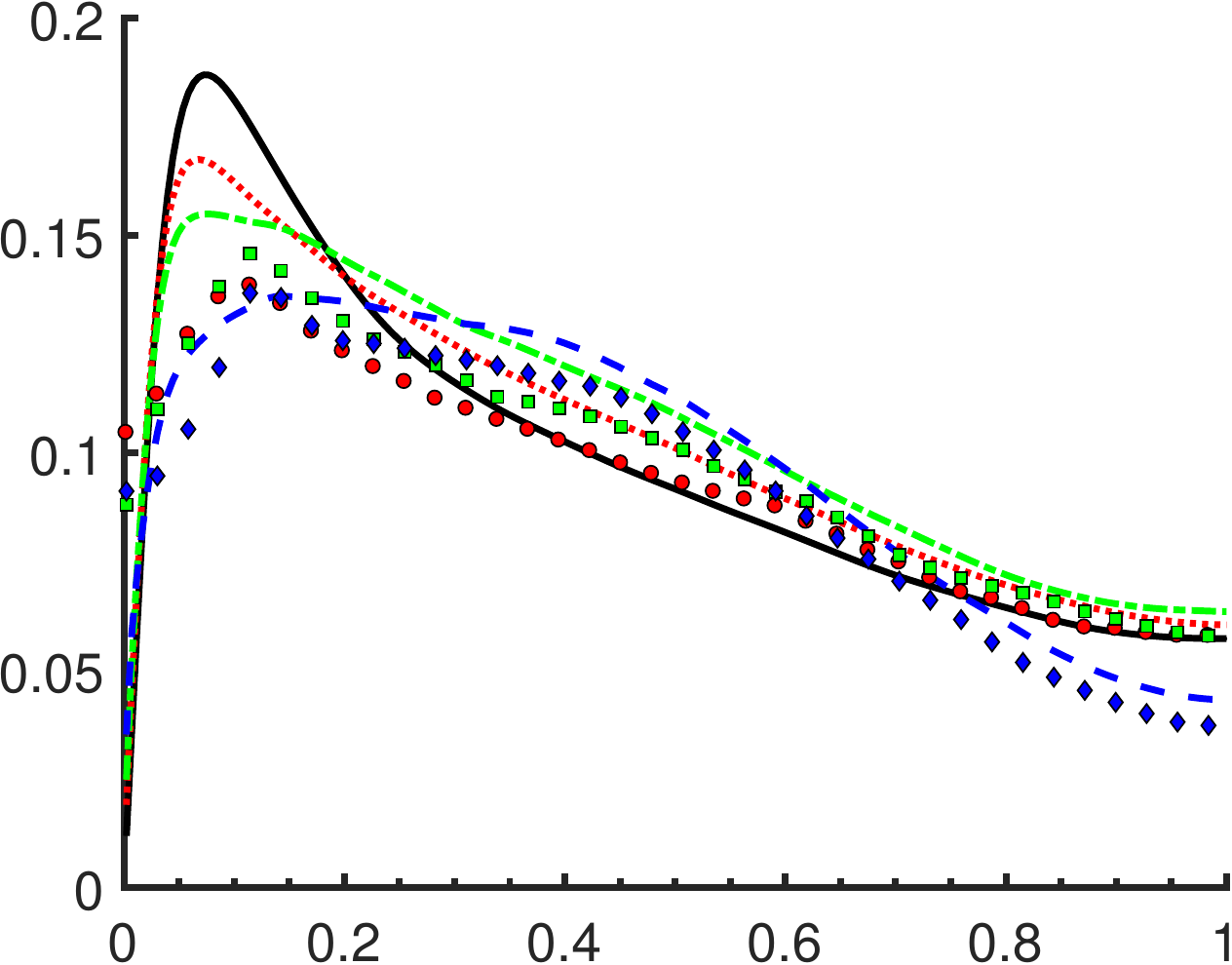}\vspace{0.2 cm}
   \put(-369,54){\rotatebox{90}{\large $u_{f/p,rms}'$}}
   \put(-267,125){\footnotesize $(a)$}
   \put(-80,125){\footnotesize $(b)$}\\ 
   
\includegraphics[width=0.41\textwidth]{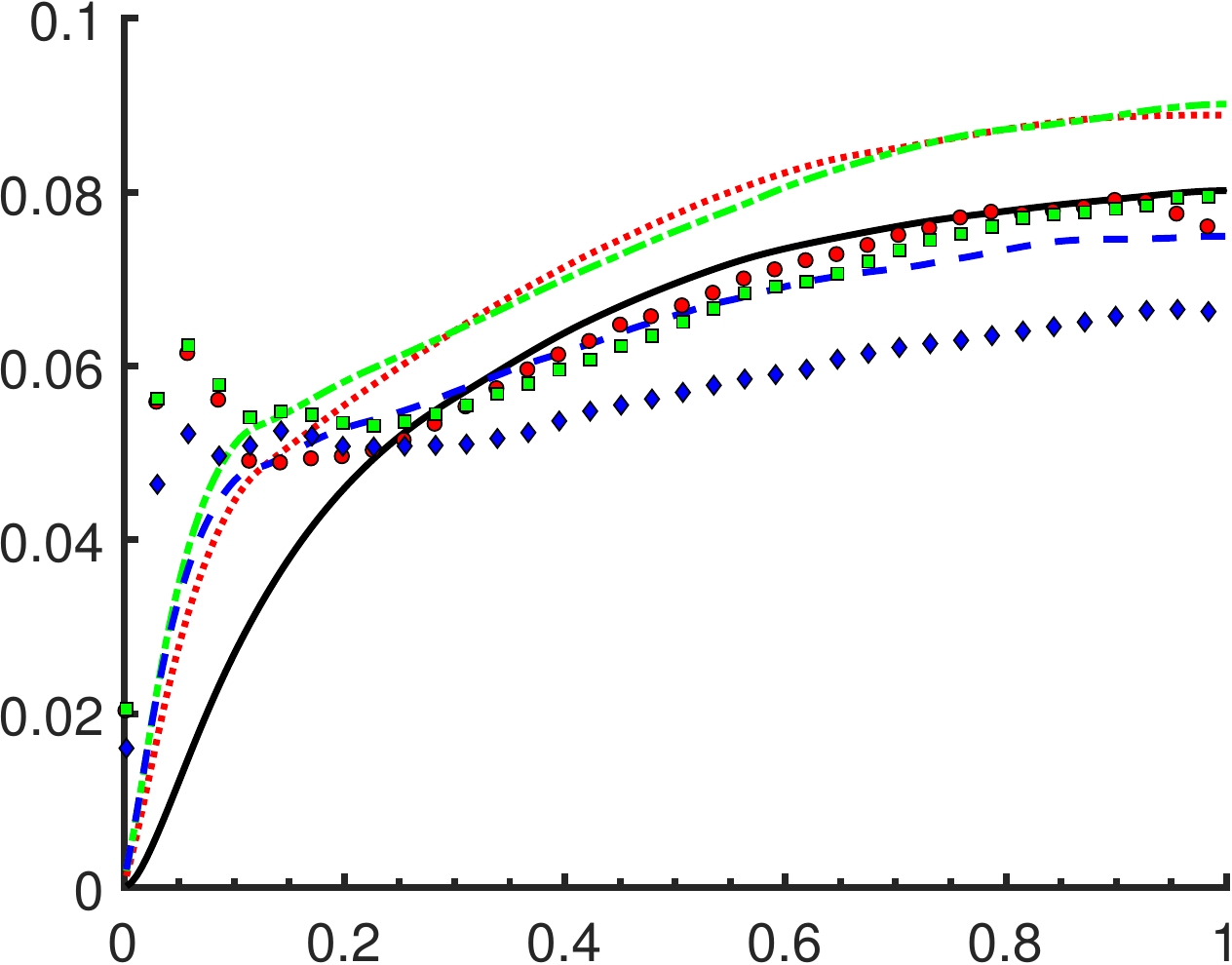}\hspace{1 cm}
   \includegraphics[width=0.41\textwidth]{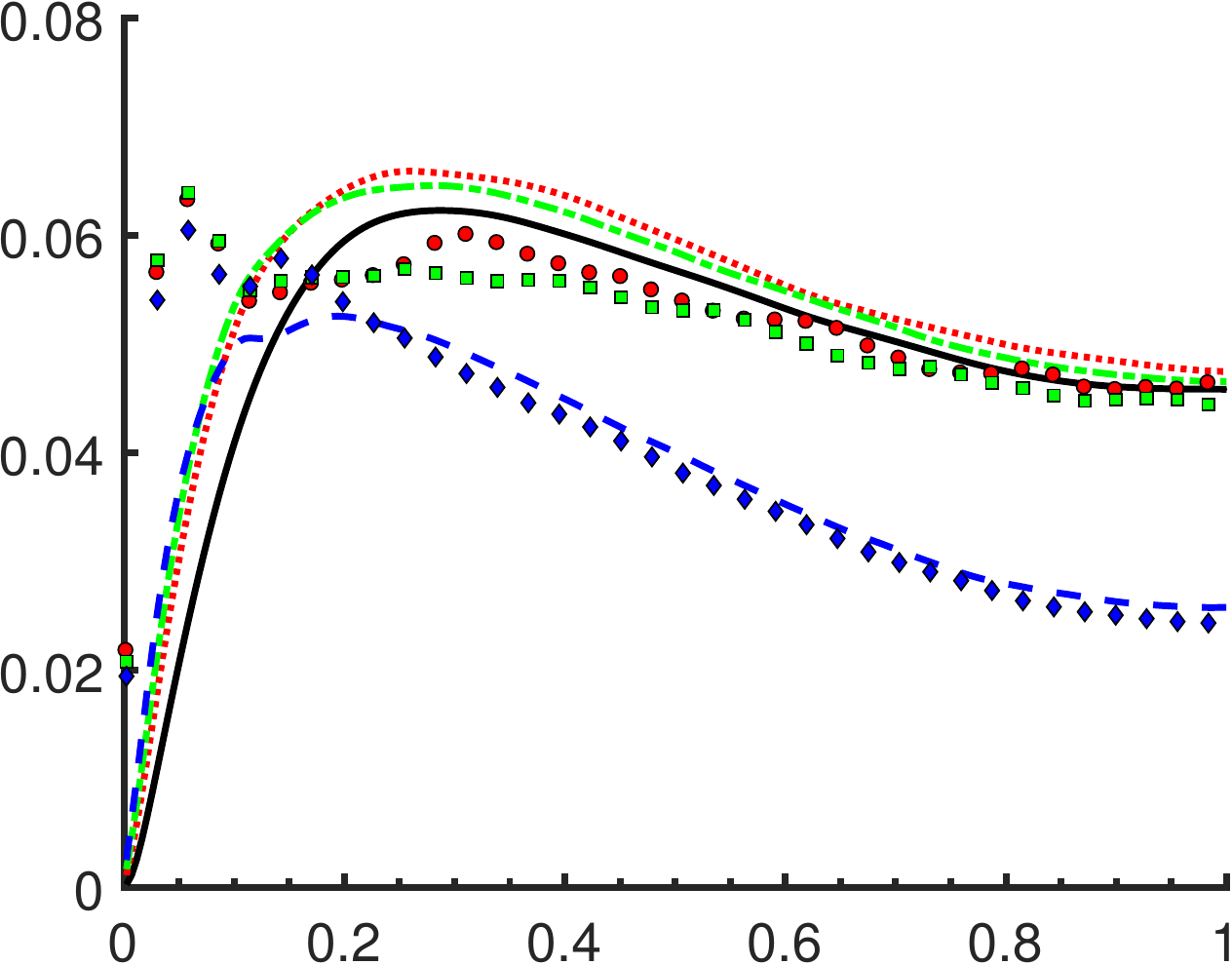}\vspace{0.2 cm}
   \put(-369,54){\rotatebox{90}{\large $v_{f/p,rms}'$}}
   \put(-267,125){\footnotesize $(c)$}
   \put(-80,125){\footnotesize $(d)$}\\ 
   
   \includegraphics[width=0.41\textwidth]{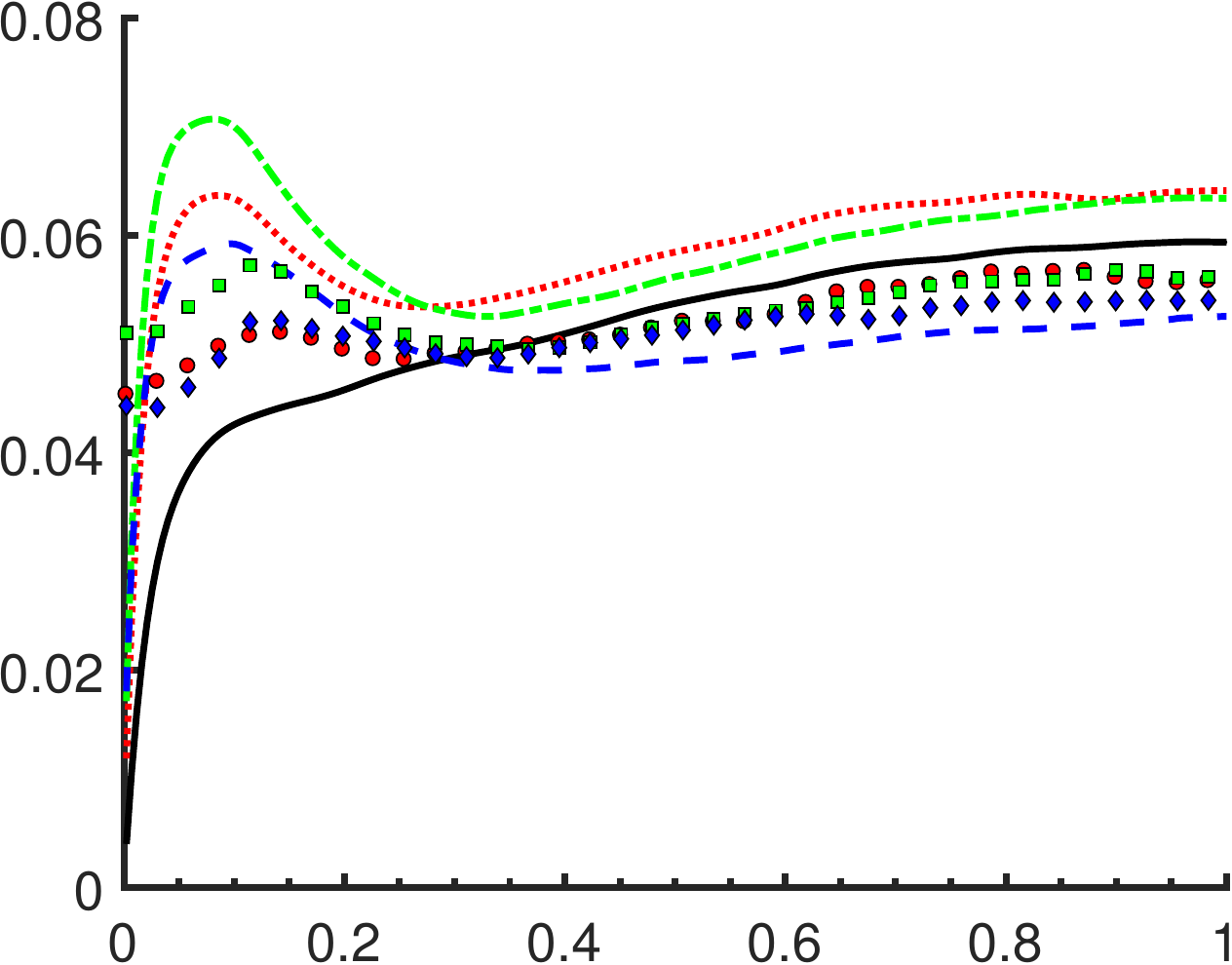}\hspace{1 cm}
   \includegraphics[width=0.41\textwidth]{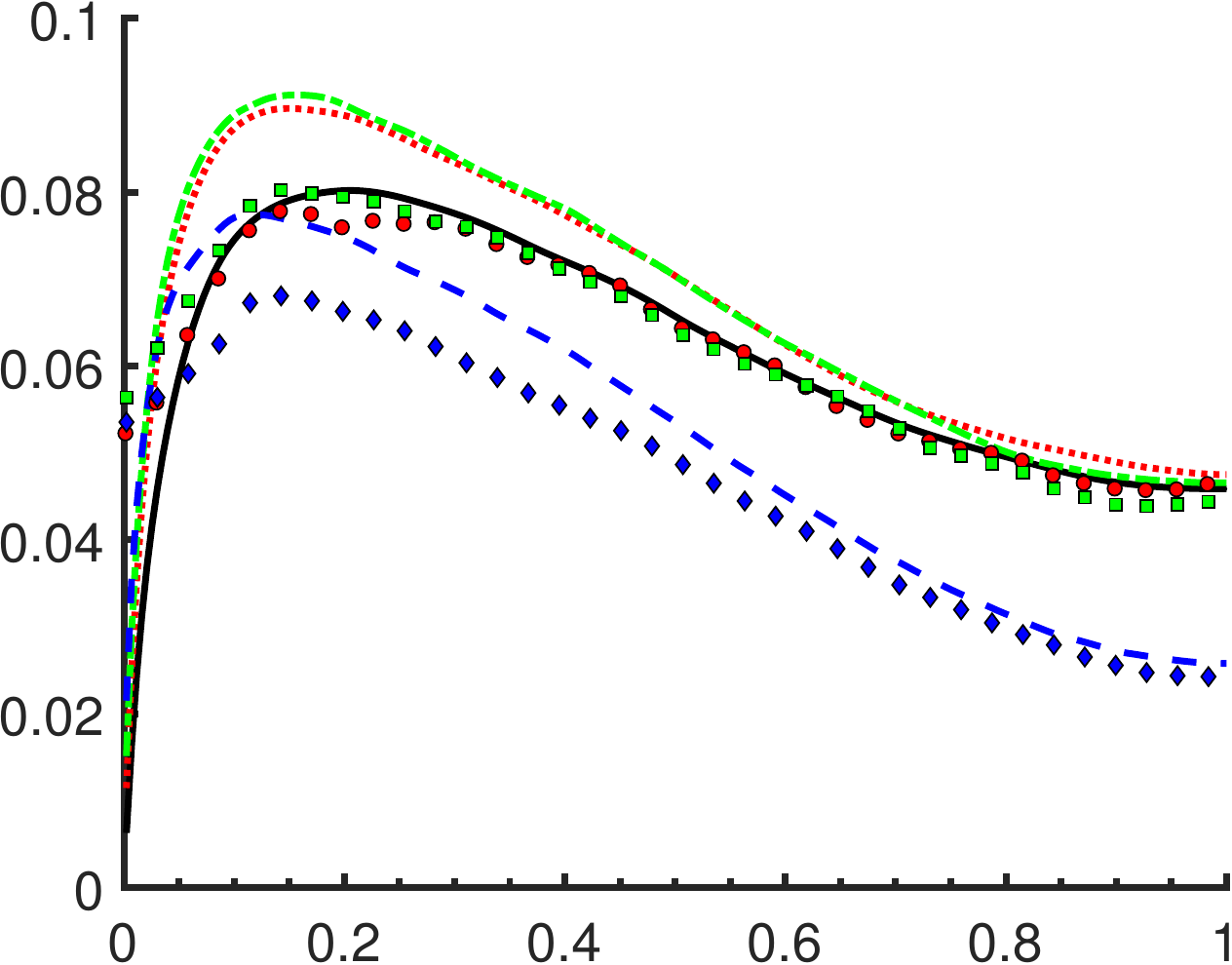}\vspace{0.2 cm}
   \put(-369,54){\rotatebox{90}{\large $w_{f/p,rms}'$}}
   \put(-269,-12){{\large $y/h$}}
   \put(-79,-12){{\large $y/h$}}
   \put(-267,125){\footnotesize $(e)$}
   \put(-80,125){\footnotesize $(f)$}\\ 
  \vspace{3pt}        
  \caption{Root-mean-square of fluid and particle velocity fluctuations: (a),(b) streamwise, 
(c),(d) wall-normal and (e),(f) spanwise ($z-$direction) components in outer units at $z/h=0.2$ (left column) and $z/h=1$ (right 
column), for all $\phi$. Lines and symbols are used for the fluid and solid phase statistics, respectively.}
\label{fig:frms2}
\end{figure}

\begin{figure}
   \centering
   \includegraphics[width=0.41\textwidth]{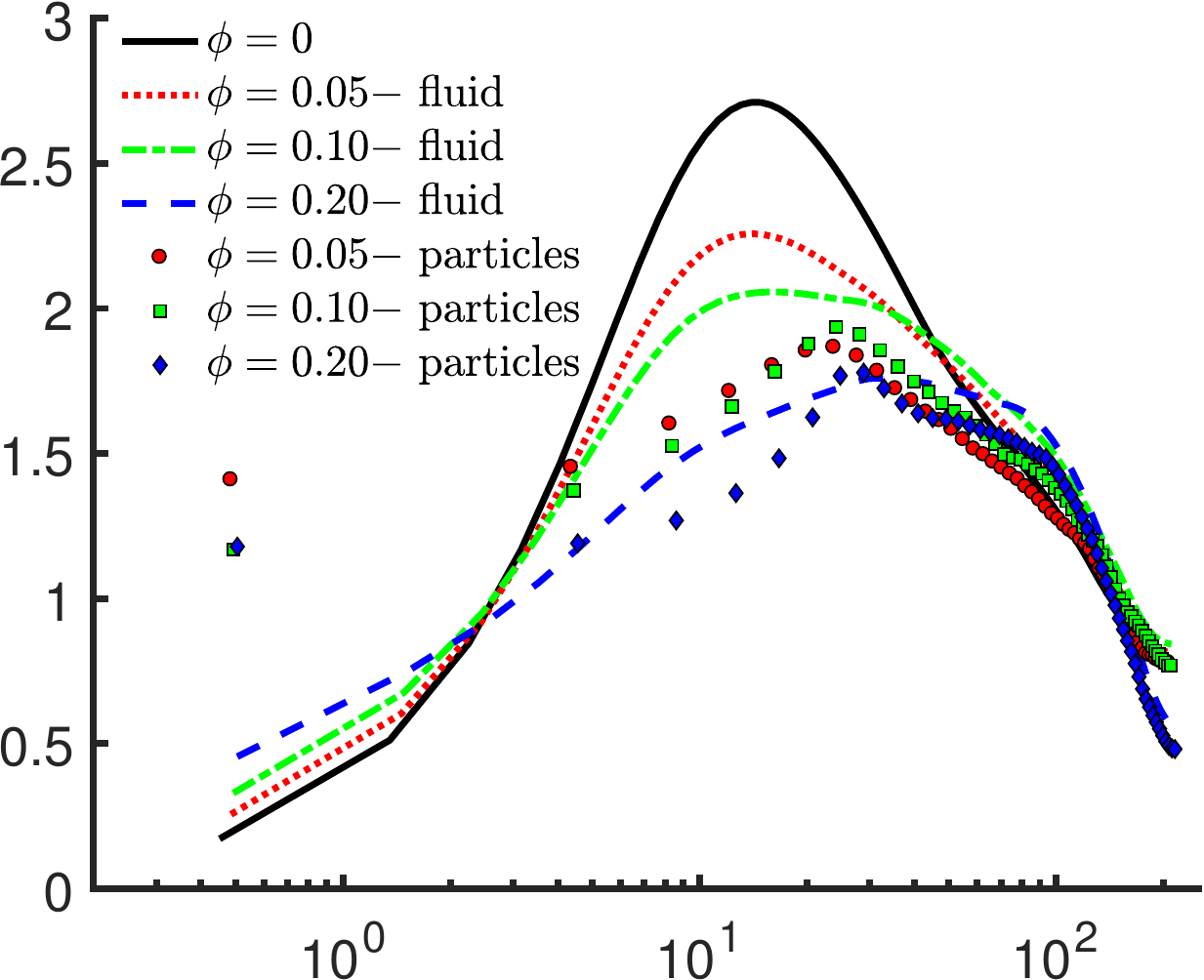}\hspace{1 cm}
   \includegraphics[width=0.41\textwidth]{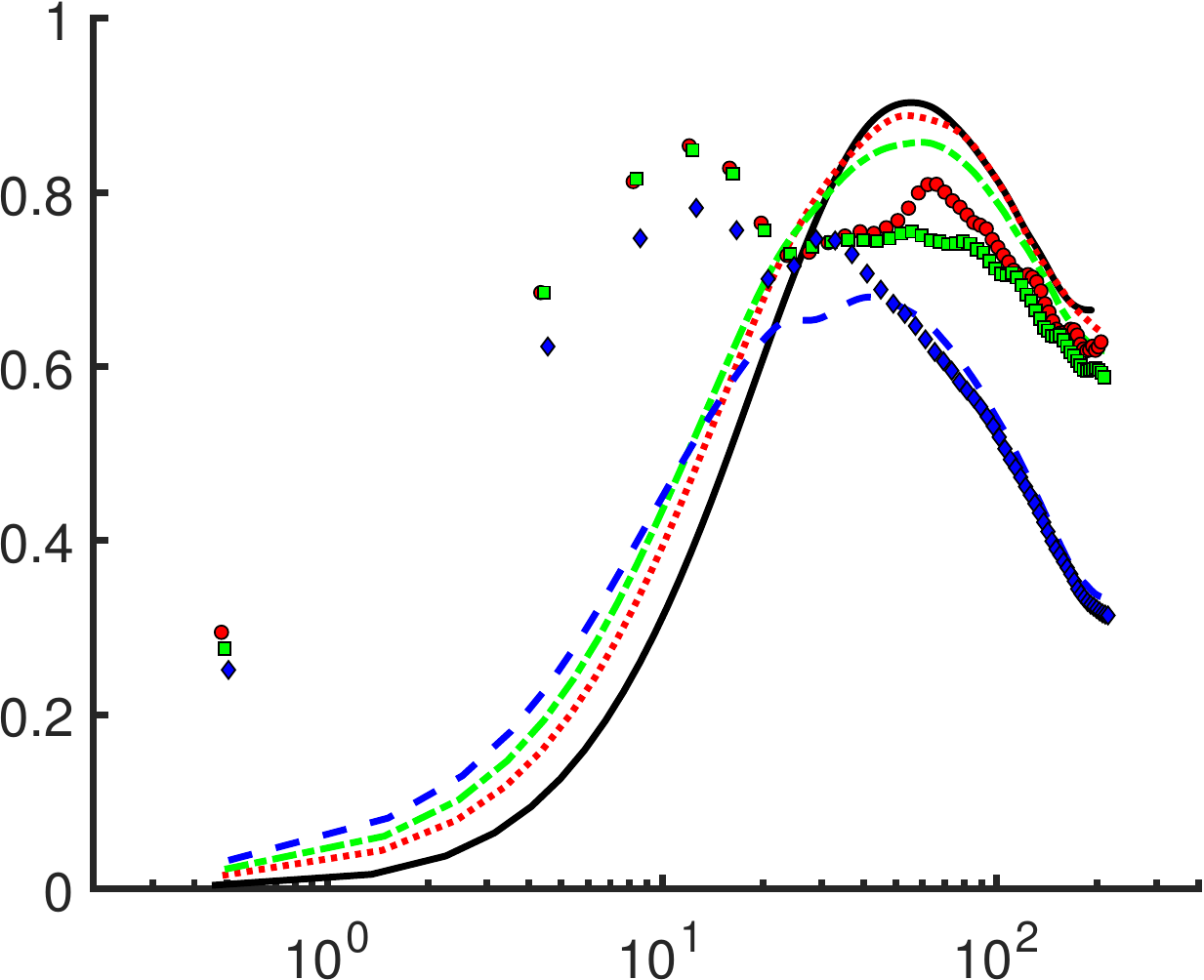}\vspace{0.4 cm}
   \put(-366,54){\rotatebox{90}{\large $u_{f/p,rms}'^+$}}
   \put(-186,54){\rotatebox{90}{\large $v_{f/p,rms}'^+$}}
   \put(-266,-12){{\large $y^+$}}
   \put(-76,-12){{\large $y^+$}}
   \put(-257,131){\footnotesize $(a)$}
   \put(-95,130){\footnotesize $(b)$}\\
   \includegraphics[width=0.41\textwidth]{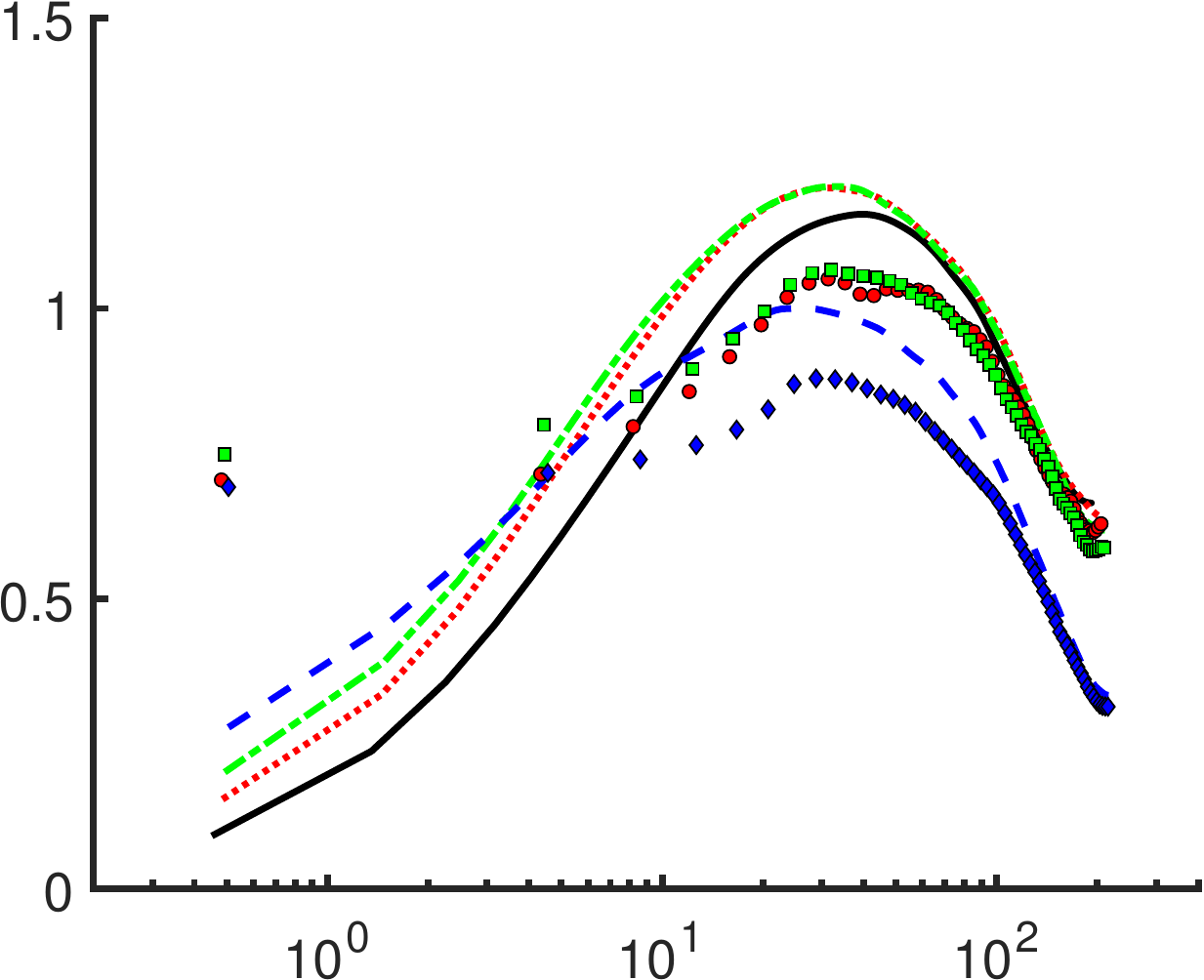}
   \put(-76,130){\footnotesize $(c)$}
   \put(-176,54){\rotatebox{90}{\large $w_{f/p,rms}'^+$}}
   \put(-76,-12){{\large $y^+$}}
  \vspace{3pt}        
  \caption{Root-mean-square of fluid and particle velocity fluctuations. (a) streamwise, 
(b) wall-normal and (c) spanwise ($z-$)direction components in inner units at $z/h=1$ for all $\phi$. 
Lines and symbols are used for the fluid and solid phase statistics as in figure~\ref{fig:frms2}.}
\label{fig:frms3}
\end{figure}

Next, we report the contours of the root-mean-square (r.m.s.) of the fluid and particle 
velocity fluctuations in outer units, see figure~\ref{fig:frms1}. Due to the symmetry around the corner-bisectors, we 
show only the contours of $v_{f/p,rms}'(y,z)$ in panels (c) and (d). Corresponding r.m.s. velocities along two lines at $z/h=0.2$ 
and $1$ are depicted in figure~\ref{fig:frms2}. Results in inner units are shown at the wall-bisector, $z/h=1$ in 
figure~\ref{fig:frms3}.

The contours of the streamwise fluid velocity fluctuations reveal that r.m.s. values are stronger near the walls (close to 
the wall-bisector), while minima are found along the corner bisectors. In this region, $u_{f,rms}'(y,z)$ is substantially 
reduced for $\phi=0.2$, as also visible from the profiles in figure~\ref{fig:frms2}(a). From figure~\ref{fig:frms2}(a) 
we also see that the local maxima of $u_{f,rms}'(y)$ close to the corners is of similar magnitude for $\phi=0.0, 0.05$ 
and $0.1$. On the other hand, at the wall-bisectors the maxima of $u_{f,rms}'(y,z)$ decrease with $\phi$, well below the value 
of the unladen case (see figure~\ref{fig:frms2}b). 
For $\phi=0.2$ the profile of $u_{f,rms}'(y)$ deeply changes. In particular, the streamwise r.m.s. velocity 
$u_{f,rms}'(y)$ is substantially smaller than in the unladen case in the core region, where the particle 
concentration and the mean velocity are high while the secondary flows are negligible. 
After the maximum value, $u_{f,rms}'(y)$ initially decreases smoothly and then sharply for $y/h 
> 0.6$.

From figure~\ref{fig:frms1}(b) and figures~\ref{fig:frms2}(a,b) we see that the streamwise r.m.s. particle velocity, 
$u_{p,rms}'(y,z)$, resembles that of the fluid phase. However, $u_{p,rms}'(y,z)$ is typically smaller than $u_{f,rms}'(y,z)$ 
in the cross-section. Exceptions are the regions close to the walls where the particle velocity does not vanish (unlike 
the fluid velocity).

From the contour of $v_{f,rms}'(y,z)$ (see figure~\ref{fig:frms1}c), we see that r.m.s. velocities are larger in the 
directions parallel to the walls, rather than in the wall-normal direction. Close to the wall-bisectors, the peak values 
of the wall-normal and parallel fluid r.m.s. velocities increase with $\phi$, again except for $\phi=0.2$ when turbulence 
is damped. From figure~\ref{fig:frms2}(c) and (e) we see instead that close to the corners, the local maxima of 
both $v_{f,rms}'(y)$ (wall-normal) and $w_{f,rms}'(y)$ (wall-parallel) increase with respect to the unladen 
case, for all $\phi$. At the wall-bisector (see figure~\ref{fig:frms2}d), wall-normal velocity fluctuations are slightly 
larger than the single-phase case for $\phi \lesssim 0.1$.

Finally, figure~\ref{fig:frms2}(f) shows profiles at the wall-bisector of the parallel component of the fluid velocity 
r.m.s, $w_{f,rms}'(y)$.
Note that the peak value of $w_{f,rms}'(y)$ increases with the volume fraction up to $\phi=0.1$ 
and moves closer to the wall. There is hence a clear redistribution of energy due to the particle 
presence towards a slightly more isotropic state.


Concerning the solid phase, the wall-normal particle r.m.s. velocity, $v_{p,rms}'$, exhibits a peak close 
to the walls where particle layers form, see figures~\ref{fig:frms1}(d) and ~\ref{fig:frms2}(d) \cite[and 
also][]{costa2016}. Differently from the channel flow analyzed by \citet{picano2015}, in the square duct flow $v_{p,rms}'$ decreases slowly 
with increasing volume fraction $\phi$ in the wall-particle layer ($y^+ \le 20$) and the maxima are similar for all $\phi$. This may be 
related to the existence of the secondary flows that pull the particles away from the walls towards the duct core. As 
for the fluid phase, the particle r.m.s. velocity parallel to the wall is greater 
than the perpendicular component close to the walls. From the profiles at the wall-bisector, see figure~\ref{fig:frms2}(f), we also see 
that the maximum of $w_{p,rms}'(y)$ is similar for $\phi=0.05$ and $0.1$, while it clearly decreases for the densest case 
simulated. More generally, we observe that particle r.m.s. velocities are similar to those of the fluid phase towards the 
core of the duct.

Both fluid and solid phase r.m.s. velocities are shown in inner units at the wall-bisector in figures~\ref{fig:frms3}(a,b,c). 
The local friction velocity has been used to normalize the velocity fluctuations. The fluid phase r.m.s. velocity 
increases with $\phi$ in the viscous sub-layer. Clearly, the presence of solid particles introduces 
additional disturbances in the fluid increasing the level of fluctuations in regions where these are typically low (in the 
unladen case). On the other hand, particle velocity fluctuations are typically smaller than the corresponding fluid r.m.s. velocity, 
except in the inner-wall region, $y^+ < 20$, where they are one order of magnitude larger.

\begin{figure}
   \centering
   \includegraphics[width=0.47\textwidth]{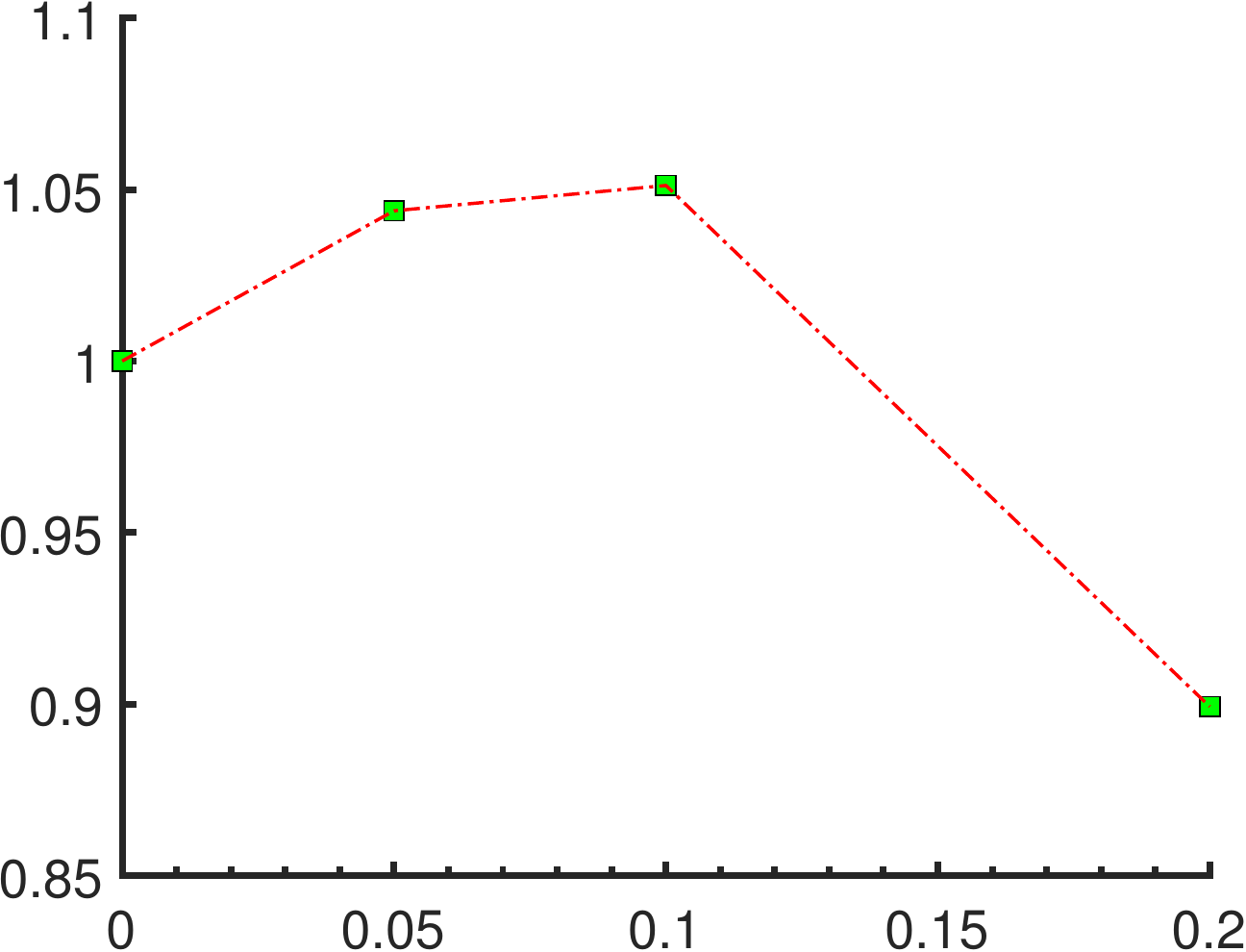}
   \put(-196,68){\rotatebox{90}{\large $I$}}
   \put(-86,-12){{\large $\phi$}}
  \vspace{3pt}        
  \caption{Mean turbulence intensity $I=\langle u' \rangle/U_b$ for all $\phi$, normalized by the value for the 
unladen case.}
\label{fig:mini}
\end{figure}

Finally, we also computed the mean turbulence intensity defined as $I=\langle u' \rangle/U_b$, with 
$u'=\sqrt{\frac{1}{3}(u_{f,rms}'^2 + v_{f,rms}'^2 + w_{f,rms}'^2)}$. This is shown is figure~\ref{fig:mini}, where results are 
normalized by the value obtained for the unladen case. As for secondary flows, we see that the turbulence intensity 
$I$ increases up to $\phi=0.1$, for which $I$ is approximately $5\%$ larger than the unladen value. For $\phi=0.2$, 
instead, the mean turbulence intensity decreases well below the single-phase value, and $I(\phi=0.2) \sim 0.9\, I
(\phi=0)$.

\subsection{Reynolds stress and mean fluid streamwise vorticity}

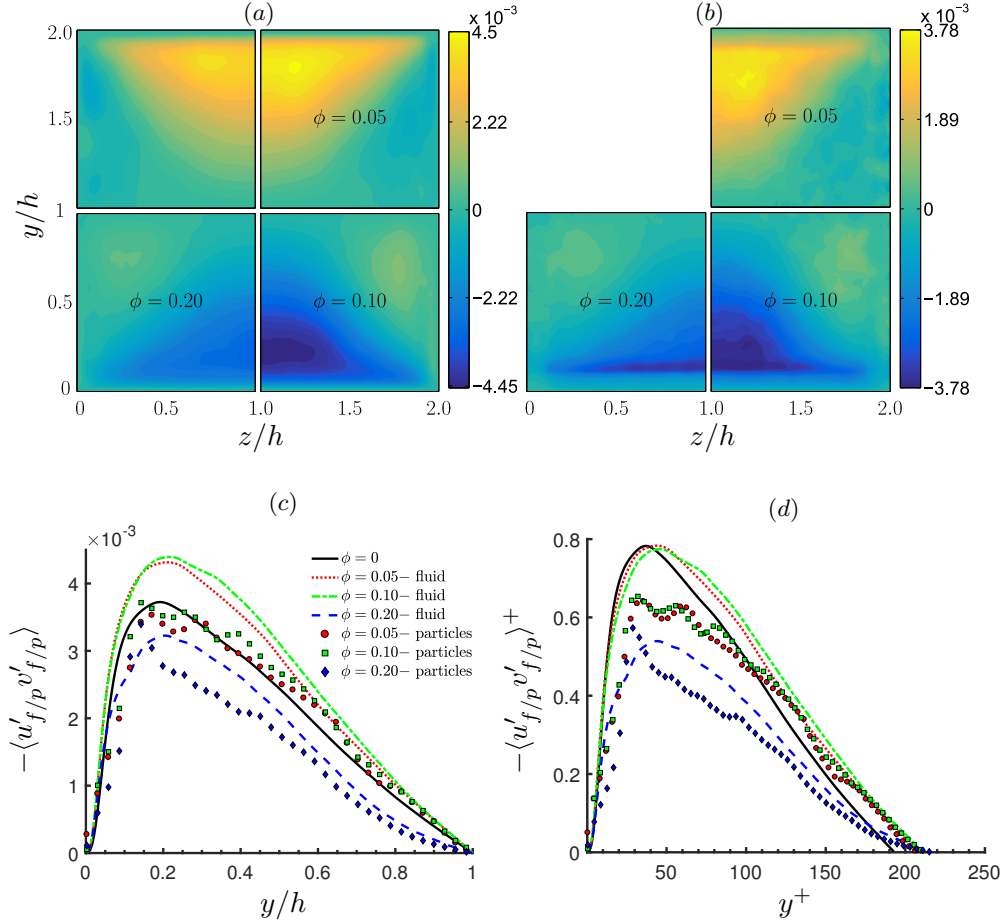
\begin{figure}
   \centering
   \includegraphics[width=0.47\textwidth]{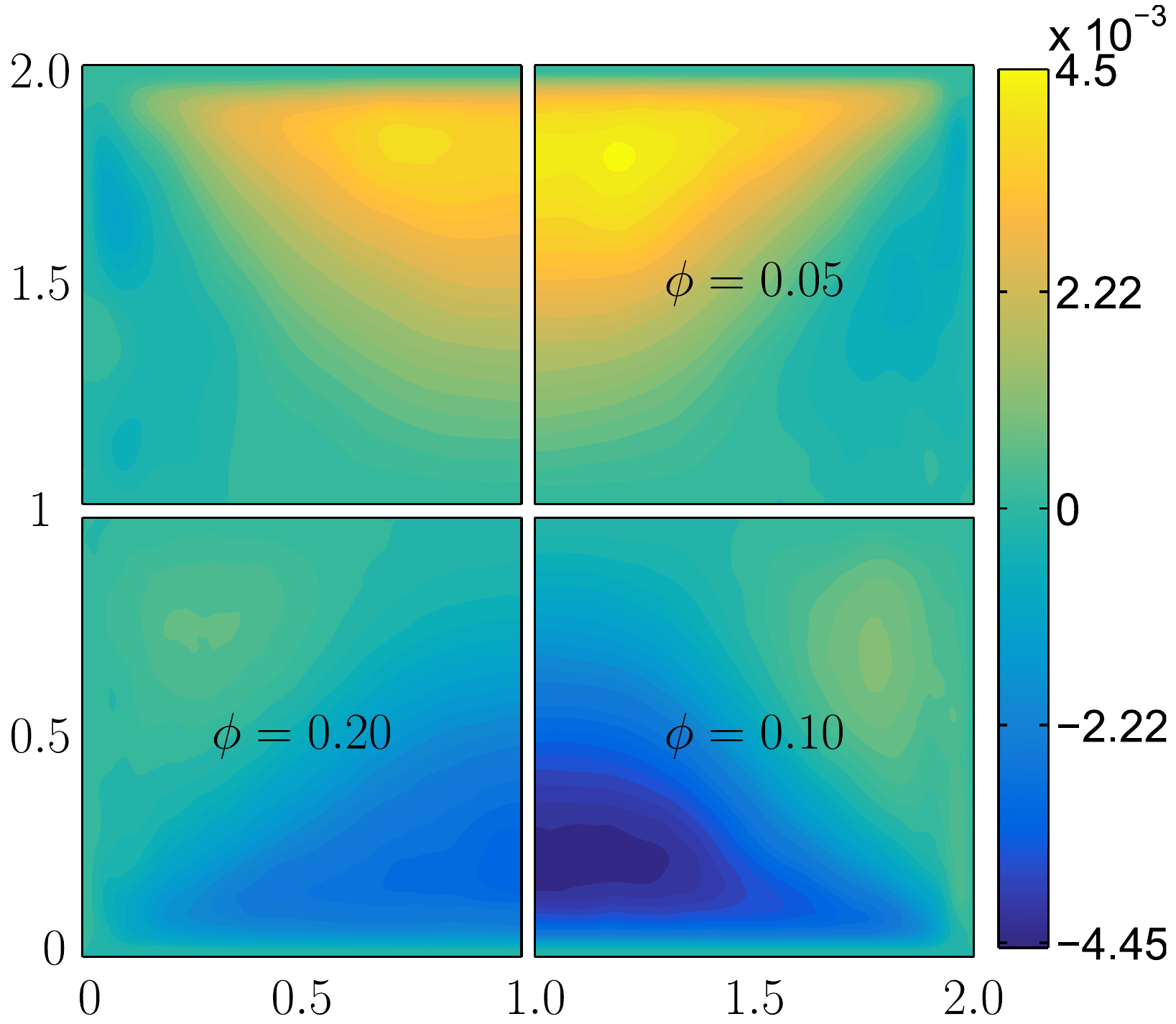}
   \put(-191,68){\rotatebox{90}{\large $y/h$}}
   \put(-107,-10){\large $z/h$}
   \put(-105,151){\footnotesize $(a)$}
   \includegraphics[width=0.445\textwidth]{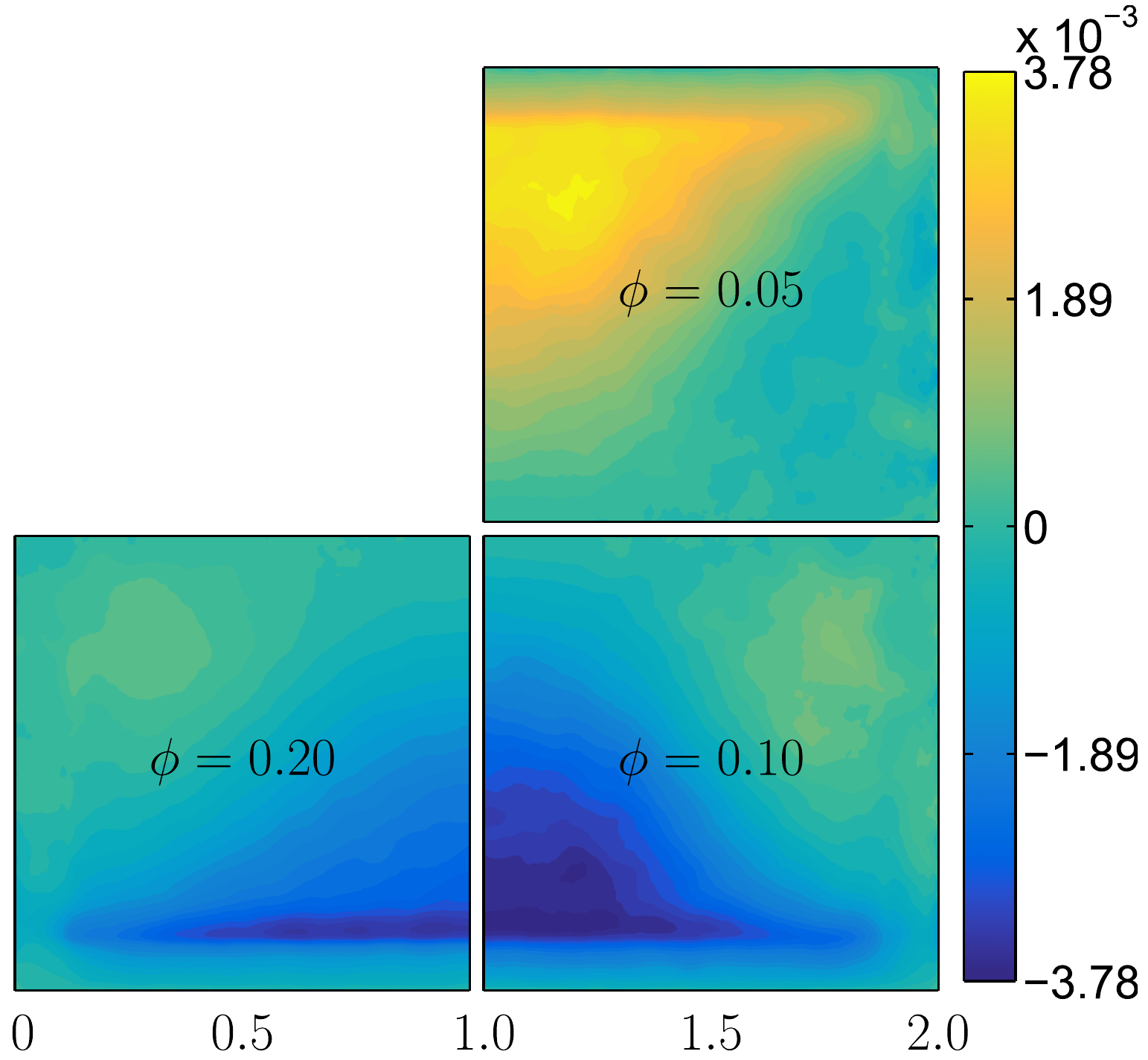}
   \put(-105,151){\footnotesize $(b)$}
   \put(-107,-10){\large $z/h$}

   \vspace{0.5 cm}\includegraphics[width=0.40\textwidth]{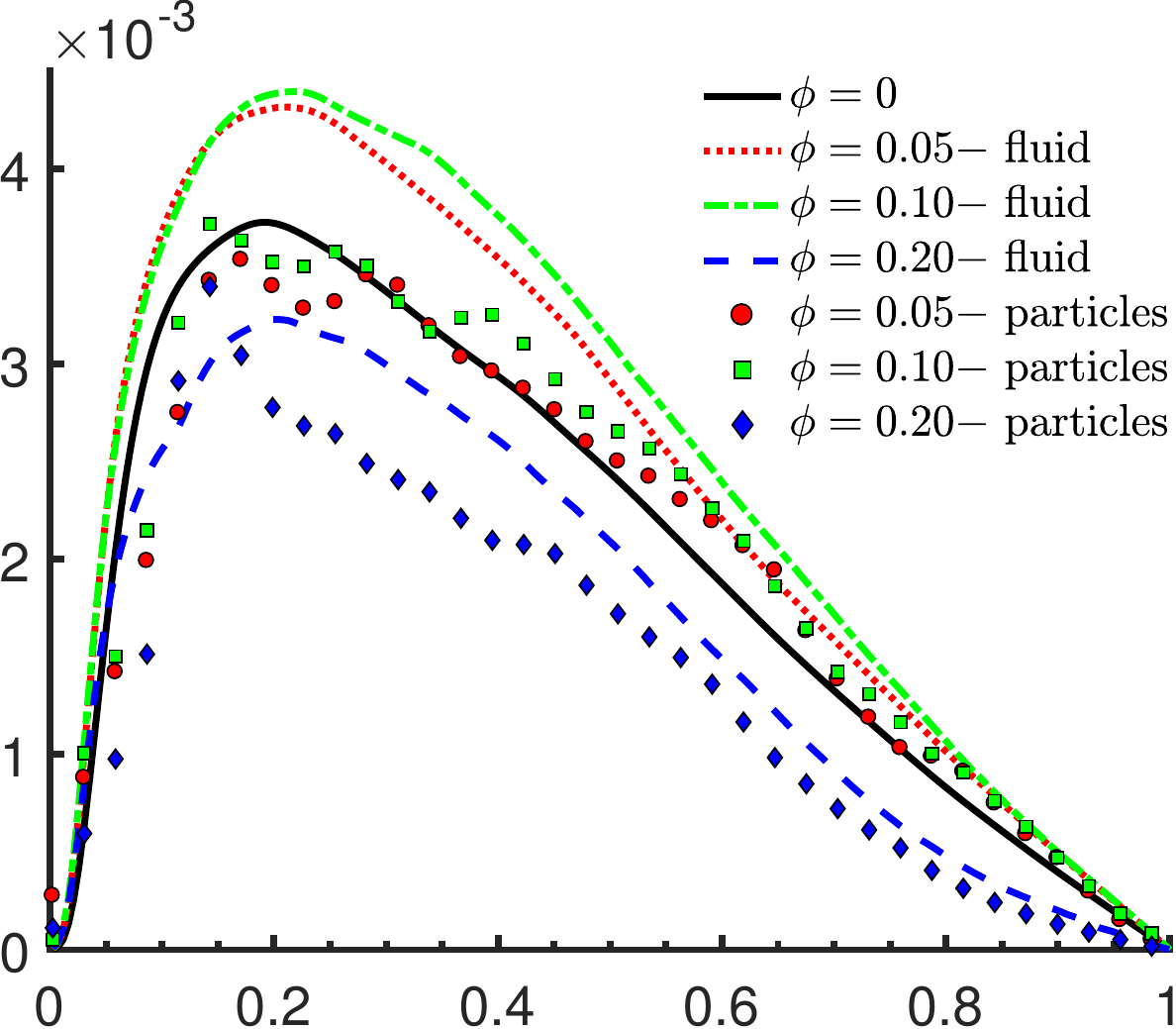}\hspace{1 cm}
  \put(-176,40){\rotatebox{90}{\large $-\langle u_{f/p}'v_{f/p}'\rangle$}}
  \put(-78,140){\footnotesize $(c)$}
   \hspace{1.05 cm}\includegraphics[width=0.44\textwidth]{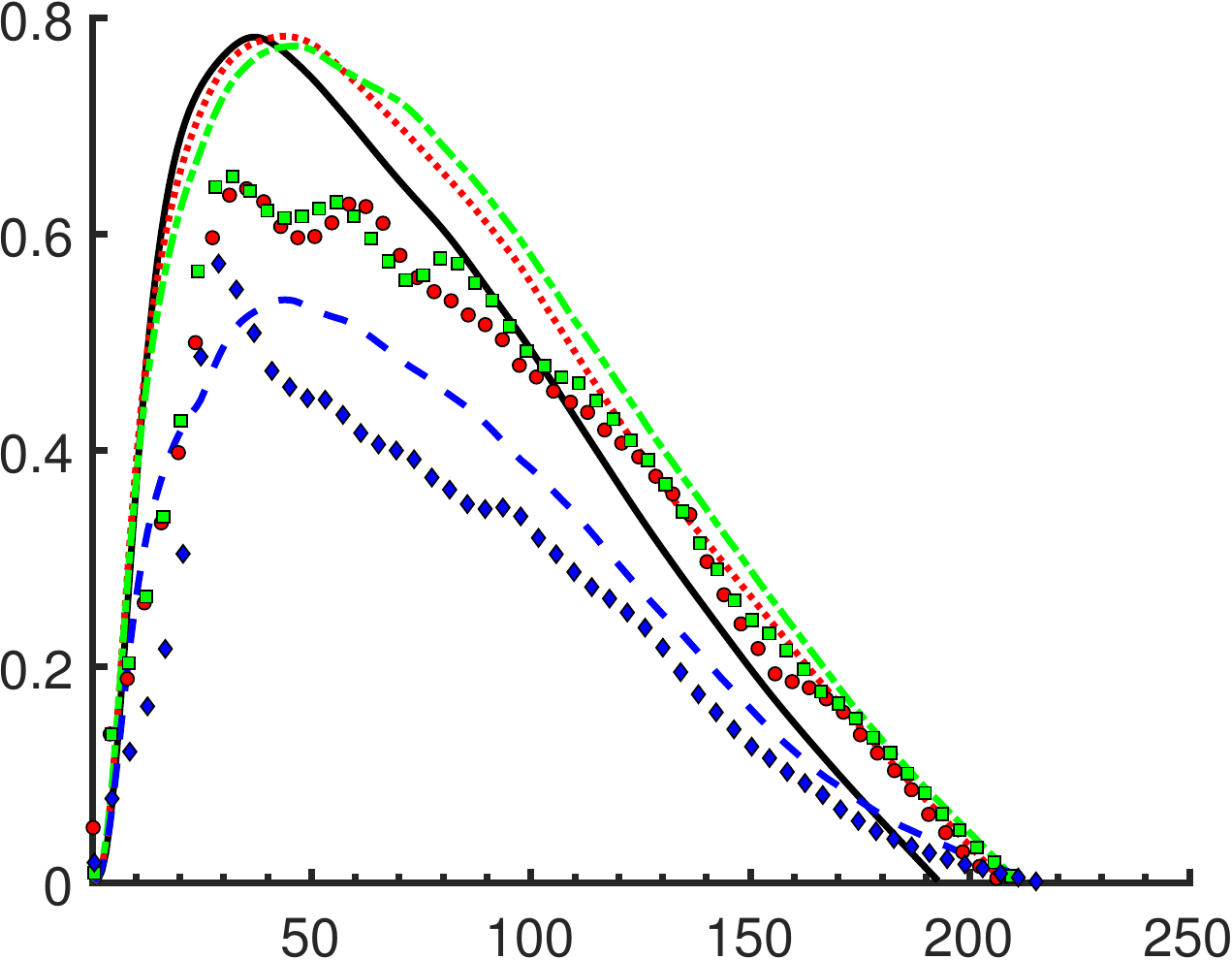}\hspace{-0.7 cm}
   \put(-168,40){\rotatebox{90}{\large $-\langle u_{f/p}'v_{f/p}'\rangle^+$}}
   \put(-64,-12){{\large $y^+$}}
   \put(-260,-12){{\large $y/h$}}
      \put(-68,138){\footnotesize $(d)$}\\
  \vspace{3pt}        
  \caption{Contours of the primary Reynolds stress $\langle u_{f/p}'v_{f/p}' \rangle$ of (a) the fluid  
and (b) particle phase for all volume fractions considered. The profiles along the wall-bisector at $z/h=1$ are shown 
in outer and inner units in panels (c) and (d). Lines and symbols are used for the fluid and solid phase statistics as 
in figure~\ref{fig:frms2}.}
\label{fig:vw1}
\end{figure}

We now turn to the discussion of the primary Reynolds stress in the duct cross-section, as this plays an important role in the 
advection of mean streamwise momentum. We show in figure~\ref{fig:vw1}(a) the $\langle u_f'v_f' \rangle$ component of the fluid 
Reynolds stress in the duct cross-section, for all $\phi$. The component $\langle u_f'w_f' \rangle$ is not shown as it 
is the $90^o$ rotation of $\langle u_f'v_f' \rangle$. We see that the maximum of $\langle u_f'v_f' 
\rangle$, located close to the wall-bisector, increases with the volume fraction up to $\phi=0.1$. The 
maximum $\langle u_f'v_f' \rangle$ then progressively decreases with $\phi$, denoting a reduction in 
turbulent activity. For $\phi=0.2$ we observe that the maximum of $\langle u_f'v_f' \rangle$ reaches values even lower than 
in the unladen case. 
 The contour of $\langle u_f'v_f' \rangle$ also changes with increasing $\phi$. 
We find that $\langle u_f'v_f' \rangle$ increases towards the corners, and also for $\phi=0.2$ it is 
larger than in the unladen case. However, the mean value of $\langle u_f'w_f' \rangle$ in one quadrant slightly 
increases up to $\phi=0.1$, while for $\phi=0.2$ the mean is $26\%$ smaller than for the unladen case.

The profiles of $\langle u_f'v_f' \rangle$ at the wall-bisector ($z/h=1$) are shown in figure~\ref{fig:vw1}(c). While the profiles for 
$\phi=0.05$ and $0.1$ are similar and assume larger values than the reference case, we see that for $\phi=0.2$ the profile is substantially lower for 
all $z/h > 0.2$. This is also different to what found in channel flow as close to the core, $\langle u_f'v_f' \rangle$ is found to be 
similar for $\phi=0.0$ and $0.2$. This is probably related to the high particle concentration at the duct core, see figure~\ref{fig:conc}(c).
Looking at the profiles in inner units, see figure~\ref{fig:vw1}(d), we also see that 
the peak values of $\langle u_f'v_f' \rangle^+$ are similar for the unladen case and for the laden cases 
with $\phi=0.05$ and $0.1$. This is in contrast to what is found in channel flows, where a reduction of the peak with 
$\phi$ is observed \citep{picano2015}. This shows that up to $\phi=0.1$, the turbulence activity is not significantly reduced by the 
presence of particles. On the other hand, for $\phi=0.2$ we observe a large reduction in the maximum 
$\langle u_f'v_f' \rangle^+$, denoting an important reduction in turbulent activity.\\
The Reynolds stress of the solid phase, $\langle u_p'v_p' \rangle$, is also shown for comparison in 
figures~\ref{fig:vw1}(b)-(d). The profiles are similar to those of the fluid phase, although $\langle 
u_p'v_p' \rangle < \langle u_f'v_f' \rangle$ for all $\phi$, except close to the walls for $\phi=0.2$. 
These local maxima may be related to the higher local particle concentration in layers close to the 
walls.

While for turbulent channel flow the cross-stream component of the Reynolds stress tensor, 
$\langle v_f'w_f' \rangle$, is negligible, in duct flow it is finite and contributes to the 
production or dissipation of mean streamwise vorticity. Hence it is directly related to the 
origin of mean secondary flows \citep{gavrilakis1992,gessner}. This can be seen from the Reynolds-averaged 
streamwise vorticity equation for a fully developed single-phase duct flow,
\begin{equation}
\label{vort}
V_f\pd{\Omega_f}{y} + W_f\pd{\Omega_f}{z} + \pdss{(\langle w_f'^2 \rangle -\langle v_f'^2 \rangle)}{y}{z} + \left(\pds{}{y}-\pds{}{z}\right) \langle v_f' w_f' \rangle - \nu \left(\pds{}{y}+\pds{}{z}\right) \Omega_f = 0
\end{equation}
where the mean vorticity
\begin{equation}
\label{vort1}
\Omega_f = \pd{W_f}{y} - \pd{V_f}{z}.
\end{equation}
The first two terms of equation (\ref{vort}) represent the convection of mean vorticity by the 
secondary flow itself and have been shown to be almost negligible \citep{gavrilakis1992}. The third term is a source of vorticity in the 
viscous sublayer due to the gradients in the anisotropy of the cross-stream normal stresses. The fourth term, 
involving the secondary Reynolds stress, acts as source or sink of vorticity. Finally, the last term 
represents the viscous diffusion of vorticity.

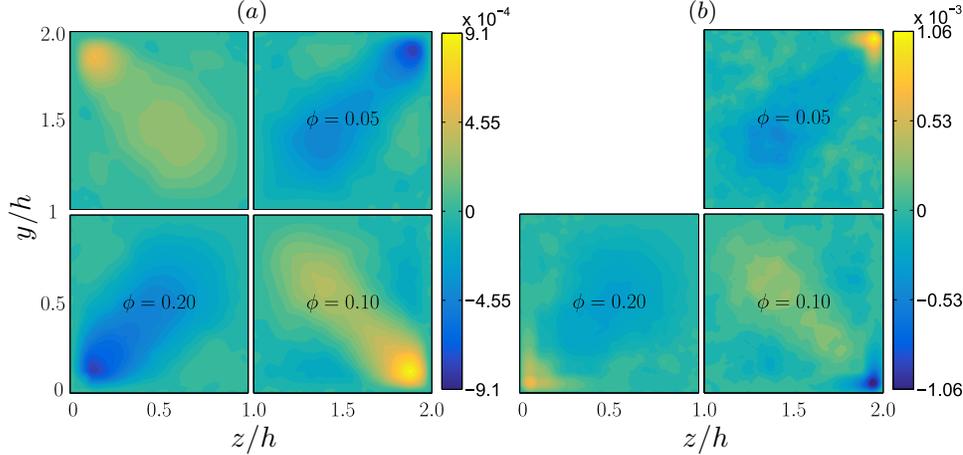
\begin{figure}
   \centering
   \includegraphics[width=0.47\textwidth]{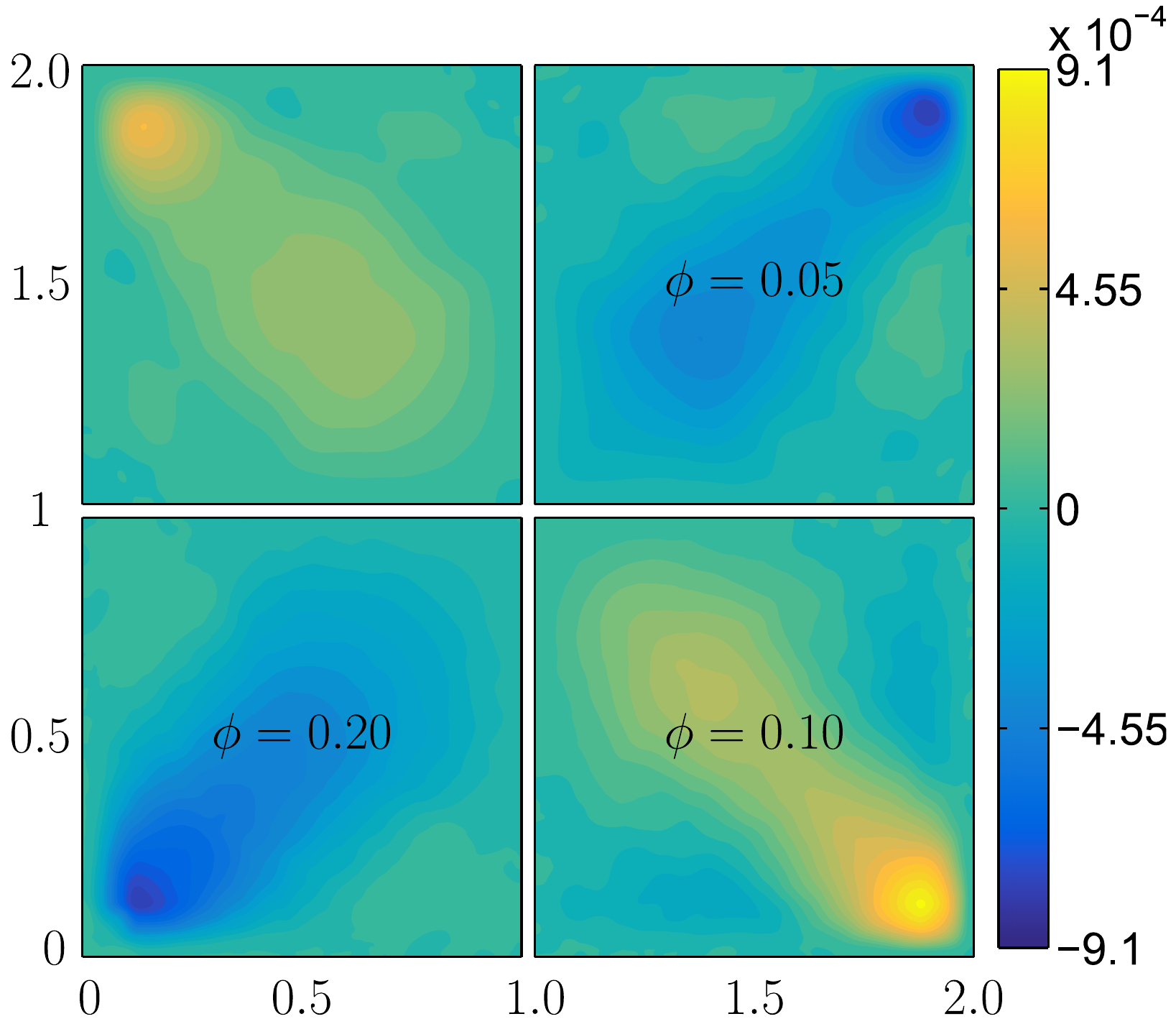}
   \put(-191,68){\rotatebox{90}{\large $y/h$}}
   \put(-107,-10){\large $z/h$}
   \put(-105,152){\footnotesize $(a)$}
   \includegraphics[width=0.445\textwidth]{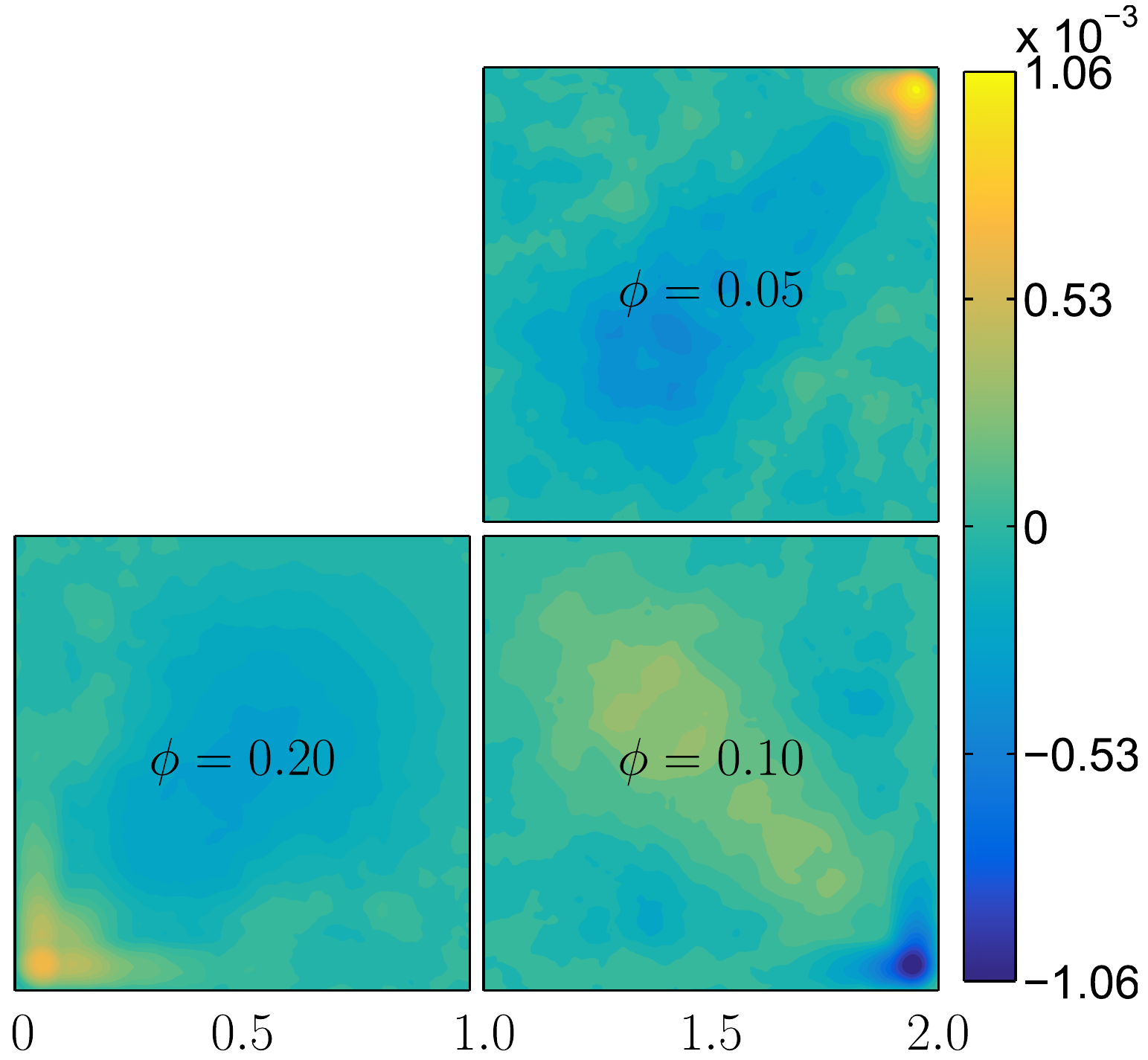}
   \put(-105,152){\footnotesize $(b)$}
   \put(-107,-10){\large $z/h$}\\
  \vspace{3pt}        
  \caption{Contours of the secondary Reynolds stress $\langle v_{f/p}'w_{f/p}' \rangle$ of  (a) the fluid 
and (b) the particle phase for all volume fractions $\phi$ under investigation.}
\label{fig:uv1}
\end{figure}

The contours of $\langle v_f'w_f' \rangle$ are shown in figure~\ref{fig:uv1} for all $\phi$. These are 
non-negligible along the corner-bisectors (being directly related to mean secondary flows), and about 
one order of magnitude smaller than the primary Reynolds stress. Interestingly, we see that the maxima 
of $\langle v_f'w_f' \rangle$ strongly increases with $\phi$ up to $\phi=0.1$ and also for $\phi=0.2$, 
the maximum value is still larger than that for $\phi=0.05$. We also notice that regions of finite 
$\langle v_f'w_f' \rangle$ become progressively broader with increasing $\phi$.
 
The secondary Reynolds stress of the solid phase, $\langle v_p'w_p' \rangle$, resembles qualitatively that of 
the fluid phase, except at the corners were a high value of opposite sign is encountered. As a 
consequence, close to the corners this term may contribute in opposite way to the production or 
dissipation of vorticity with respect to $\langle v_f'w_f' \rangle$.

\begin{figure}
   \centering
   \includegraphics[width=0.47\textwidth]{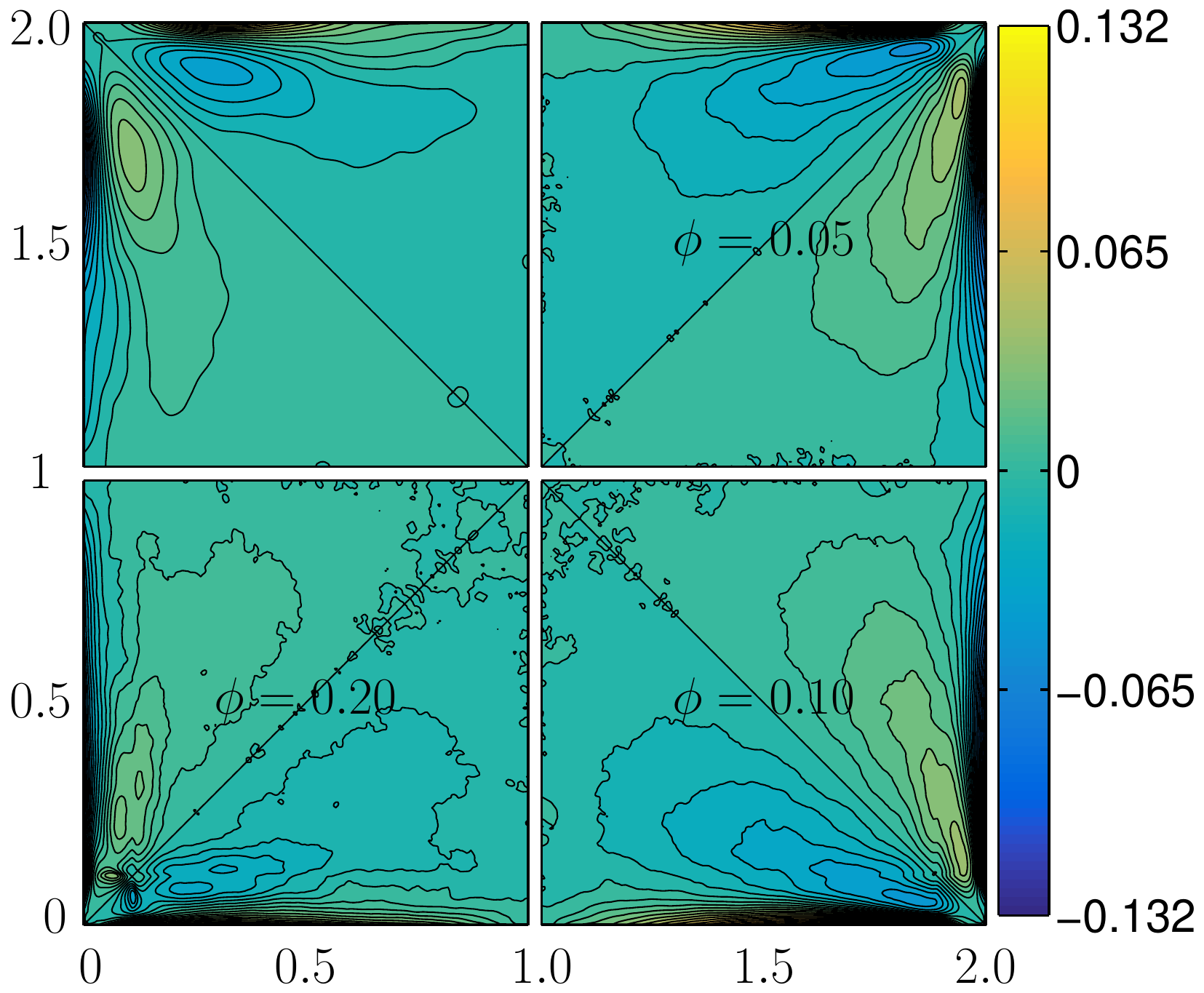}
   \put(-191,68){\rotatebox{90}{\large $y/h$}}
   \put(-107,-10){\large $z/h$}
  \vspace{3pt}
  \vspace{3pt}        
  \caption{Contours of the mean fluid streamwise vorticity $\Omega_f$
for all $\phi$ under investigation.} 
\label{fig:vort}
\end{figure}

We conclude this section by showing in figure~\ref{fig:vort} the mean streamwise fluid vorticity $\Omega_f$ for all 
$\phi$.
The region of maximum vorticity at the wall is found between $z/h=0.2$ and $z/h=0.5$ for 
the unladen case and it extends closer to the corner for increasing $\phi$. We also find that the maximum $\Omega_f$ 
initially increases with the solid volume fraction (for $\phi=0.1$ the maximum $\Omega_f$ is just slightly smaller 
than for $\phi=0.05$). This is expected as also the intensity of the secondary motions increase. The contours of 
$\Omega_f$ become more noisy for $\phi=0.2$, with a maximum value below that of the single-phase case. Also, for 
$\phi=0.2$ the location of maximum $\Omega_f$ is similar to that found for the unladen case.

The mean streamwise vorticity changes sign further away from the walls. For the unladen case, the magnitude of 
the maximum vorticity at the walls is $2.5$ times larger than the magnitude of the maximum vorticity of opposite sign 
found closer to the corner-bisector. 
As reference and for clarity, we will consider the vorticity to be positive at 
the wall, and negative further away, as found for the top and bottom walls (for the lateral walls the situation is 
reversed). 
We find that the vorticity minimum approaches progressively the walls as $\phi$ increases up to $\phi=0.1$, and that the 
ratio between the magnitudes of the maximum and minimum of vorticity increases up to $\sim 3$. On the other hand, for 
$\phi=0.2$ we clearly notice several local vorticity minima. One minimum is further away from the walls, at a location 
similar to what found for the unladen case. Another minimum is further away towards the corner bisector. The other 
minimum is instead close to the corners. As we have previously shown, at the largest $\phi$ there is a significant mean 
particle concentration exactly at the corners and correspondingly we see two intense spots of vorticity around this 
location, antisymmetric with respect to the corner-bisector.

Although equation~(\ref{vort}) is valid for single-phase duct flow, we have calculated the convective, source/sink 
and diffusive terms for the cases of $\phi=0, 0.05$ and $0.1$. For $\phi=0.2$ there is a strong coupling between the 
dynamics of both fluid and solid phases, and it is therefore difficult to draw conclusions only by estimating the terms in 
equation~(\ref{vort}).

As discussed by \citet{gavrilakis1992}, the production of vorticity within the viscous sublayer is the main 
responsible for the presence of vorticity in the bulk of the flow. The main contribution to the production of vorticity 
is given by the term involving the gradients of the cross-stream normal stresses. For the unladen case, the maxima of 
this term are located close to the corners \cite[at $y/h=0.016$, $z/h=0.16$ from the bottom-left corner, similarly to 
what found by][]{gavrilakis1992}. Another positive, almost negligible contribution to the production of mean streamwise 
vorticity is given by the convective term. On the other hand, the diffusive term and the term involving the gradients 
of the secondary Reynolds stress give a negative contribution to the generation of vorticity in this inner-wall region. 
Note that the largest negative contribution due to the latter term is found close to the maximum production ($(y/h, z/h)
=(0.016, 0.15)$ for $\phi=0$). Generally, we observe that the maximum production and dissipation increase and approach 
the corner as the volume fraction increases up to $\phi=0.1$ (not shown). In order to understand how the presence of particles 
contributes to the generation of vorticity, we calculate the ratio between the overall production and dissipation at the 
location of the maximum of the normal stress term (i.e. the summation of the convective term and the normal stress term, 
divided by the absolute value of the summation between the diffusive and secondary Reynolds stress terms). For the unladen 
case we have a good balance and the ratio is approximately $1$. For $\phi=0.05$ and $0.1$, the ratio is $0.96$ and $0.95$. Hence, 
the presence of the solid phase gives an additional contribution of about $5\%$ to the generation of mean streamwise 
vorticity in the near-wall region at the location of maximum production. The increased production due to larger gradients 
of the normal stress difference and due to the additional contribution by particles, leads globally to the larger $\Omega_f$ 
observed at $\phi=0.05$ and $0.1$.

\subsection{Quadrant analysis and two-point velocity correlations}

In this section we employ the quadrant analysis to identify the contribution from so-called ejection and sweep 
events to the production of $\langle u_f'v_f' \rangle$ and $\langle u_f'w_f' \rangle$. In single-phase 
wall-bounded turbulent flows, it is known that these phenomena are associated to pairs of counter-rotating streamwise vortices 
that exist in the shear layer near the wall. These force low-momentum fluid at the wall towards the 
high-speed core of the flow. This is typically referred to as a Q2 or ejection event, with negative 
$u_f'$ and positive $v_f'$. On the other hand, events with positive $u_f'$ and negative $v_f'$ are 
associated with the inrush of high-momentum fluid towards the walls and are known as sweeps or Q4 events. 
Events Q1 (positive $u_f'$ and positive $v_f'$) and Q3 (negative $u_f'$ and negative $v_f'$) are not 
associated with any particular turbulent structure when there is only one inhomogeneous direction. 
However, as discussed by \citet{huser93} in turbulent duct flows these also contribute to the total 
turbulence production.

\begin{figure}
   \centering
   \includegraphics[width=0.49\textwidth]{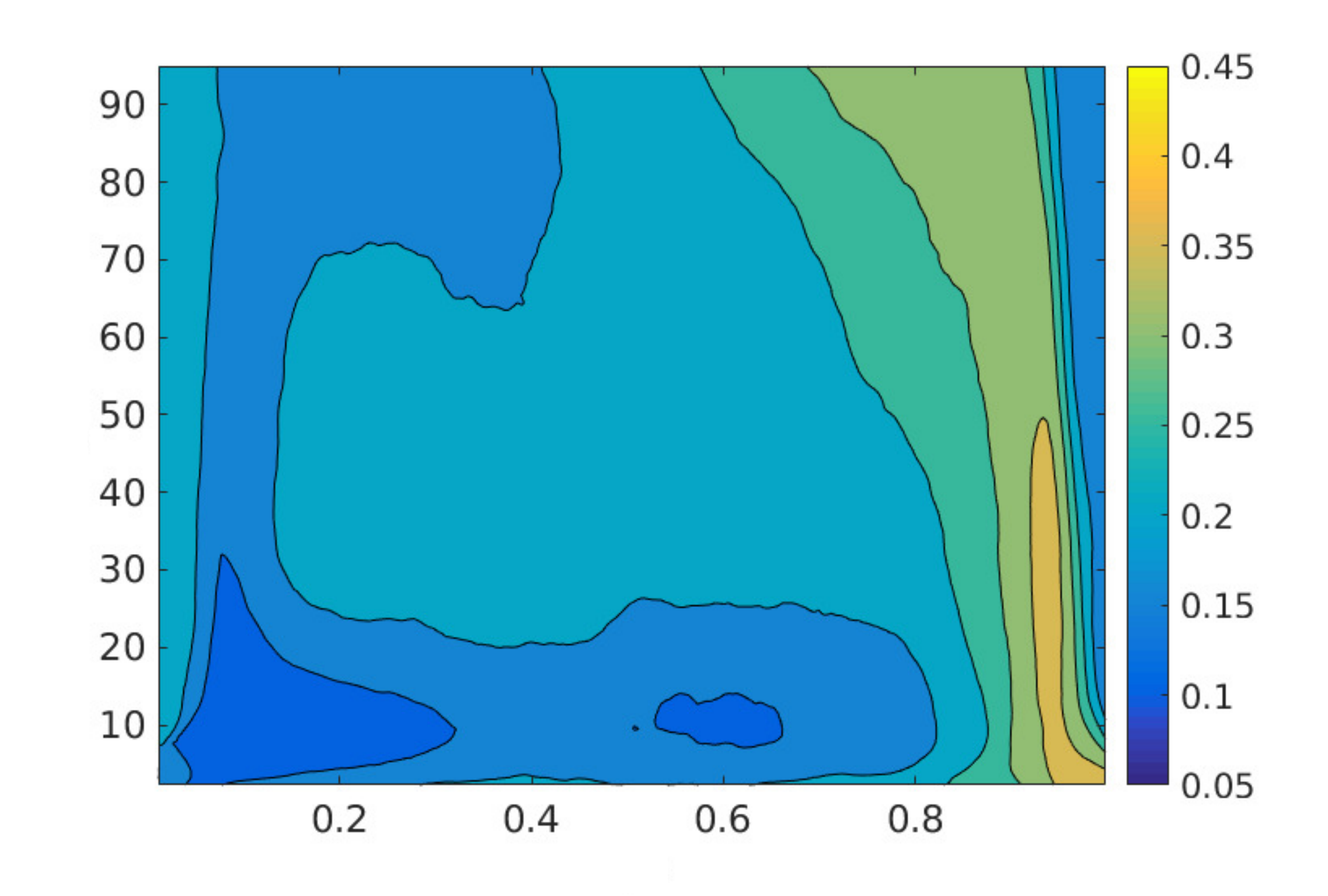}
   \includegraphics[width=0.49\textwidth]{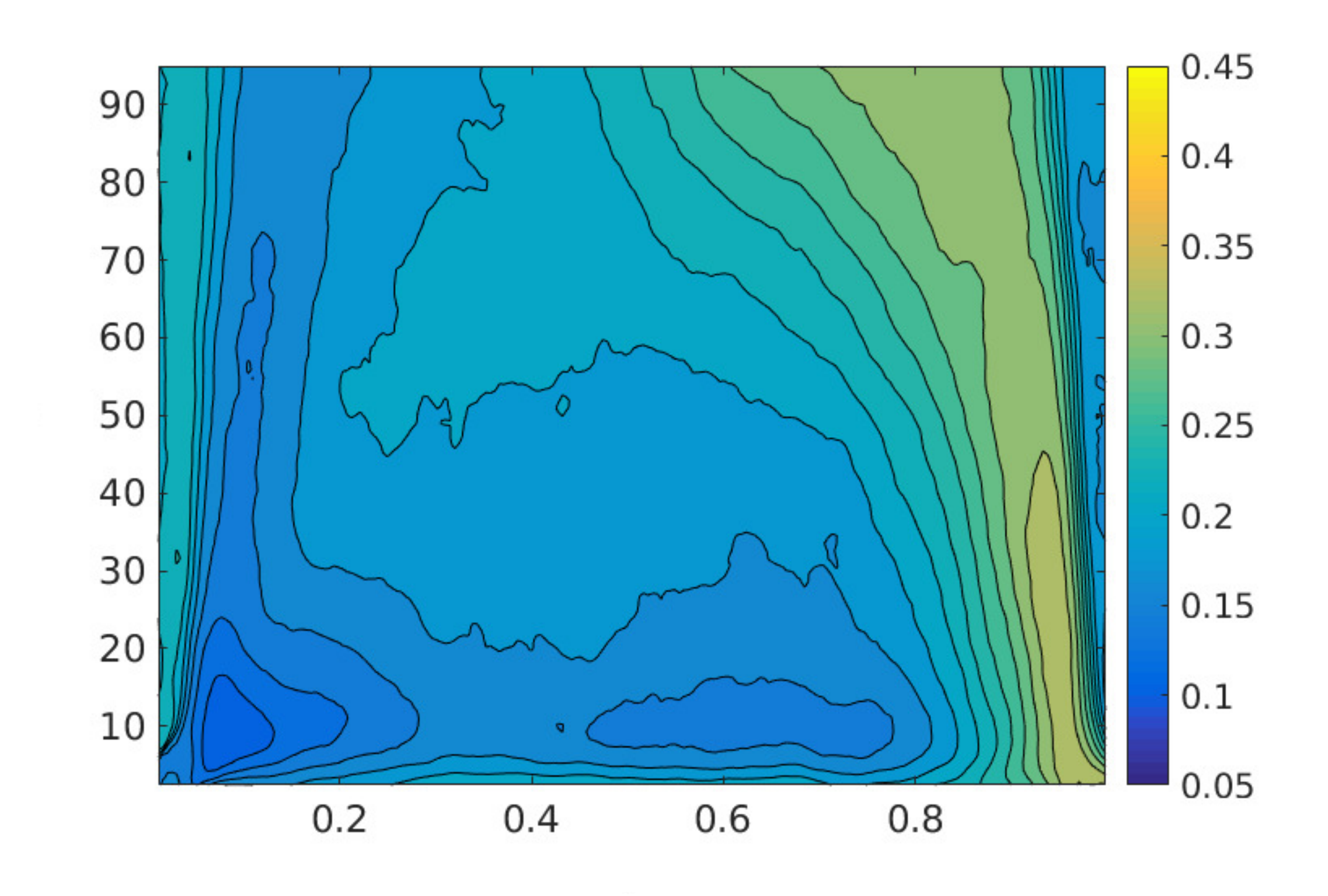}
   \put(-380,62){\rotatebox{90}{\large $y^+$}}
   \put(-222,125){{$Q_1$}}
   \put(-30,125){{$Q_1$}}
   \put(-298,-2){{\large $z/h$}}
   \put(-108,-2){{\large $z/h$}}
   \put(-295,125){\footnotesize $(a)$}
   \put(-105,125){\footnotesize $(b)$}\\ 
   \includegraphics[width=0.49\textwidth]{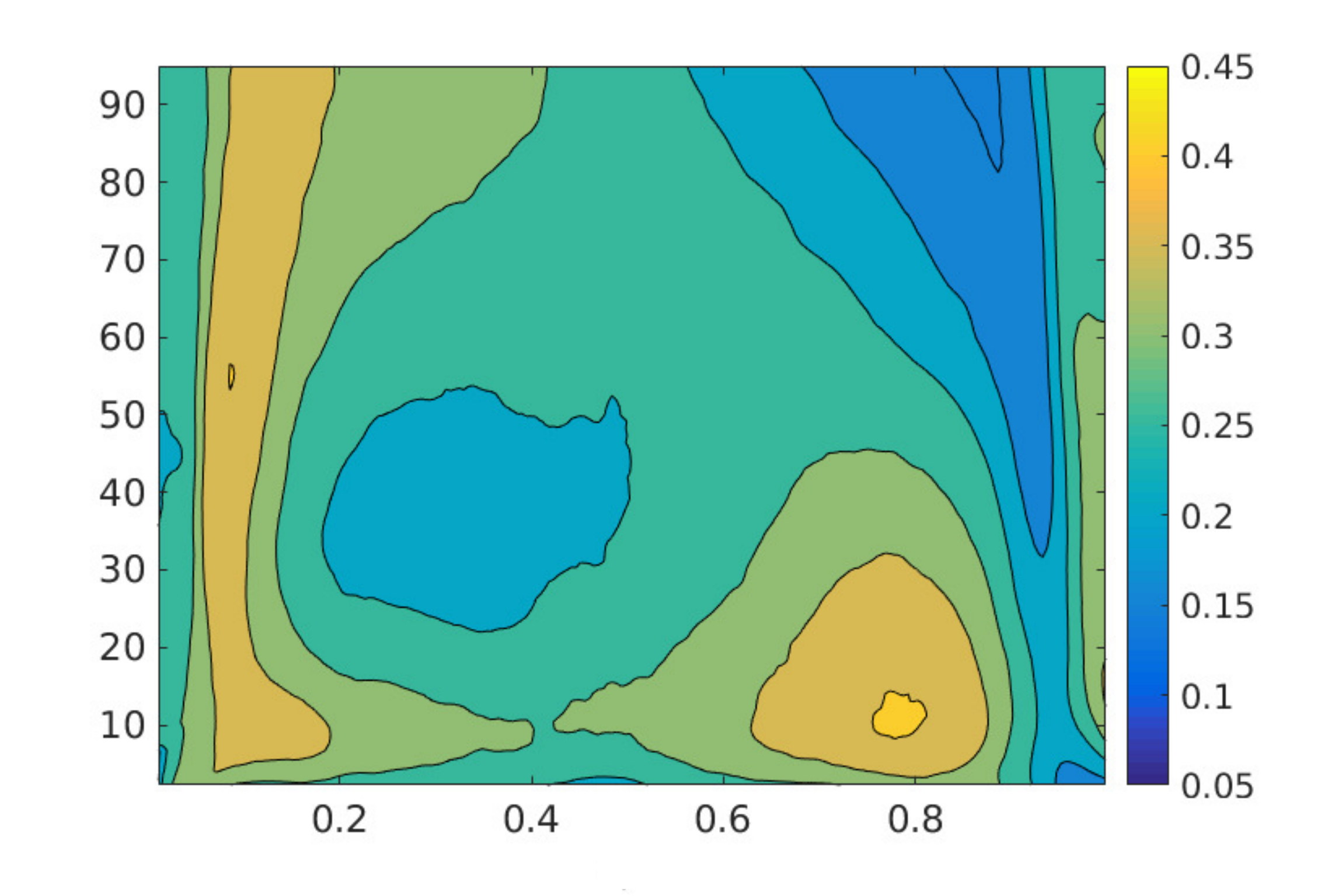}
   \includegraphics[width=0.49\textwidth]{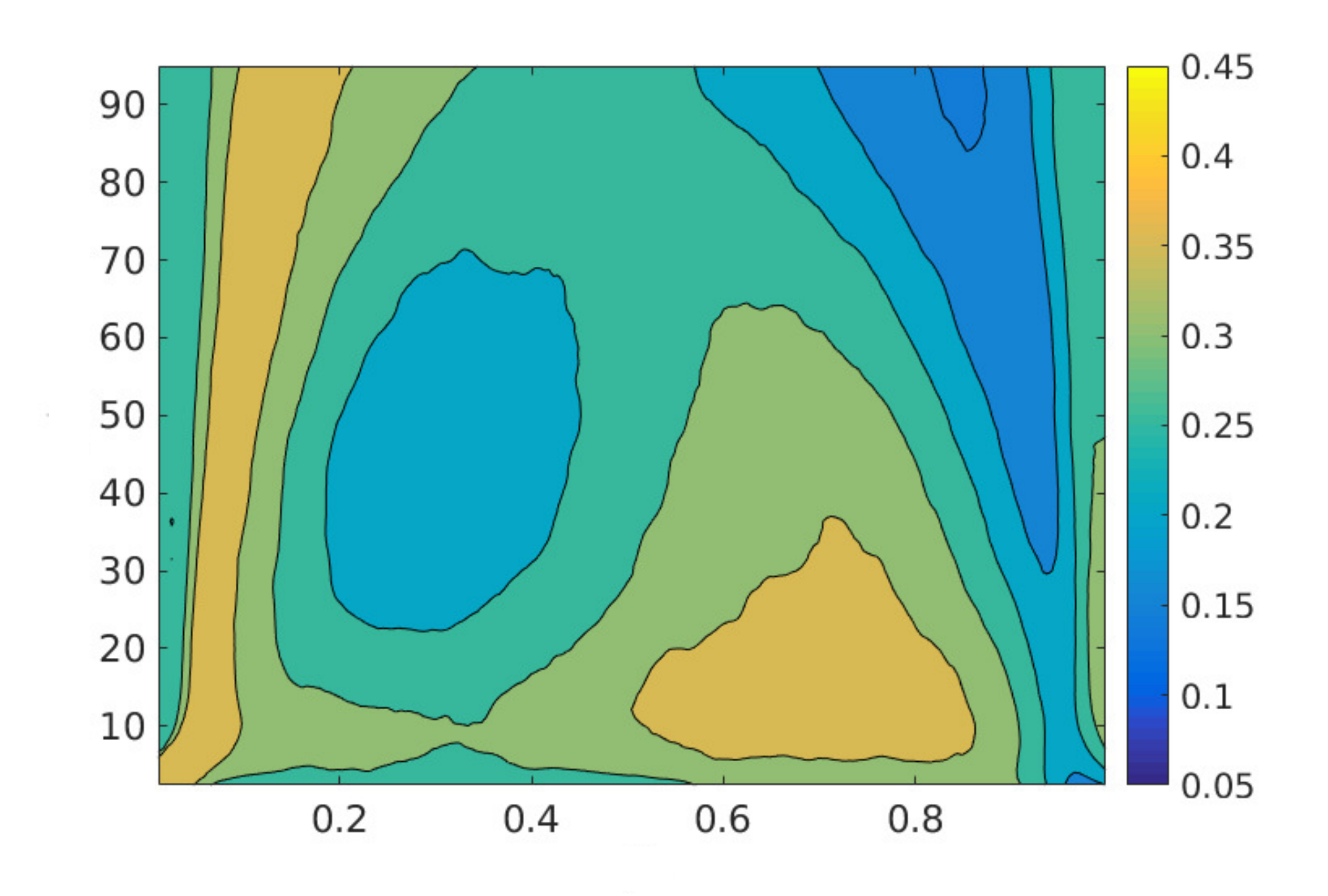}
   \put(-380,62){\rotatebox{90}{\large $y^+$}}
   \put(-222,125){{$Q_2$}}
   \put(-30,125){{$Q_2$}}
   \put(-298,-2){{\large $z/h$}}
   \put(-108,-2){{\large $z/h$}}
   \put(-295,125){\footnotesize $(c)$}
   \put(-105,125){\footnotesize $(d)$}\\ 
   \includegraphics[width=0.49\textwidth]{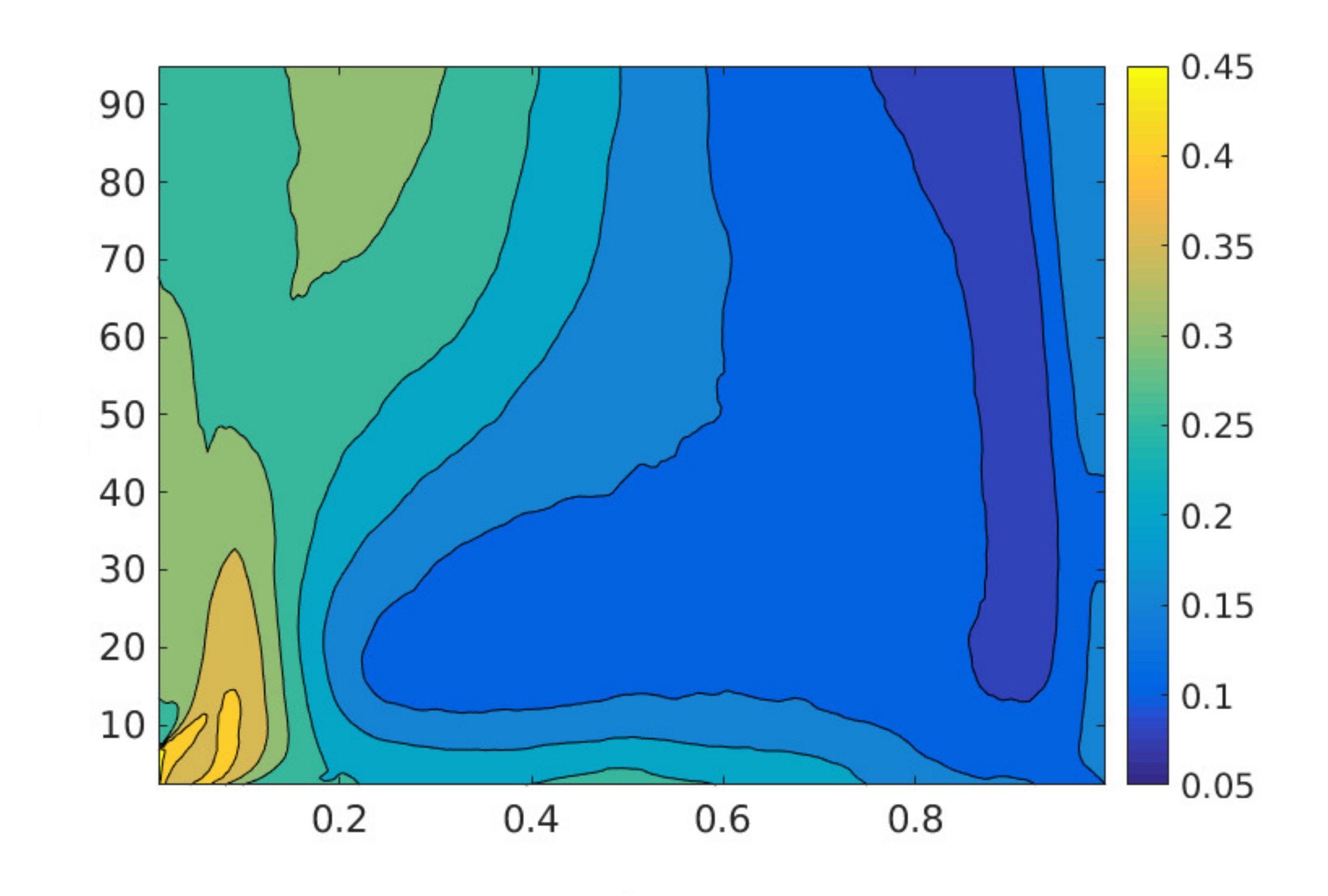}
   \includegraphics[width=0.49\textwidth]{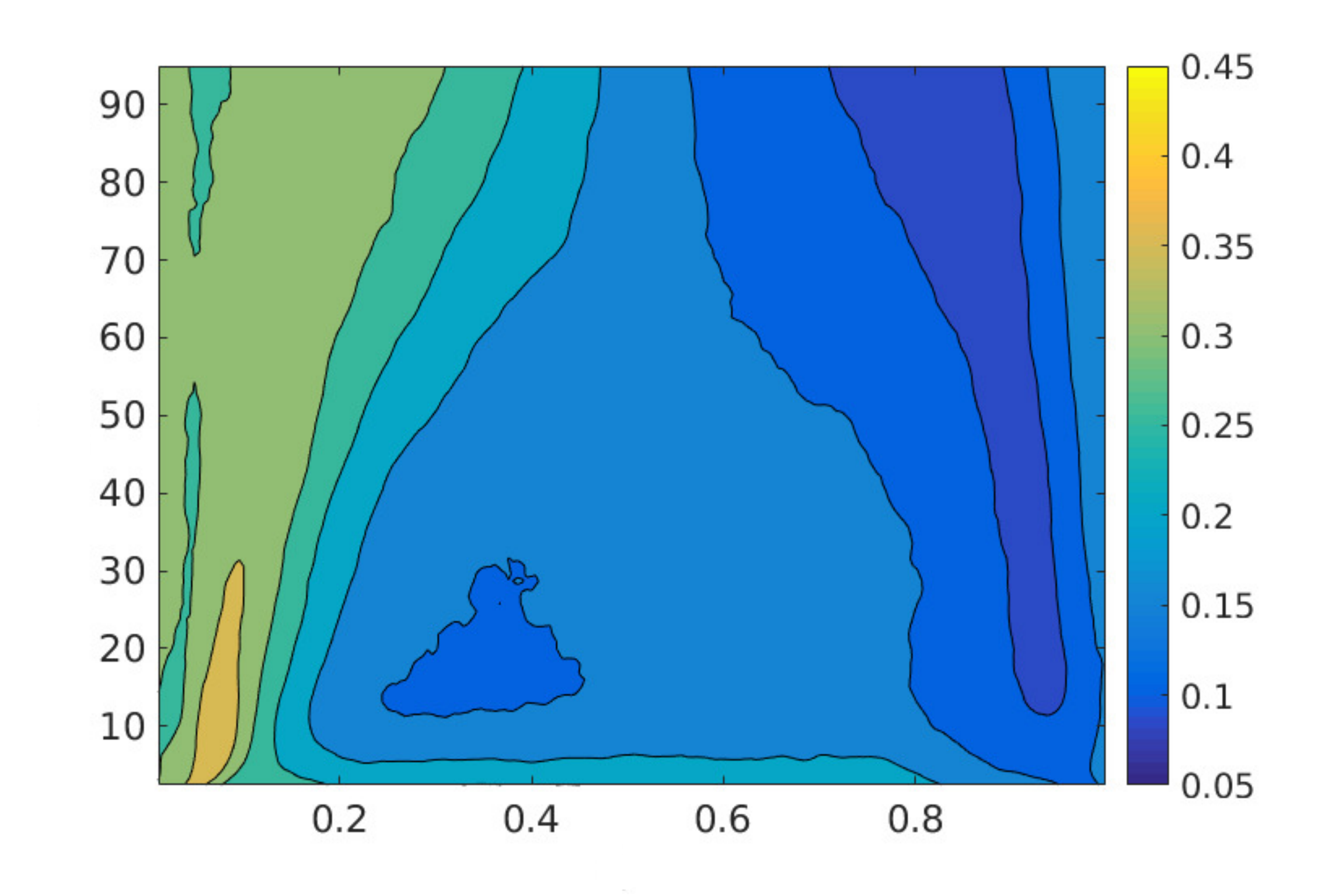}
   \put(-380,62){\rotatebox{90}{\large $y^+$}}
   \put(-222,125){{$Q_3$}}
   \put(-30,125){{$Q_3$}}
   \put(-298,-2){{\large $z/h$}}
   \put(-108,-2){{\large $z/h$}}
   \put(-295,125){\footnotesize $(e)$}
   \put(-105,125){\footnotesize $(f)$}\\ 
   \includegraphics[width=0.49\textwidth]{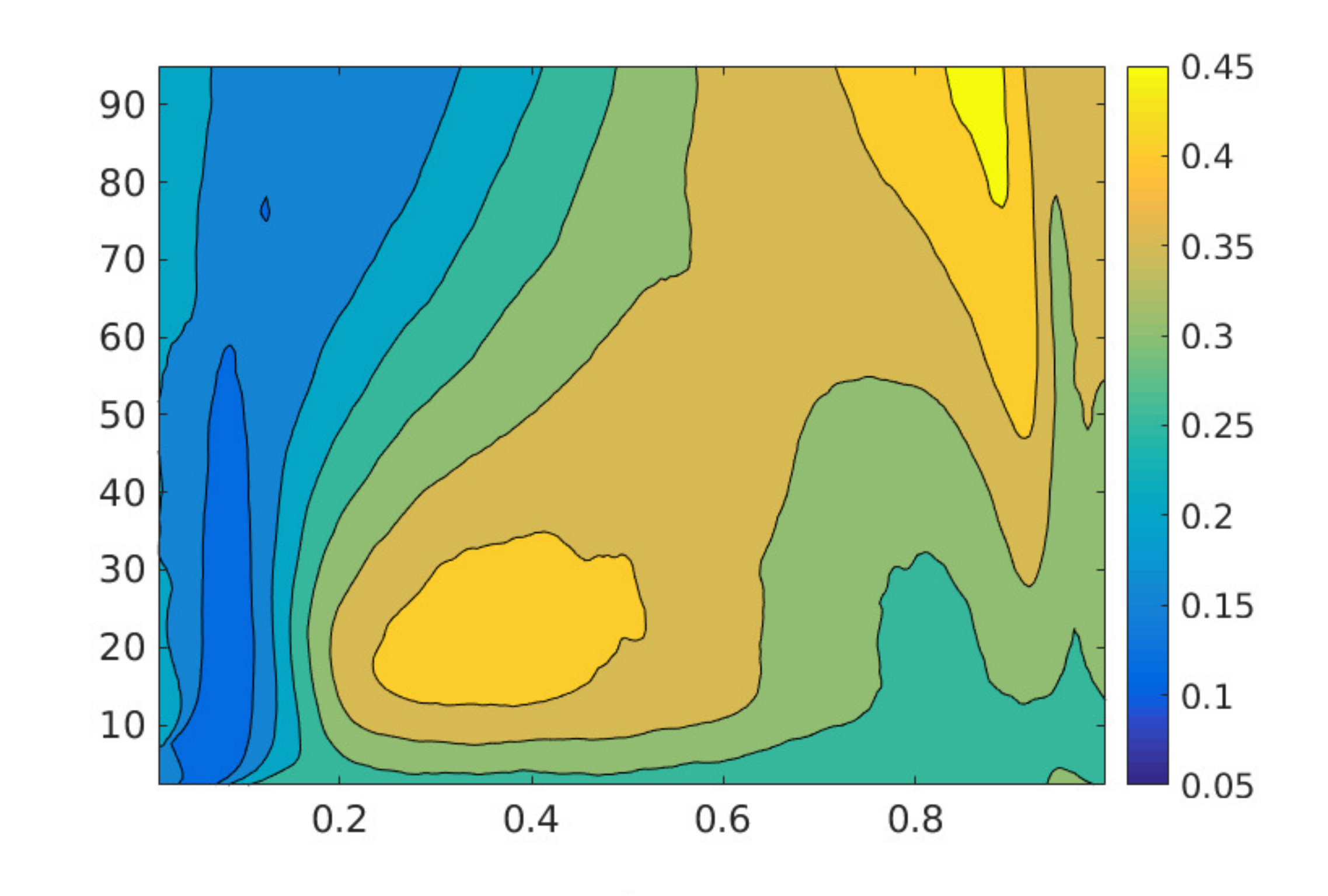}
   \includegraphics[width=0.49\textwidth]{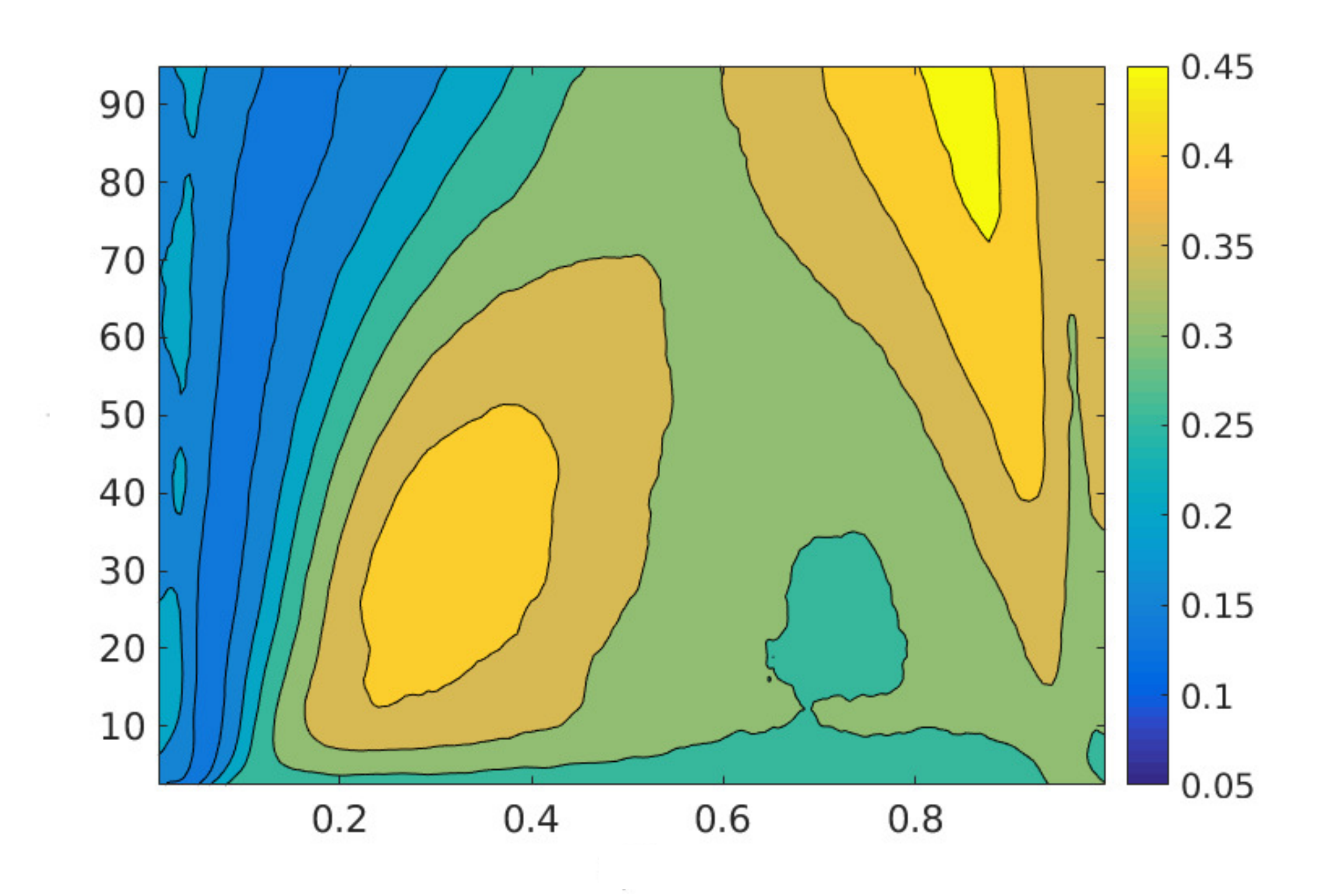}
   \put(-380,62){\rotatebox{90}{\large $y^+$}}
   \put(-222,125){{$Q_4$}}
   \put(-30,125){{$Q_4$}}
   \put(-298,-2){{\large $z/h$}}
   \put(-108,-2){{\large $z/h$}}
   \put(-295,125){\footnotesize $(g)$}
   \put(-105,125){\footnotesize $(h)$}\\ 
  \vspace{3pt}        
  \caption{Maps of probability of Q1, Q2, Q3, Q4 events in panels (a), (c), (e), (g) and (b), (d), (f), (h) for particle 
volume fractions $\phi=0.0$ and $0.05$, respectively.}
\label{fig:quad1}
\end{figure}

\begin{figure}
   \centering
   \includegraphics[width=0.49\textwidth]{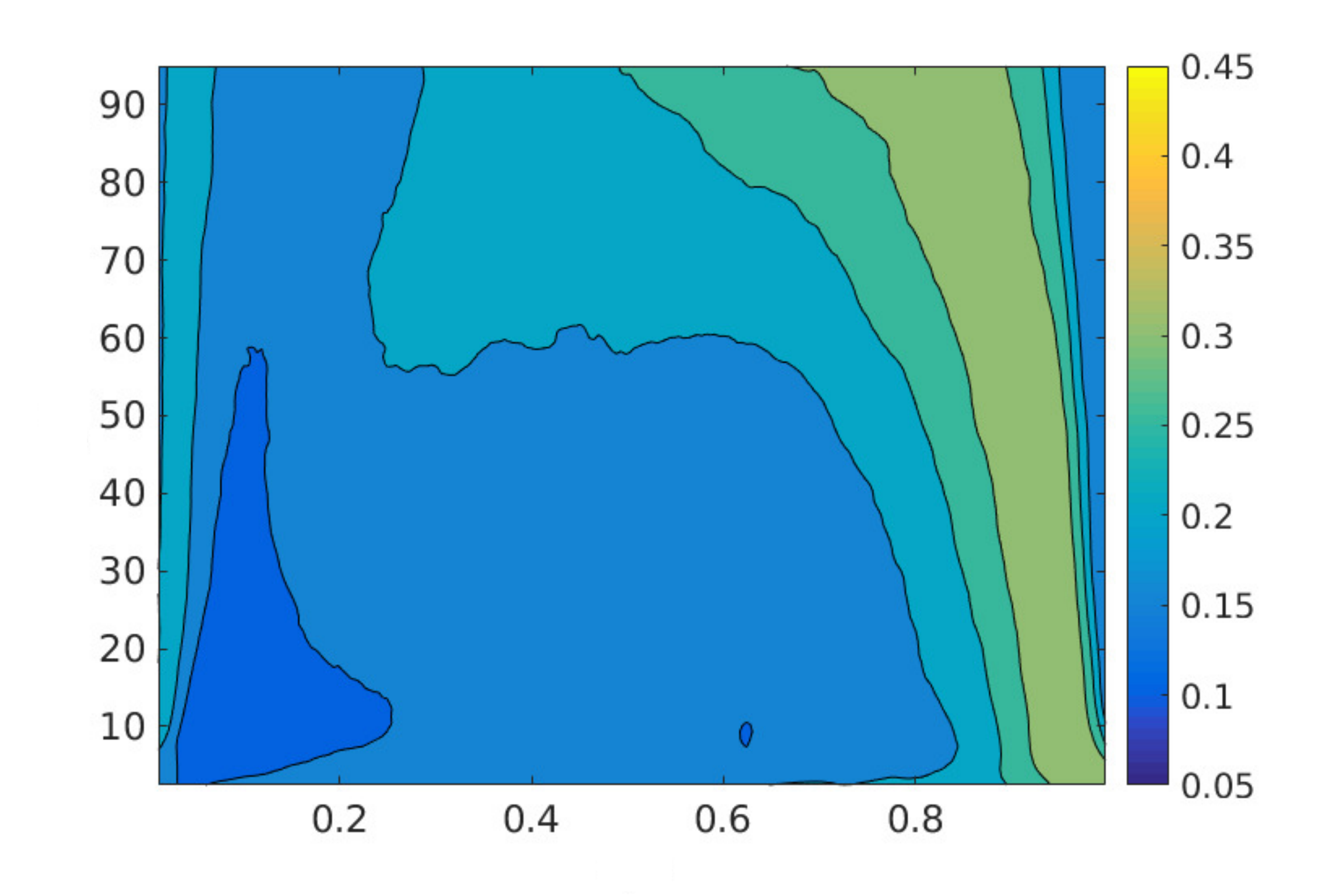}
   \includegraphics[width=0.49\textwidth]{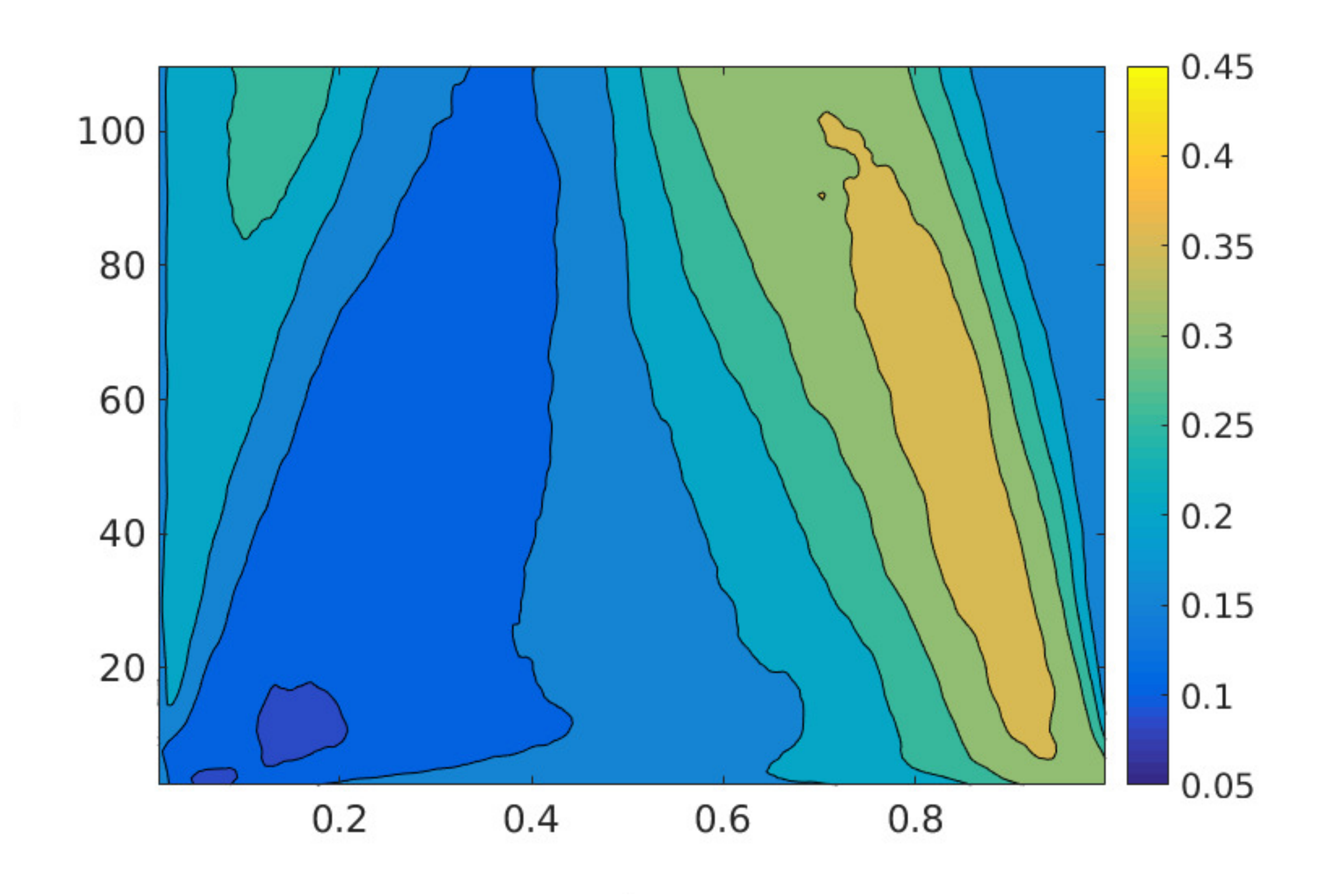}
   \put(-380,62){\rotatebox{90}{\large $y^+$}}
   \put(-222,125){{$Q_1$}}
   \put(-30,125){{$Q_1$}}
   \put(-298,-2){{\large $z/h$}}
   \put(-108,-2){{\large $z/h$}}
   \put(-295,125){\footnotesize $(a)$}
   \put(-105,125){\footnotesize $(b)$}\\ 
   \includegraphics[width=0.49\textwidth]{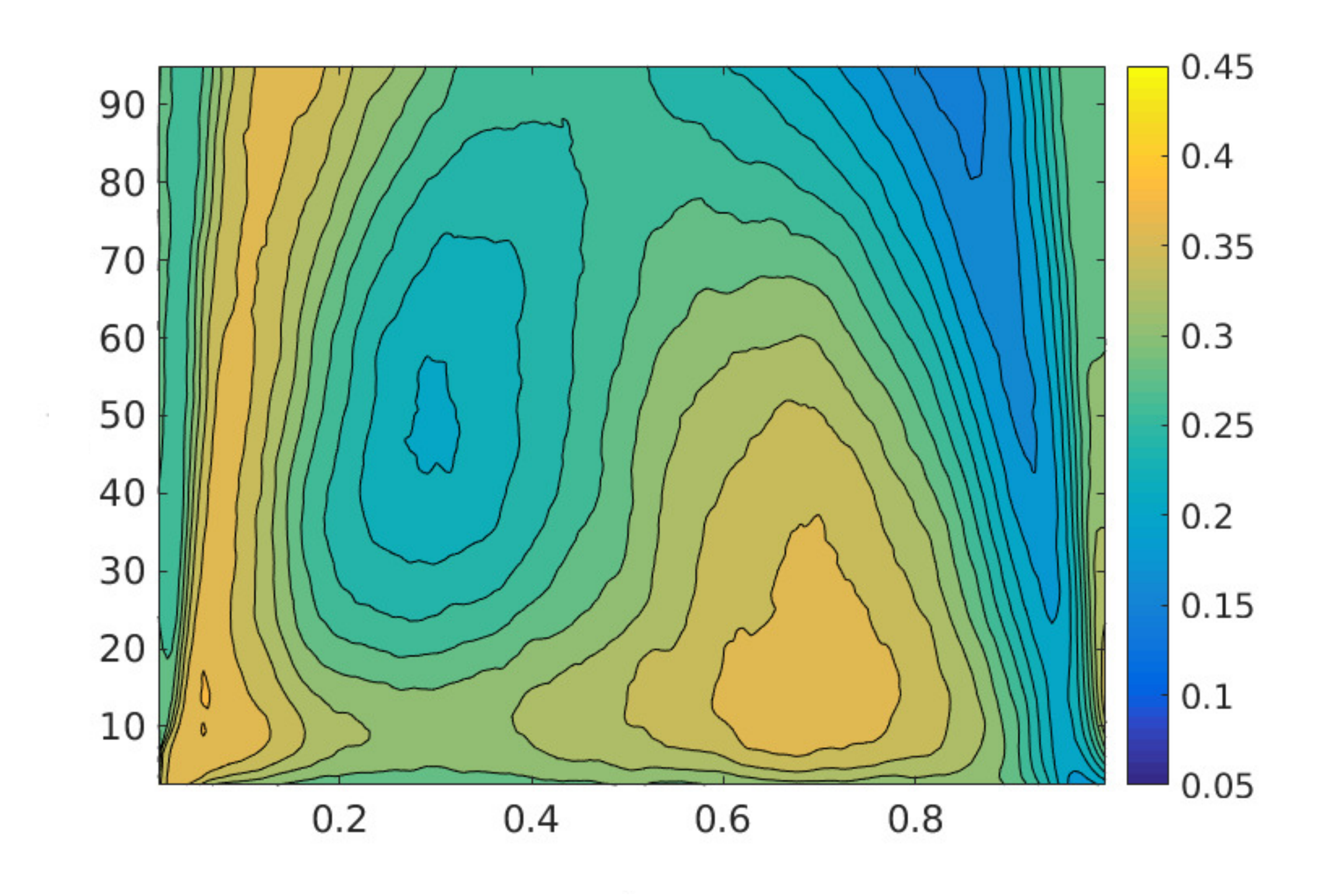}
   \includegraphics[width=0.49\textwidth]{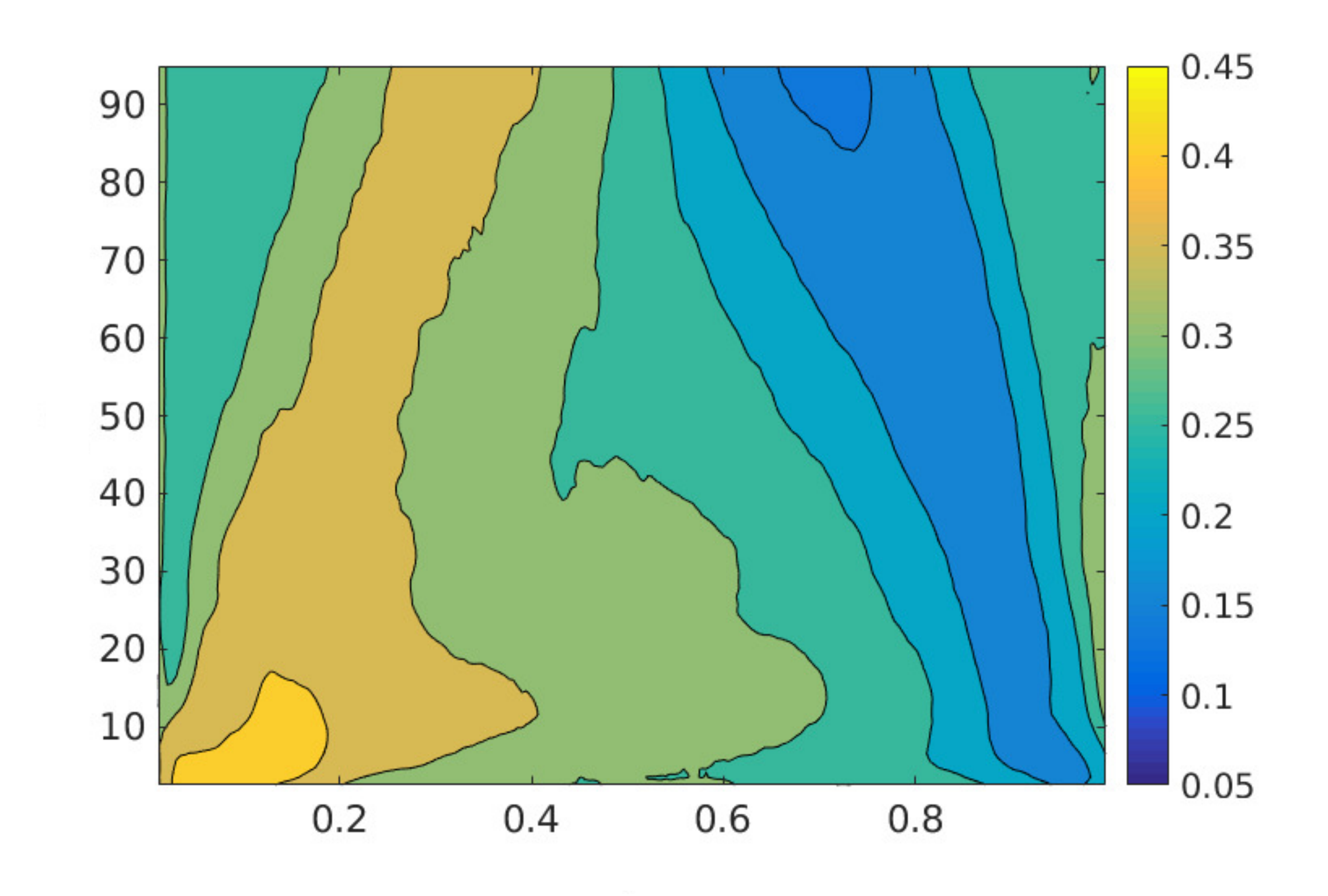}
   \put(-380,62){\rotatebox{90}{\large $y^+$}}
   \put(-222,125){{$Q_2$}}
   \put(-30,125){{$Q_2$}}
   \put(-298,-2){{\large $z/h$}}
   \put(-108,-2){{\large $z/h$}}
   \put(-295,125){\footnotesize $(c)$}
   \put(-105,125){\footnotesize $(d)$}\\ 
   \includegraphics[width=0.49\textwidth]{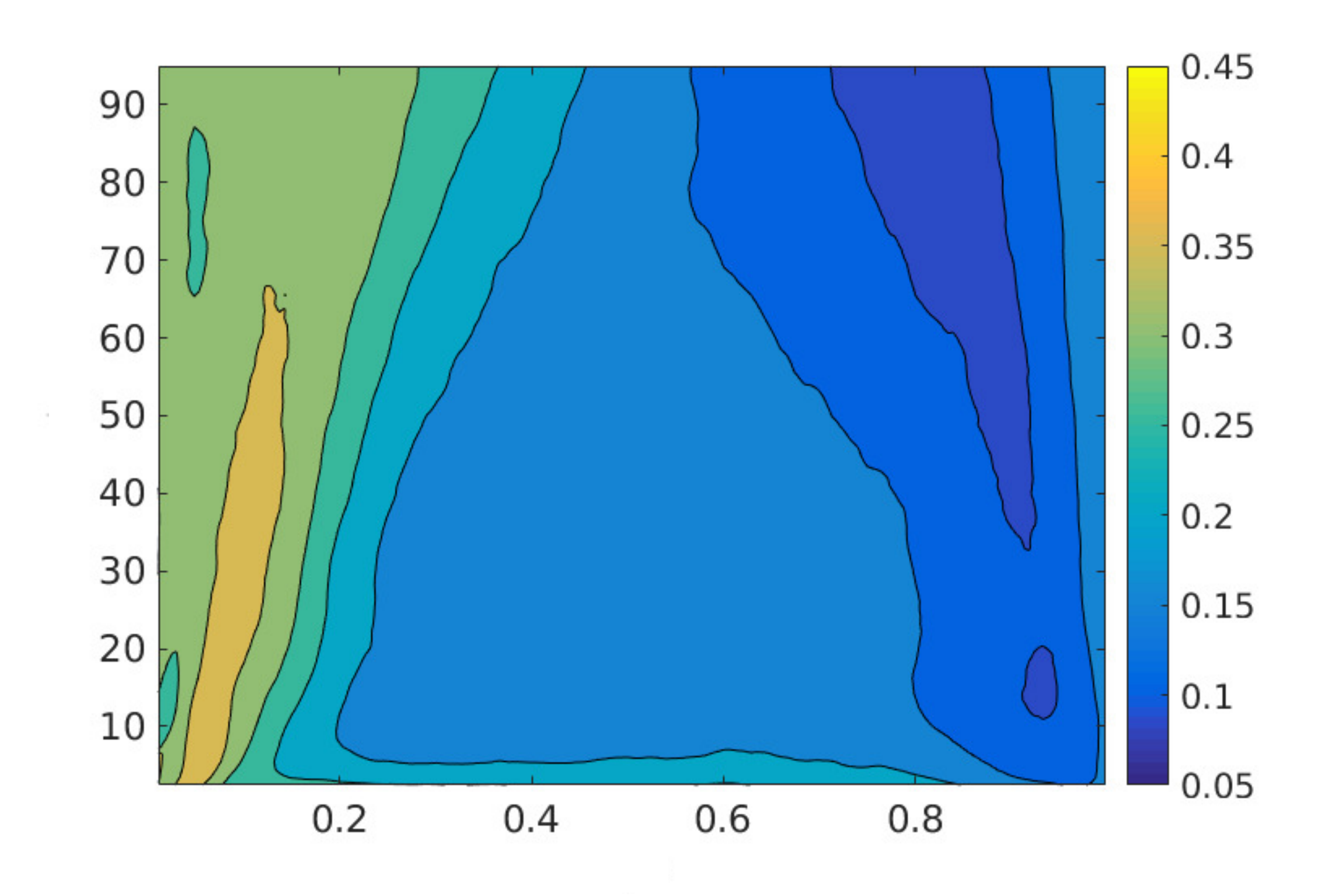}
   \includegraphics[width=0.49\textwidth]{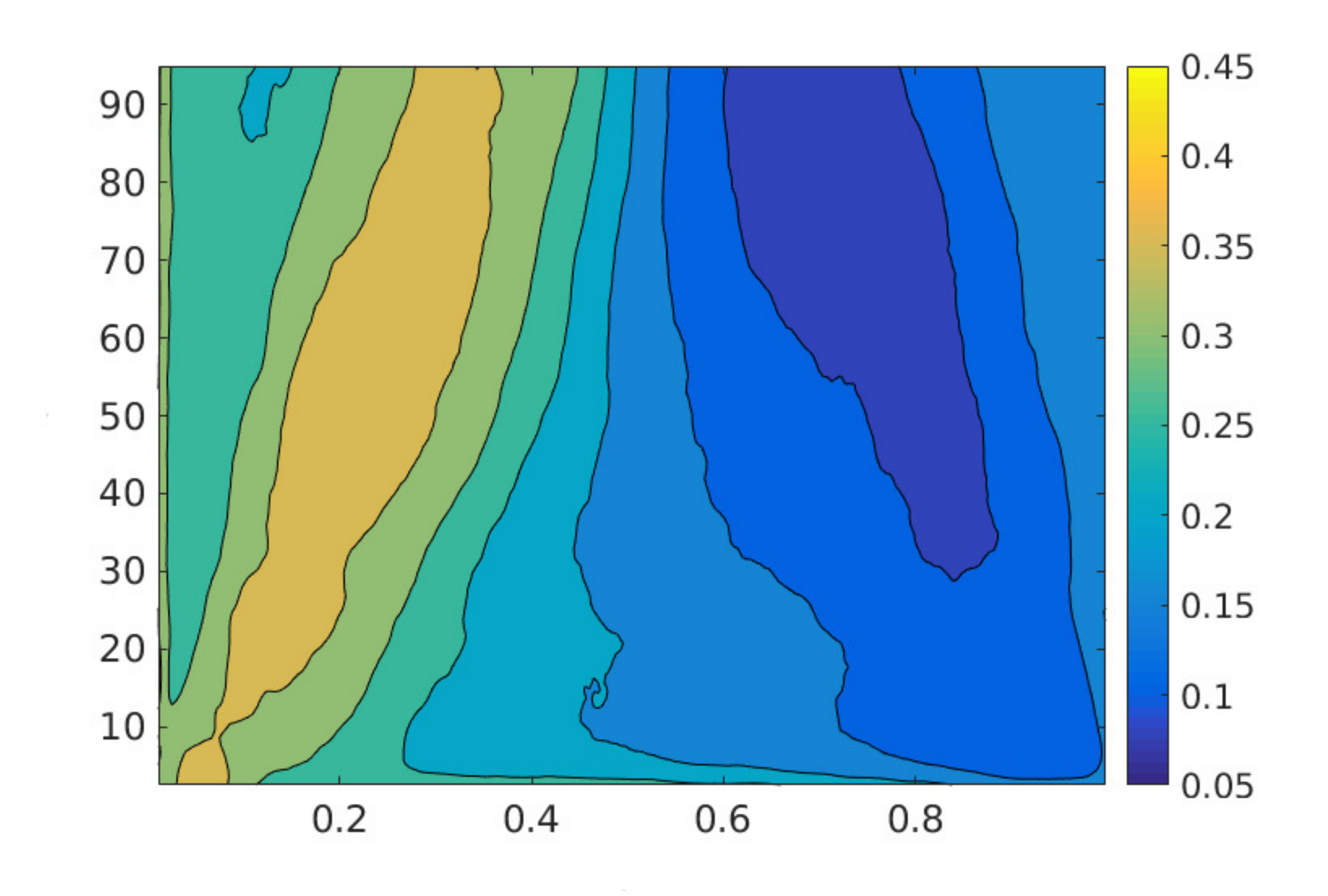}
   \put(-380,62){\rotatebox{90}{\large $y^+$}}
   \put(-222,125){{$Q_3$}}
   \put(-30,125){{$Q_3$}}
   \put(-298,-2){{\large $z/h$}}
   \put(-108,-2){{\large $z/h$}}
   \put(-295,125){\footnotesize $(e)$}
   \put(-105,125){\footnotesize $(f)$}\\ 
   \includegraphics[width=0.49\textwidth]{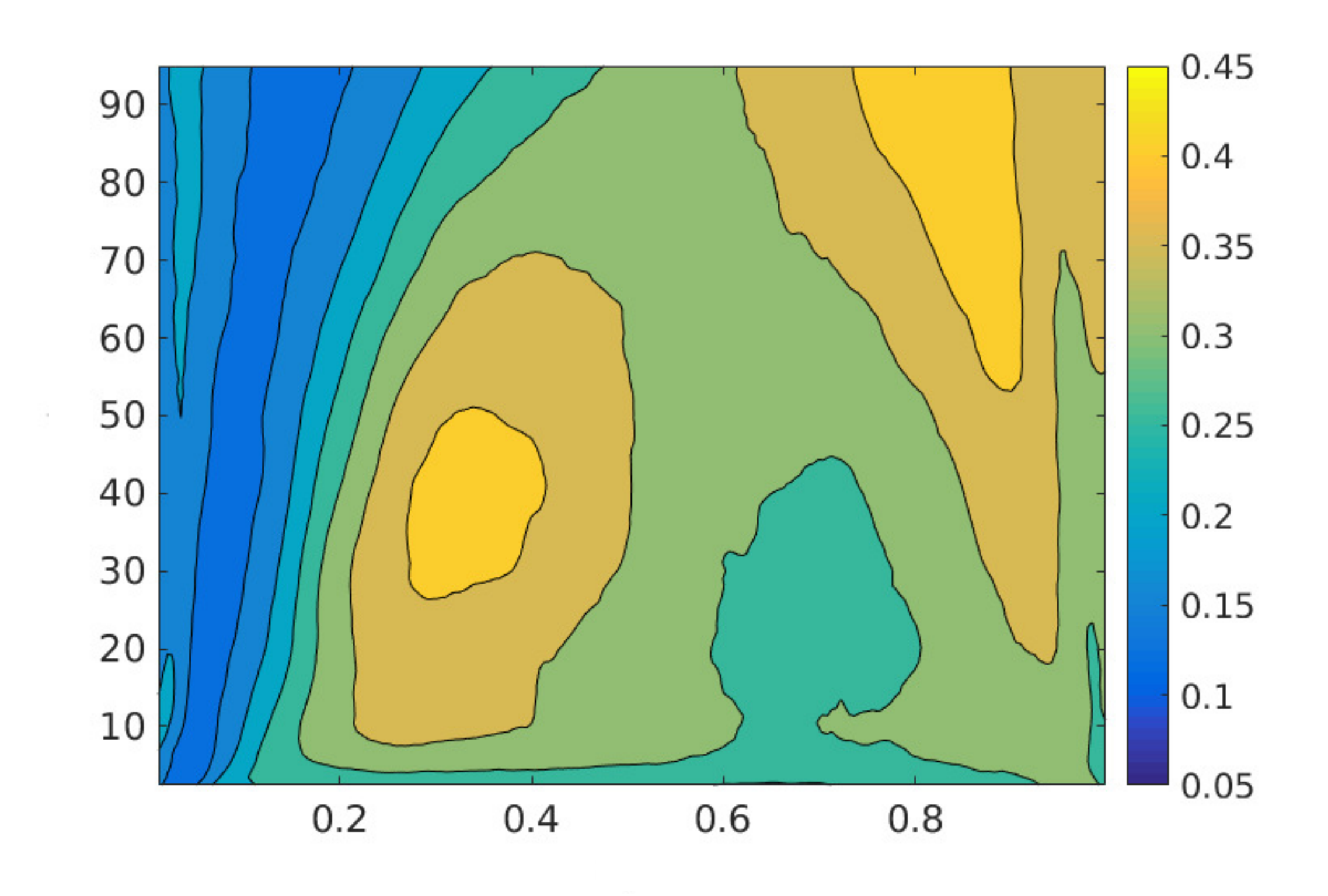}
   \includegraphics[width=0.49\textwidth]{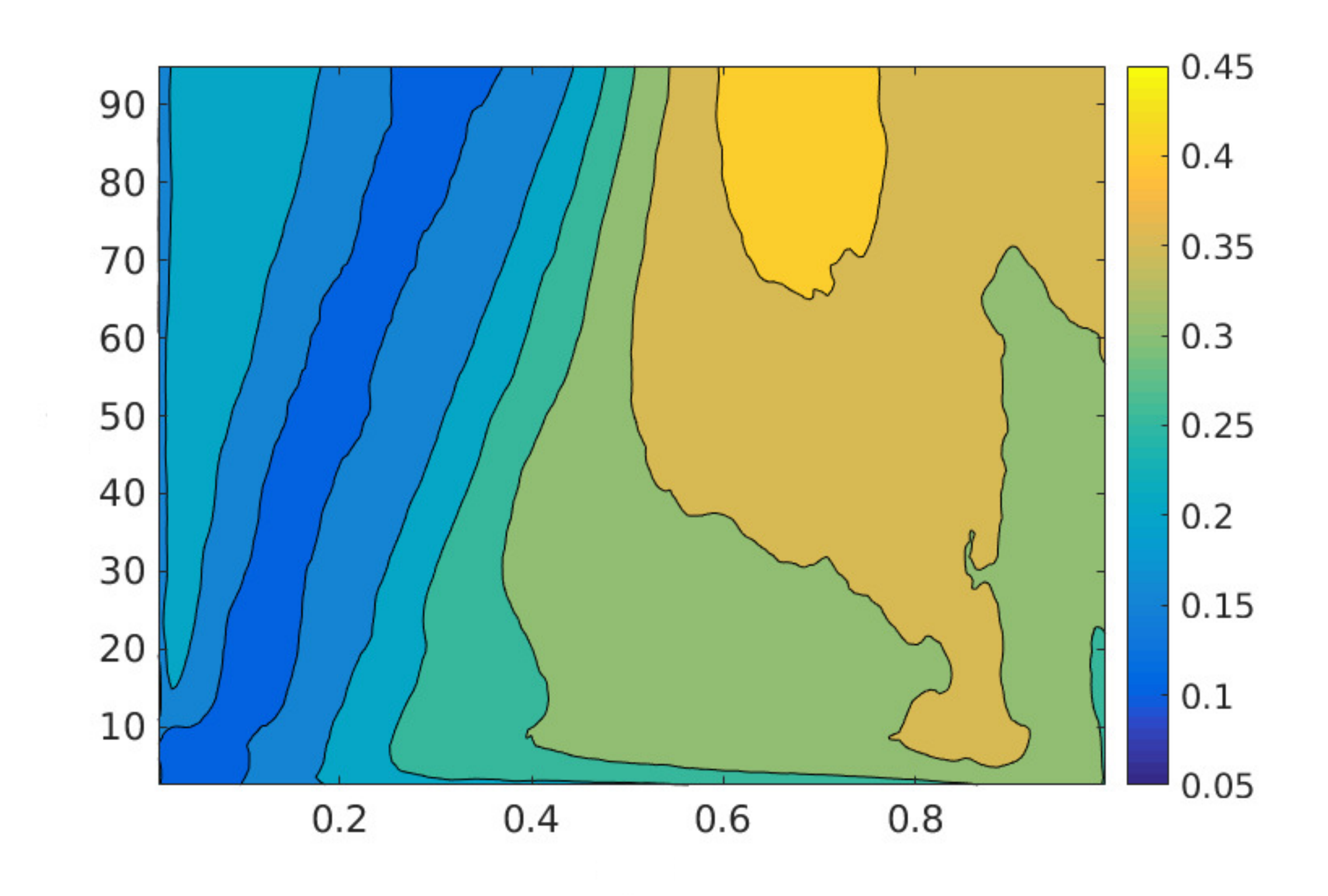}
   \put(-380,62){\rotatebox{90}{\large $y^+$}}
   \put(-222,125){{$Q_4$}}
   \put(-30,125){{$Q_4$}}
   \put(-298,-2){{\large $z/h$}}
   \put(-108,-2){{\large $z/h$}}
   \put(-295,125){\footnotesize $(g)$}
   \put(-105,125){\footnotesize $(h)$}\\ 
  \vspace{3pt}        
  \caption{Maps of probability of Q1, Q2, Q3, Q4 events in panels (a), (c), (e), (g) and (b), (d), (f), (h) for particle volume 
fractions $\phi=0.1$ and $0.2$, respectively.}
\label{fig:quad2}
\end{figure}

We first show contours of the probability of finding Q1, Q2, Q3, Q4 events for $\phi=0.0$ and $0.05$ in 
figure~\ref{fig:quad1}, and for $\phi=0.1$ and $0.2$ in figure~\ref{fig:quad2}. The maximum probability 
of an event is $1$, so that the sum Q1+Q2+Q3+Q4=$1$ at each point. Since these events 
are typically important close to the walls, the $y$-coordinate is reported in inner units (the 
viscous length at the wall-bisector is chosen), while the wall-parallel $z$-coordinate spans 
half of the duct. The contours of the probability of the different events directly enables us to easily compare and 
discuss all cases. 

Results for the unladen case are substantially in agreement with those by \citet{huser93} and \citet{joung2007}. 
Ejection (Q2) events are important at the wall-bisector ($y/h=1$) and around $y/h\sim0.8$. \citet{joung2007} 
report strong Q2 events already $y/h=0.62$, and this is probably due to the fact that their $Re_b$ was 
smaller  than ours ($Re_b=4410$ instead of 5600). Q4 events are instead dominant closer to the vertical walls, around 
$y/h=0.3$ \cite[as also reported by][]{joung2007}. Q3 events are found when both $u_f'$ and $v_f'$ are 
negative. As also shown by \citet{huser93}, we see that these increase from the wall-bisector towards 
the vertical walls, being dominant at the corners. Consequently, these events are created by ejections 
from the vertical wall (and particularly from the corners), resulting in the increase of $\langle u_f'v_f' 
\rangle$ in this region.

The general picture is similar for $\phi=0.05$ and $0.1$. The probability maps of Q1, Q2, Q3 and Q4 events are only 
slightly changed with respect to those of the unladen case. 
Concerning the probability of Q3 events we see that it is still higher at the corner and that it increases 
with $\phi$ along the vertical wall and along the corner-bisector. 
It is clear from figure~\ref{fig:quad2} that the turbulence activity is strongly reduced for $\phi=0.2$. 
The probability of ejection and sweep events close to the walls is drastically reduced. The probability of 
sweep (Q4) events is around $35\%$ only in a small region close to the wall bisector ($y/h \sim 0.85$). 
Instead, the probabilities of Q2 (ejections) and Q3 events are above $35\%$ in a region around the corner 
bisector, where we recall there is high mean particle concentration ($y/h \in (0;0.4)$). At this volume 
fraction the probability of these events is therefore mostly correlated to fluid-particle interactions. The 
probability contours for $\phi=0.2$ are indeed very different from those at lower $\phi$ and from those 
of the unladen case. The absence of high probability regions of Q2 and Q4 events around $z/h \sim 0.8$ and 
$z/h \sim 0.3$ denotes the disruption of the coherent streamwise vortical structures.

The presence of two inhomogeneous directions allows us to extend the quadrant analysis to the secondary shear stress 
$\langle v_f'w_f' \rangle$, which contributes to the production, dissipation and transport of mean streamwise vorticity as discussed above. 
As in \citet{huser93}, to distinguish between the events related to the primary and secondary Reynolds stresses, we will 
refer to the latter as $Q1_s, Q2_s, Q3_s$ and $Q4_s$ events. These authors showed that as the corners are approached, 
contributions of the $Q1_s$ and $Q3_s$ events progressively decrease. On the other hand, $Q2_s$ and $Q4_s$ events 
are found to be stronger than the latter close to the corners. This is due to the interaction between ejections from both walls 
that tilt the ejection stem toward the perpendicular wall. We looked at probability maps of these events as previously 
done for the events related to the primary Reynolds stress. For the unladen case we see indeed that close to the corners 
the probability of having $Q2_s$ and $Q4_s$ events is substantially larger than that of $Q1_s, \, Q3_s$ events (not shown). 
As we increase the volume fraction $\phi$ the probability of these events remains high until the highest $\phi=0.2$ is 
reached. In this case, we find that there is an approximately equal share of probability between all type of events 
indicating again that there is a substantial alteration of the turbulence and less structured flow in the presence of many particles. 

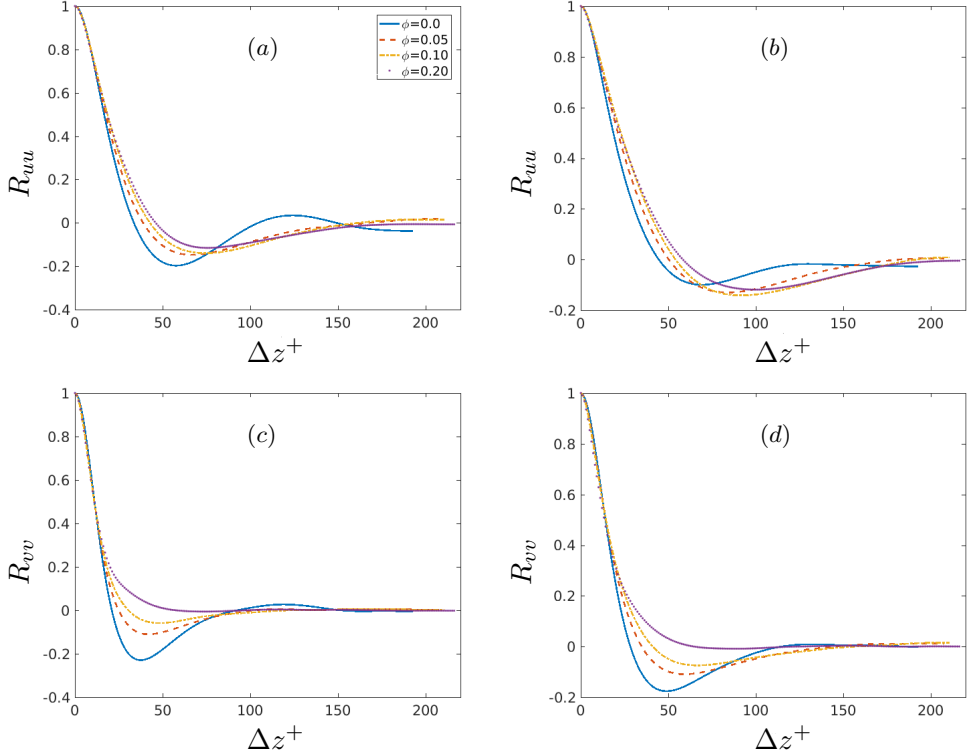
\begin{figure}
   \centering
   \includegraphics[width=0.49\textwidth]{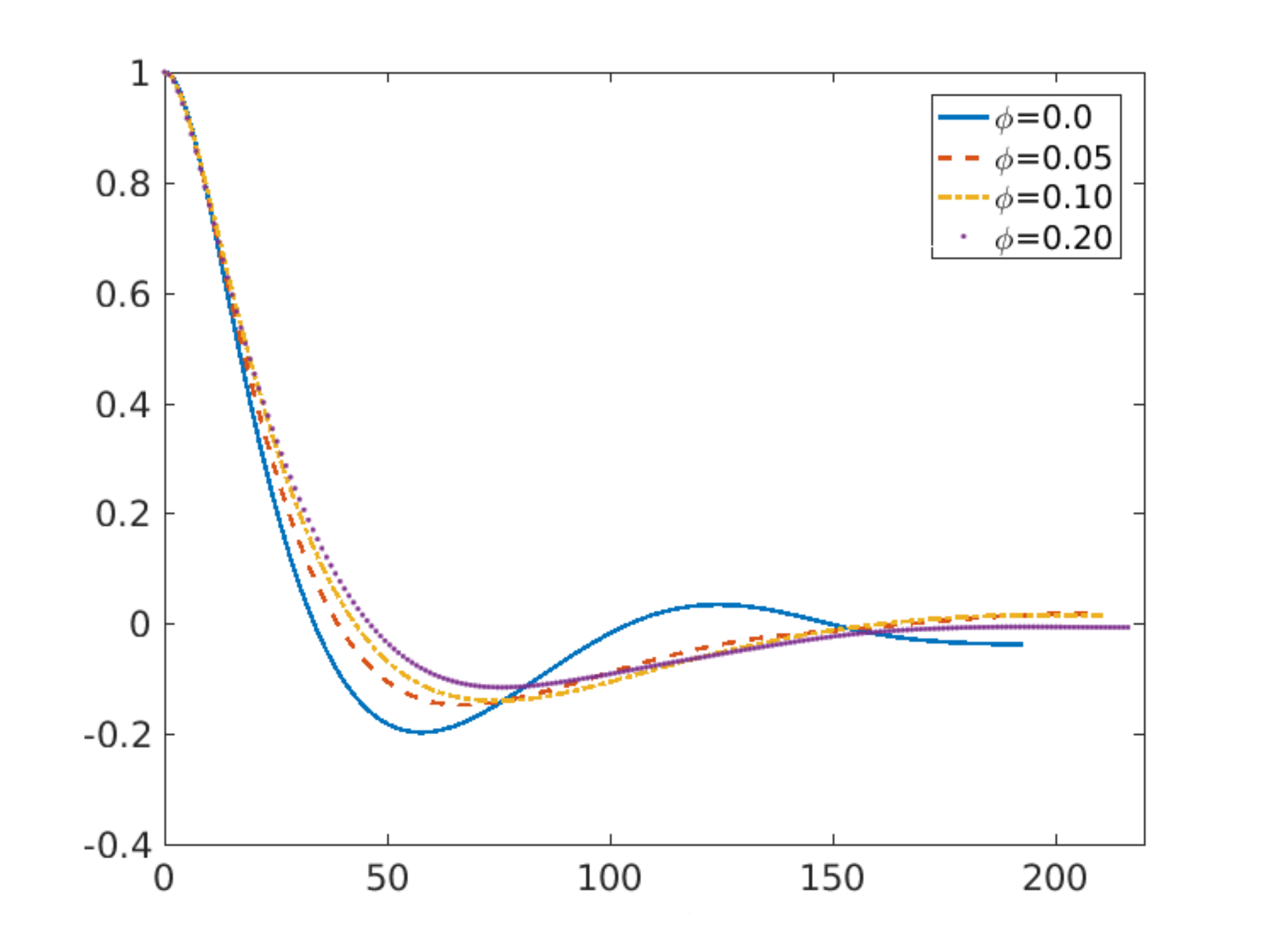}
   \includegraphics[width=0.49\textwidth]{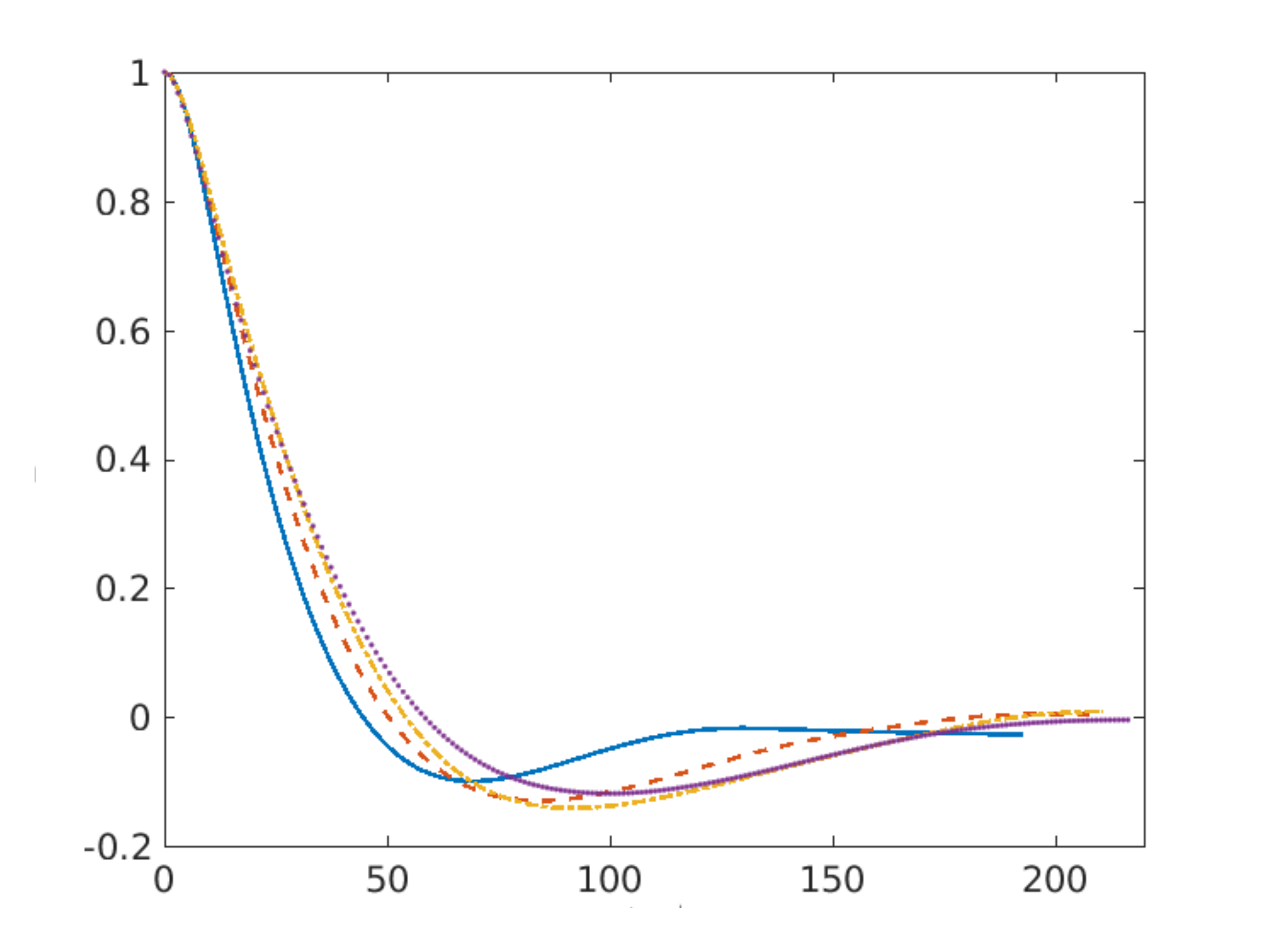}
   \put(-380,58){\rotatebox{90}{\large $R_{uu}$}}
   \put(-290,-4){{\large $\Delta z^+$}}
   \put(-190,58){\rotatebox{90}{\large $R_{uu}$}}
   \put(-98,-4){{\large $\Delta z^+$}}
   \put(-290,112){\footnotesize $(a)$}
   \put(-96,112){\footnotesize $(b)$}\\ 
   \includegraphics[width=0.49\textwidth]{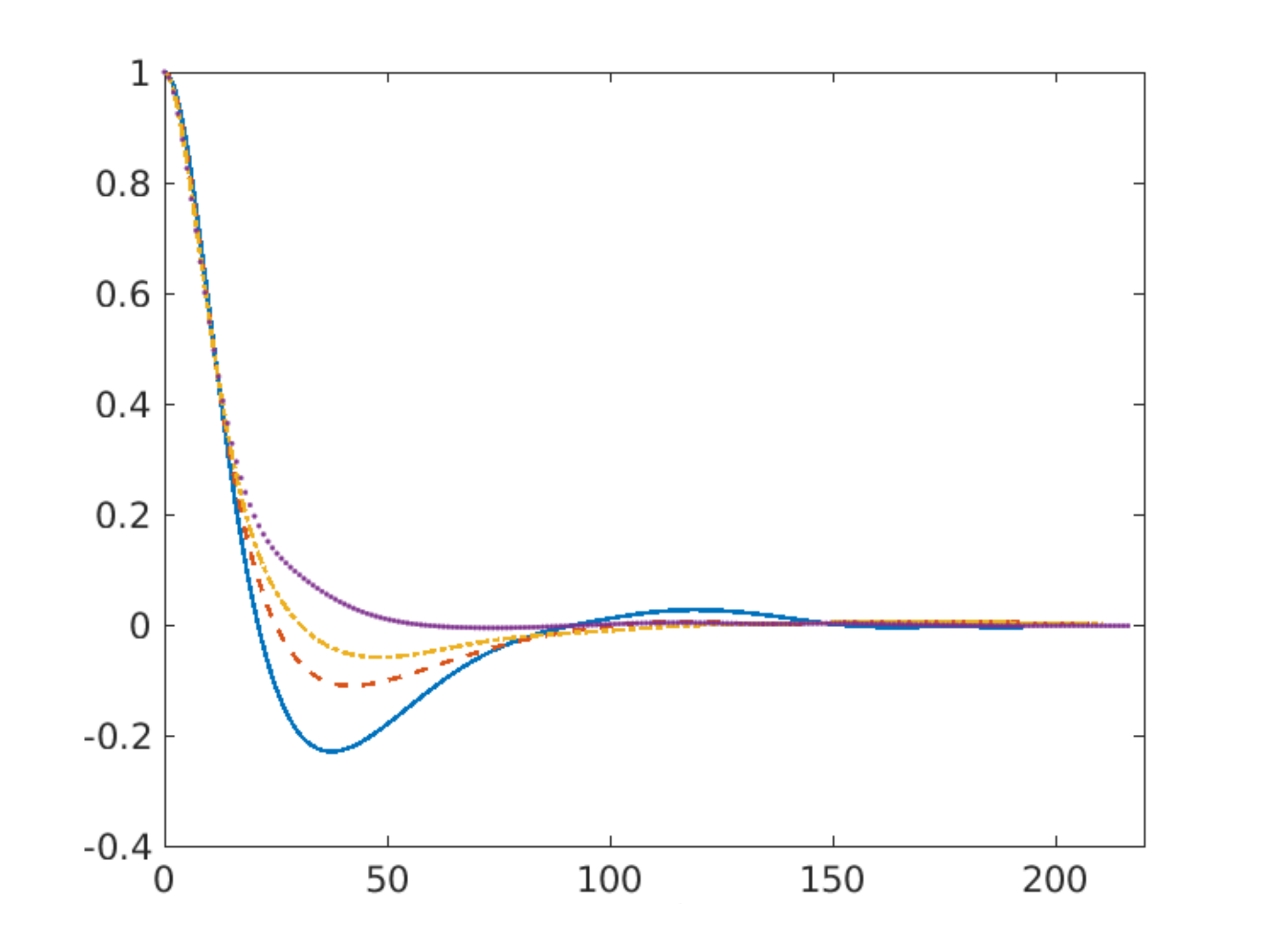}
   \includegraphics[width=0.49\textwidth]{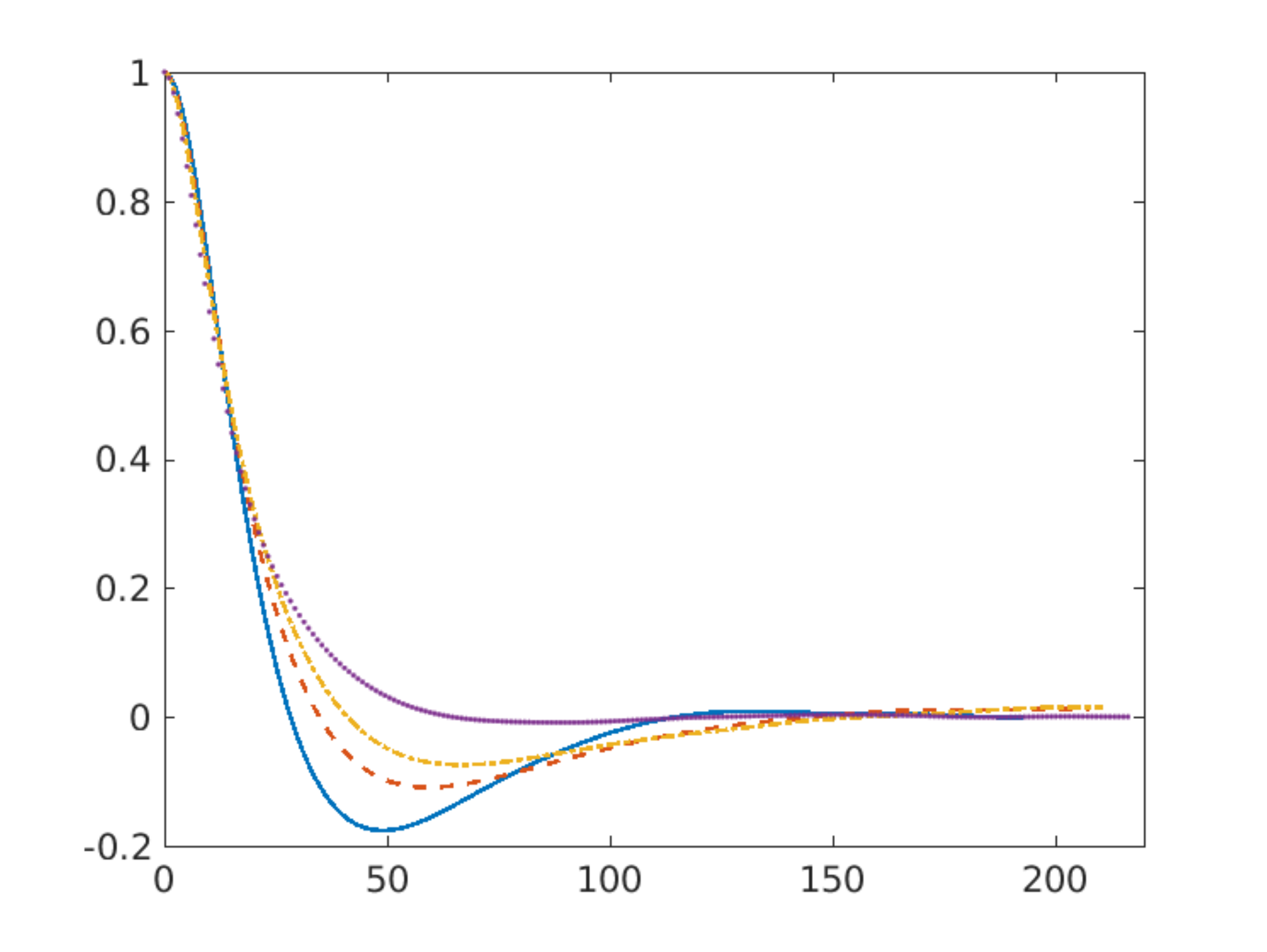}
   \put(-380,58){\rotatebox{90}{\large $R_{vv}$}}
   \put(-290,-4){{\large $\Delta z^+$}}
   \put(-190,58){\rotatebox{90}{\large $R_{vv}$}}
   \put(-98,-4){{\large $\Delta z^+$}}
   \put(-290,112){\footnotesize $(c)$}
   \put(-96,112){\footnotesize $(d)$}\\ 
  \vspace{3pt}        
  \caption{Two-point correlations of the velocity fluctuations versus the spanwise separation $\Delta z^+$ for all $\phi$. 
Streamwise-streamwise component $R_{uu}$ at (a) $y=2a=h/9 \simeq 20\delta_*$ and (b) $y=4a=2h/9 \simeq 40\delta_*$ ($a$ is the 
particle radius). Wall-normal component $R_{vv}$ at (c) $y=a=h/9 \simeq 20\delta_*$ and (d) $y=4a=2h/9 \simeq 40\delta_*$.}
\label{fig:corr}
\end{figure}

To further understand the effect of the solid particles on the turbulence features we report in figure~\ref{fig:corr} 
the two-point autocorrelations of streamwise and wall-normal velocities along the spanwise direction
\begin{align}
R_{uu}(y,\Delta z) =& \frac{\langle u'(x,y,z,t)u'(x,y,z+\Delta z,t)\rangle}{u_{rms}'^2}\\
R_{vv}(y,\Delta z) =& \frac{\langle v'(x,y,z,t)v'(x,y,z+\Delta z,t)\rangle}{v_{rms}'^2}.
\end{align}
These are shown at two wall-normal distances corresponding to 1 and 2 particles diameters ($y=2a \simeq 20\delta_*$ and 
$y=4a \simeq 40\delta_*$) in figures~\ref{fig:corr}(a,c) and figures~\ref{fig:corr}(b,d), respectively. The correlations 
are here calculated for the combined phase, and $\Delta z$ is normalized by the viscous length at the wall-bisector.
 
It is well-known that for single-phase turbulent flows these correlations exhibit negative minimum values in the 
near-wall region and that the corresponding $\Delta z^+$ is indicative of the width of the typical structures of 
wall-bounded turbulence that sustain the turbulence process \cite[i.e. quasi-streamwise vortices and low-speed 
streaks,][]{kim1987turbulence,waleffe,pope2000}. The streamwise autocorrelation $R_{uu}$ evaluated at 
$y=2a \simeq 20\delta_*$, see figure~\ref{fig:corr}(a), reveals that the minimum is smoothened with $\phi$, while 
the separation distance increases. The smoothening of the minimum indicates that the width of the near-wall 
structures is less evident.
Also at $y=4a \simeq 40\delta_*$, figure~\ref{fig:corr}(b), the separation 
distance increases with $\phi$. However, we also see that the minimum value decreases with the volume fraction 
until $\phi=0.1$. 
Hence, at this distance from the wall it is possible to identify wider streamwise velocity streaks with respect to those 
of the unladen case.

The autocorrelation of the wall-normal velocity fluctuations at the same wall-normal locations, are instead shown 
in figure~\ref{fig:corr}(c) and (d). The data show that the negative minima of the unladen case progressively disappears by 
increasing solid volume fraction $\phi$. Therefore, as $\phi$ increases, the 
turbulence structures become less organised in coherent structures as also observed by \citet{picano2015} for channel flow.

\section{Final remarks}

We have studied turbulent duct flows laden with suspensions of finite-size particles, particles larger than 
the smallest flow structures. We have considered 
neutrally-buoyant rigid spheres of size $a=h/18 \sim 10 \delta_*$ and three solid volume fractions $\phi
=0.05, 0.1, 0.2$. The bulk Reynolds number has been set to $Re_b=5600$ to compare results with those found 
for turbulent channel flow of similar $h/a$ and $Re_b$. For the unladen duct, this choice of $Re_b$ 
results in a mean friction Reynolds number $Re_{\tau}=185$ and friction factor of $0.035$ \cite[same value 
that is obtained via the empirical formula of][]{jones1976}.
\sloppy

One of the main findings concerns the effect of the particle presence on the so-called secondary/cross-flow 
velocities of the fluid phase. In single-phase turbulent duct flows their magnitude is about $2\%$ of the 
bulk velocity. We have found that the intensity of these secondary flows progressively increases with solid 
volume fraction up to $\phi=0.1$. Above this volume fraction, a strong turbulence activity attenuation is 
found and correspondingly the maximum value of the cross-flow velocity magnitude sharply drops below the 
value of the unladen case. Interestingly, we have found a lag between fluid and particle cross-flow 
velocities. In particular, at the corner bisectors the fluid cross-flow velocity is larger than that 
of the solid phase. On the contrary, the fluid and particle cross-flow velocities are similar at the walls away 
from the corners. It is well-known that at the wall-bisectors these secondary flows convect low-momentum fluid 
from the walls towards the duct core. This induces a convexity in the mean fluid streamwise velocity isotach 
(i.e. at equal distance from the walls, the mean streamwise velocity is larger at the corner-bisector than at 
the wall-bisector). As the cross-flow velocity increases with $\phi$, so does the convexity of the mean 
streamwise velocity contours. For $\phi=0.2$ the convexity is less than in the unladen case. For all 
$\phi$, the mean streamwise velocity of the solid phase is similar to that of the fluid, except very close to 
the walls were the particle velocity is not subjected to the no-slip condition.

At the wall-bisector, the mean streamwise velocity profiles are found to be similar to those in channel flows. 
In particular, as $\phi$ increases, the velocity decreases closer to the walls and increases towards the core. 
For $\phi=0.2$ we have found a more abrupt increase in velocity than what was previously found in channel flows. 
We have also reported the mean streamwise velocity in inner units and calculated the von K\'arm\'an constant, 
$\kappa$, and the additive coefficient $B$. For the unladen case we have obtained the same $\kappa$ found by 
\citet{gavrilakis1992} for $Re_b=4410$. Increasing $\phi$ both $\kappa$ and $B$ are found to decrease denoting 
a contrasting behavior in terms of drag reduction or enhancement. The friction Reynolds number calculated at 
the wall-bisector, $Re_{\tau,bis}$ is found to increase for all $\phi$, as in channel flow, although 
the increase is smaller than in the latter flow case at equal $\phi$.
Interestingly,
for $\phi=0.1$ the mean $Re_{\tau}$ and that calculated at the bisector almost coincide and, for larger $\phi$ 
the mean $Re_{\tau}$ is then found to decrease. Hence, 
the importance of particle-induced stresses on the overall drag is less in a more realistic geometry such as the 
one studied here, compared to channel flow.

Other important observations concern the mean particle concentration. Particles tend to form a stable 
layer very close to the walls for all $\phi$. However, for $\phi \le 0.1$, the higher particle 
concentration is on the corner-bisectors close to the corner (at a distance of about $0.27h$). 
The high particle concentration at the corners is probably related to the existence of the secondary flows. 
Eventually, it results from the observed lag between the fluid and solid phases cross-flow velocities. 
It should be noted that in laminar duct flows of similar $h/a$ and $Re_b \sim 
500$ particles are also found to mostly accumulate at the corners \citep{kazerooni}. In the latter case 
this is a result of the particle inertial migration away from the duct core. 
For $\phi=0.2$, the particle 
concentration close to the corners is still high, but the maximum concentration is found at the duct core. This 
observation is in contrast to what found in channel flow, revealing the importance of confinement on particle dynamics. 
From the data available for channel flow, we have indeed found that the there is a substantial number of particles 
crossing the periodic boundaries (i.e. a significant particle diffusion in the spanwise direction). The additional 
confinement due to the vertical walls induces therefore a different particle concentration.

Examining the fluid and particles velocity fluctuations, we see that close to the walls there is a redistribution 
of energy towards a more isotropic state (i.e. decrease of the streawise r.m.s. velocity with $\phi$ and an 
increase of the fluctuation velocities in the cross-stream directions). 
For $\phi=0.2$ all components of the r.m.s. velocity are found to be smaller than those of the unladen case, except for the spanwise and 
wall-normal components close to the corners. 
Looking at the primary Reynolds stress, we have seen that it increases 
throughout the cross-section up to $\phi=0.1$. It is interesting to note that for turbulent channel flow, 
\citet{picano2015} found that the maximum $\langle u_f'v_f' \rangle^+$ decreases with $\phi$. On the other hand, we 
have here found that at the wall-bisector the maximum is approximately constant up to $\phi=0.1$. So the turbulence 
activity reduction becomes important for volume fractions larger than $\phi=0.1$ as can be seen from the results 
obtained for $\phi=0.2$. 

Interestingly, close to the corners the secondary Reynolds stress $\langle v_f'w_f' \rangle$, is larger than in the 
single-phase case for all $\phi$. The secondary Reynolds stress of the solid phase, $\langle v_p'w_p' \rangle$, resemble 
those of the fluid, except at the corners where they change sign. Therefore they may represent an additional source of mean 
streamwise vorticity. Indeed, close to the corners and in the viscous sublayer, the gradients of the fluid secondary 
Reynolds stress typically act as a sink of mean streamwise vorticity (the production of vorticity within the viscous 
sublayer is mainly responsible for the presence of vorticity in the bulk of the flow). The opposite sign may 
indicate that this new contribution due to the solid phase actually acts as a source term.

Finally, we have performed quadrant analyses of both primary and secondary Reynolds stresses, looking at the 
occurrence probability of Q1, Q2, Q3 and Q4 events in the cross-section. The probabilities are everywhere similar up 
to $\phi=0.1$, again indicating that the turbulence is not strongly altered up to this solid volume fraction. 
However, for $\phi=0.2$ the probability maps change drastically denoting a strong reduction of the turbulence 
activity. The velocity autocorrelations confirm these results as for the largest $\phi$ the typical negative 
minima disappear, denoting a weaker organisation of the turbulence in coherent structures.

\begin{acknowledgments}
This work was supported by the European Research Council Grant No.\ ERC-2013-CoG-616186, TRITOS.
Computer time was provided by SNIC (Swedish National Infrastructure for Computing). The Authors also 
aknowledge the support from the COST Action MP1305: \emph{Flowing matter}.
\end{acknowledgments}


\begin{thebibliography}{64}
\expandafter\ifx\csname natexlab\endcsname\relax\def\natexlab#1{#1}\fi

\bibitem[Abbas {\em et~al.\/}(2014)Abbas, Magaud, Gao \& Geoffroy]{abbas2014}
{\sc Abbas, M, Magaud, P, Gao, Y \& Geoffroy, S} 2014 Migration of finite sized
  particles in a laminar square channel flow from low to high Reynolds numbers.
  {\em Physics of Fluids (1994-present)\/} {\bf 26}~(12), 123301.

\bibitem[Amini {\em et~al.\/}(2012)Amini, Sollier, Weaver \& Di~Carlo]{amini}
{\sc Amini, Hamed, Sollier, Elodie, Weaver, Westbrook~M \& Di~Carlo, Dino} 2012
  Intrinsic particle-induced lateral transport in microchannels. {\em
  Proceedings of the National Academy of Sciences\/} {\bf 109}~(29),
  11593--11598.

\bibitem[Ardekani {\em et~al.\/}(2017)Ardekani, Costa, Breugem, Picano \&
  Brandt]{niazi}
{\sc Ardekani, M~Niazi, Costa, Pedro, Breugem, W-P, Picano, Francesco \&
  Brandt, Luca} 2017 Drag reduction in turbulent channel flow laden with
  finite-size oblate spheroids. {\em Journal of Fluid Mechanics\/} {\bf 816},
  43--70.

\bibitem[Balachandar \& Eaton(2010)]{balach-rev2010}
{\sc Balachandar, S \& Eaton, John~K} 2010 Turbulent dispersed multiphase flow.
  {\em Annual Review of Fluid Mechanics\/} {\bf 42}, 111--133.

\bibitem[Brenner(1961)]{brenner1961}
{\sc Brenner, Howard} 1961 The slow motion of a sphere through a viscous fluid
  towards a plane surface. {\em Chemical engineering science\/} {\bf 16}~(3-4),
  242--251.

\bibitem[Breugem \& Boersma(2005)]{breugem2005}
{\sc Breugem, WP \& Boersma, BJ} 2005 Direct numerical simulations of turbulent
  flow over a permeable wall using a direct and a continuum approach. {\em
  Physics of Fluids (1994-present)\/} {\bf 17}~(2), 025103.

\bibitem[Breugem(2012)]{breugem2012}
{\sc Breugem, Wim-Paul} 2012 A second-order accurate immersed boundary method
  for fully resolved simulations of particle-laden flows. {\em Journal of
  Computational Physics\/} {\bf 231}~(13), 4469--4498.

\bibitem[Breugem {\em et~al.\/}(2014)Breugem, Van~Dijk \& Delfos]{breugem2014}
{\sc Breugem, Wim-Paul, Van~Dijk, Vincent \& Delfos, Ren{\'e}} 2014 Flows
  through real porous media: X-ray computed tomography, experiments, and
  numerical simulations. {\em Journal of Fluids Engineering\/} {\bf 136}~(4),
  040902.

\bibitem[Chun \& Ladd(2006)]{chun2006}
{\sc Chun, B \& Ladd, AJC} 2006 Inertial migration of neutrally buoyant
  particles in a square duct: An investigation of multiple equilibrium
  positions. {\em Physics of Fluids (1994-present)\/} {\bf 18}~(3), 031704.

\bibitem[Costa {\em et~al.\/}(2015)Costa, Boersma, Westerweel \&
  Breugem]{costa2015}
{\sc Costa, Pedro, Boersma, Bendiks~Jan, Westerweel, Jerry \& Breugem,
  Wim-Paul} 2015 Collision model for fully resolved simulations of flows laden
  with finite-size particles. {\em Physical Review E\/} {\bf 92}~(5), 053012.

\bibitem[Costa {\em et~al.\/}(2016)Costa, Picano, Brandt \& Breugem]{costa2016}
{\sc Costa, Pedro, Picano, Francesco, Brandt, Luca \& Breugem, Wim-Paul} 2016
  Universal scaling laws for dense particle suspensions in turbulent
  wall-bounded flows. {\em Physical Review Letters\/} {\bf 117}~(13), 134501.

\bibitem[Doyeux {\em et~al.\/}(2016)Doyeux, Priem, Jibuti, Farutin, Ismail \&
  Peyla]{doyeux}
{\sc Doyeux, Vincent, Priem, Stephane, Jibuti, Levan, Farutin, Alexander,
  Ismail, Mourad \& Peyla, Philippe} 2016 Effective viscosity of
  two-dimensional suspensions: Confinement effects. {\em Physical Review
  Fluids\/} {\bf 1}~(4), 043301.

\bibitem[Fornari {\em et~al.\/}(2016{\natexlab{{\em a\/}}})Fornari, Brandt,
  Chaudhuri, Lopez, Mitra \& Picano]{fornariPRL}
{\sc Fornari, Walter, Brandt, Luca, Chaudhuri, Pinaki, Lopez, Cyan~Umbert,
  Mitra, Dhrubaditya \& Picano, Francesco} 2016{\natexlab{{\em a\/}}} Rheology
  of confined non-brownian suspensions. {\em Physical Review Letters\/} {\bf
  116}~(1), 018301.

\bibitem[Fornari {\em et~al.\/}(2016{\natexlab{{\em b\/}}})Fornari, Formenti,
  Picano \& Brandt]{forn2016}
{\sc Fornari, Walter, Formenti, Alberto, Picano, Francesco \& Brandt, Luca}
  2016{\natexlab{{\em b\/}}} The effect of particle density in turbulent
  channel flow laden with finite size particles in semi-dilute conditions. {\em
  Physics of Fluids (1994-present)\/} {\bf 28}~(3), 033301.

\bibitem[Fornari {\em et~al.\/}(2016{\natexlab{{\em c\/}}})Fornari, Picano \&
  Brandt]{fornari2016JFM}
{\sc Fornari, Walter, Picano, Francesco \& Brandt, Luca} 2016{\natexlab{{\em
  c\/}}} Sedimentation of finite-size spheres in quiescent and turbulent
  environments. {\em Journal of Fluid Mechanics\/} {\bf 788}, 640--669.

\bibitem[Fornari {\em et~al.\/}(2018)Fornari, Picano \& Brandt]{forn2018}
{\sc Fornari, Walter, Picano, Francesco \& Brandt, Luca} 2018 The effect of
  polydispersity in a turbulent channel flow laden with finite-size particles.
  {\em European Journal of Mechanics - B/Fluids\/} {\bf 67}, 54 -- 64.

\bibitem[Gavrilakis(1992)]{gavrilakis1992}
{\sc Gavrilakis, S} 1992 Numerical simulation of low-Reynolds-number turbulent
  flow through a straight square duct. {\em Journal of Fluid Mechanics\/} {\bf
  244}, 101--129.

\bibitem[Gessner(1973)]{gessner}
{\sc Gessner, FB} 1973 The origin of secondary flow in turbulent flow along a
  corner. {\em Journal of Fluid Mechanics\/} {\bf 58}~(01), 1--25.

\bibitem[Guazzelli \& Morris(2011)]{guazz2011}
{\sc Guazzelli, Elisabeth \& Morris, Jeffrey~F} 2011 {\em A physical
  introduction to suspension dynamics\/}, , vol.~45. Cambridge University
  Press.

\bibitem[Henningson \& Kim(1991)]{henningson1991}
{\sc Henningson, Dan~S \& Kim, John} 1991 On turbulent spots in plane
  poiseuille flow. {\em Journal of fluid mechanics\/} {\bf 228}, 183--205.

\bibitem[Huser \& Biringen(1993)]{huser93}
{\sc Huser, Asmund \& Biringen, Sedat} 1993 Direct numerical simulation of
  turbulent flow in a square duct. {\em Journal of Fluid Mechanics\/} {\bf
  257}, 65--95.

\bibitem[Jones(1976)]{jones1976}
{\sc Jones, OC} 1976 An improvement in the calculation of turbulent friction in
  rectangular ducts. {\em Journal of Fluids Engineering\/} {\bf 98}~(2),
  173--180.

\bibitem[Joung {\em et~al.\/}(2007)Joung, Choi \& Choi]{joung2007}
{\sc Joung, Younghoon, Choi, Sung-Uk \& Choi, Jung-Il} 2007 Direct numerical
  simulation of turbulent flow in a square duct: analysis of secondary flows.
  {\em Journal of engineering mechanics\/} {\bf 133}~(2), 213--221.

\bibitem[Kazerooni {\em et~al.\/}(2017)Kazerooni, Fornari, Hussong \&
  Brandt]{kazerooni}
{\sc Kazerooni, H.~Tabaei, Fornari, W., Hussong, J. \& Brandt, L.} 2017
  Inertial migration in dilute and semidilute suspensions of rigid particles in
  laminar square duct flow. {\em Physical Review Fluids\/} {\bf 2}, 084301.

\bibitem[Kim {\em et~al.\/}(1987)Kim, Moin \& Moser]{kim1987turbulence}
{\sc Kim, John, Moin, Parviz \& Moser, Robert} 1987 Turbulence statistics in
  fully developed channel flow at low Reynolds number. {\em Journal of fluid
  mechanics\/} {\bf 177}, 133--166.

\bibitem[Koh {\em et~al.\/}(1994)Koh, Hookham \& Leal]{koh1994}
{\sc Koh, Christopher~J, Hookham, Philip \& Leal, LG} 1994 An experimental
  investigation of concentrated suspension flows in a rectangular channel. {\em
  Journal of Fluid Mechanics\/} {\bf 266}, 1--32.

\bibitem[Kulick {\em et~al.\/}(1994)Kulick, Fessler \& Eaton]{kulick1994}
{\sc Kulick, Jonathan~D, Fessler, John~R \& Eaton, John~K} 1994 Particle
  response and turbulence modification in fully developed channel flow. {\em
  Journal of Fluid Mechanics\/} {\bf 277}~(1), 109--134.

\bibitem[Kulkarni \& Morris(2008)]{Morrispof08}
{\sc Kulkarni, P.M. \& Morris, J.F.} 2008 Suspension properties at finite
  Reynolds number from simulated shear flow. {\em Phys. Fluids\/} {\bf
  20}~(040602).

\bibitem[Lashgari {\em et~al.\/}(2017{\natexlab{{\em a\/}}})Lashgari, Ardekani,
  Banerjee, Russom \& Brandt]{lash2017}
{\sc Lashgari, Iman, Ardekani, Mehdi~Niazi, Banerjee, Indradumna, Russom, Aman
  \& Brandt, Luca} 2017{\natexlab{{\em a\/}}} Inertial migration of spherical
  and oblate particles in straight ducts. {\em Journal of Fluid Mechanics\/}
  {\bf 819}, 540--561.

\bibitem[Lashgari {\em et~al.\/}(2015)Lashgari, Picano \& Brandt]{lashg2015}
{\sc Lashgari, Iman, Picano, Francesco \& Brandt, Luca} 2015 Transition and
  self-sustained turbulence in dilute suspensions of finite-size particles.
  {\em Theoretical and Applied Mechanics Letters\/} .

\bibitem[Lashgari {\em et~al.\/}(2014)Lashgari, Picano, Breugem \&
  Brandt]{lashgari2014}
{\sc Lashgari, Iman, Picano, Francesco, Breugem, Wim-Paul \& Brandt, Luca} 2014
  Laminar, turbulent, and inertial shear-thickening regimes in channel flow of
  neutrally buoyant particle suspensions. {\em Physical Review Letters\/} {\bf
  113}~(25), 254502.

\bibitem[Lashgari {\em et~al.\/}(2016)Lashgari, Picano, Breugem \&
  Brandt]{lashg2016}
{\sc Lashgari, Iman, Picano, Francesco, Breugem, Wim~Paul \& Brandt, Luca} 2016
  Channel flow of rigid sphere suspensions: Particle dynamics in the inertial
  regime. {\em International Journal of Multiphase Flow\/} {\bf 78}, 12--24.

\bibitem[Lashgari {\em et~al.\/}(2017{\natexlab{{\em b\/}}})Lashgari, Picano,
  Costa, Breugem \& Brandt]{lashgari2017}
{\sc Lashgari, Iman, Picano, Francesco, Costa, Pedro, Breugem, Wim-Paul \&
  Brandt, Luca} 2017{\natexlab{{\em b\/}}} Turbulent channel flow of a dense
  binary mixture of rigid particles. {\em Journal of Fluid Mechanics\/} {\bf
  818}, 623--645.

\bibitem[Lin {\em et~al.\/}(2017)Lin, Shao, Yu \& Wang]{Lin}
{\sc Lin, Zhao-wu, Shao, Xue-ming, Yu, Zhao-sheng \& Wang, Lian-ping} 2017
  Effects of finite-size heavy particles on the turbulent flows in a square
  duct. {\em Journal of Hydrodynamics, Ser. B\/} {\bf 29}~(2), 272--282.

\bibitem[Loisel {\em et~al.\/}(2013)Loisel, Abbas, Masbernat \&
  Climent]{Loisel2013}
{\sc Loisel, Vincent, Abbas, Micheline, Masbernat, Olivier \& Climent, Eric}
  2013 The effect of neutrally buoyant finite-size particles on channel flows
  in the laminar-turbulent transition regime. {\em Physics of Fluids
  (1994-present)\/} {\bf 25}~(12), 123304.

\bibitem[Matas {\em et~al.\/}(2003)Matas, Morris \& Guazzelli]{matas2003}
{\sc Matas, J-P, Morris, Jeffrey~F \& Guazzelli, E} 2003 Transition to
  turbulence in particulate pipe flow. {\em Physical Review Letters\/} {\bf
  90}~(1), 014501.

\bibitem[Matas {\em et~al.\/}(2004)Matas, Morris \& Guazzelli]{matas}
{\sc Matas, Jean-Philippe, Morris, Jeffrey~F \& Guazzelli, {\'E}lisabeth} 2004
  Inertial migration of rigid spherical particles in poiseuille flow. {\em
  Journal of Fluid Mechanics\/} {\bf 515}, 171--195.

\bibitem[Maxey(2017)]{maxey2017}
{\sc Maxey, Martin} 2017 Simulation methods for particulate flows and
  concentrated suspensions. {\em Annual Review of Fluid Mechanics\/} {\bf 49},
  171--193.

\bibitem[Morita {\em et~al.\/}(2017)Morita, Itano \& Sugihara-Seki]{morita}
{\sc Morita, Yusuke, Itano, Tomoaki \& Sugihara-Seki, Masako} 2017 Equilibrium
  radial positions of neutrally buoyant spherical particles over the circular
  cross-section in poiseuille flow. {\em Journal of Fluid Mechanics\/} {\bf
  813}, 750--767.

\bibitem[Morris(2009)]{morris2009}
{\sc Morris, Jeffrey~F} 2009 A review of microstructure in concentrated
  suspensions and its implications for rheology and bulk flow. {\em Rheologica
  acta\/} {\bf 48}~(8), 909--923.

\bibitem[Morris \& Haddadi(2014)]{Morris2014}
{\sc Morris, J.~F. \& Haddadi, H.} 2014 Microstructure and rheology of finite
  inertia neutrally buoyant suspensions. {\em Journal of Fluid Mechanics\/} {\bf 749},
  431--459.

\bibitem[Nakagawa {\em et~al.\/}(2015)Nakagawa, Yabu, Otomo, Kase, Makino,
  Itano \& Sugihara-Seki]{nakagawa}
{\sc Nakagawa, Naoto, Yabu, Takuya, Otomo, Ryoko, Kase, Atsushi, Makino,
  Masato, Itano, Tomoaki \& Sugihara-Seki, Masako} 2015 Inertial migration of a
  spherical particle in laminar square channel flows from low to high Reynolds
  numbers. {\em Journal of Fluid Mechanics\/} {\bf 779}, 776.

\bibitem[Noorani {\em et~al.\/}(2016)Noorani, Vinuesa, Brandt \&
  Schlatter]{noorani}
{\sc Noorani, Azad, Vinuesa, Ricardo, Brandt, Luca \& Schlatter, Philipp} 2016
  Aspect ratio effect on particle transport in turbulent duct flows. {\em
  Physics of Fluids\/} {\bf 28}~(11), 115103.

\bibitem[Picano {\em et~al.\/}(2015)Picano, Breugem \& Brandt]{picano2015}
{\sc Picano, Francesco, Breugem, Wim-Paul \& Brandt, Luca} 2015 Turbulent
  channel flow of dense suspensions of neutrally buoyant spheres. {\em Journal
  of Fluid Mechanics\/} {\bf 764}, 463--487.

\bibitem[Picano {\em et~al.\/}(2013)Picano, Breugem, Mitra \&
  Brandt]{picano2013}
{\sc Picano, Francesco, Breugem, Wim-Paul, Mitra, Dhrubaditya \& Brandt, Luca}
  2013 Shear thickening in non-brownian suspensions: an excluded volume effect.
  {\em Physical Review Letters\/} {\bf 111}~(9), 098302.

\bibitem[Pinelli {\em et~al.\/}(2010)Pinelli, Uhlmann, Sekimoto \&
  Kawahara]{pinelli2010}
{\sc Pinelli, Alfredo, Uhlmann, Markus, Sekimoto, Atsushi \& Kawahara, Genta}
  2010 Reynolds number dependence of mean flow structure in square duct
  turbulence. {\em Journal of fluid mechanics\/} {\bf 644}, 107--122.

\bibitem[Pope(2000)]{pope2000}
{\sc Pope, Stephen~B} 2000 {\em Turbulent flows\/}. Cambridge university press.

\bibitem[Pourquie {\em et~al.\/}(2009)Pourquie, Breugem \&
  Boersma]{pourquie2009}
{\sc Pourquie, MBJM, Breugem, WP \& Boersma, Bendiks~Jan} 2009 Some issues
  related to the use of immersed boundary methods to represent square
  obstacles. {\em International Journal for Multiscale Computational
  Engineering\/} {\bf 7}~(6).

\bibitem[Prandtl(1963)]{prandtl}
{\sc Prandtl, Ludwig} 1963 {\em The essentials of fluid dynamics\/}. Blackie \&
  Son Limited.

\bibitem[Reeks(1983)]{reeks1983}
{\sc Reeks, MW} 1983 The transport of discrete particles in inhomogeneous
  turbulence. {\em Journal of aerosol science\/} {\bf 14}~(6), 729--739.

\bibitem[Samanta {\em et~al.\/}(2015)Samanta, Vinuesa, Lashgari, Schlatter \&
  Brandt]{samanta2015}
{\sc Samanta, Arghya, Vinuesa, Ricardo, Lashgari, Iman, Schlatter, Philipp \&
  Brandt, Luca} 2015 Enhanced secondary motion of the turbulent flow through a
  porous square duct. {\em Journal of Fluid Mechanics\/} {\bf 784}, 681--693.

\bibitem[Sardina {\em et~al.\/}(2011)Sardina, Picano, Schlatter, Brandt \&
  Casciola]{sardina2011}
{\sc Sardina, Gaetano, Picano, Francesco, Schlatter, Philipp, Brandt, Luca \&
  Casciola, Carlo~Massimo} 2011 Large scale accumulation patterns of inertial
  particles in wall-bounded turbulent flow. {\em Flow, turbulence and
  combustion\/} {\bf 86}~(3-4), 519--532.

\bibitem[Sardina {\em et~al.\/}(2012)Sardina, Schlatter, Brandt, Picano \&
  Casciola]{sardina2012}
{\sc Sardina, G, Schlatter, Philipp, Brandt, Luca, Picano, F \& Casciola, CM}
  2012 Wall accumulation and spatial localization in particle-laden wall flows.
  {\em Journal of Fluid Mechanics\/} {\bf 699}, 50--78.

\bibitem[Segre \& Silberberg(1962)]{segre1962}
{\sc Segre, G \& Silberberg, A} 1962 Behaviour of macroscopic rigid spheres in
  poiseuille flow part 2. experimental results and interpretation. {\em Journal
  of Fluid Mechanics\/} {\bf 14}~(01), 136--157.

\bibitem[Sharma \& Phares(2006)]{Sharma}
{\sc Sharma, Gaurav \& Phares, Denis~J} 2006 Turbulent transport of particles
  in a straight square duct. {\em International Journal of Multiphase Flow\/}
  {\bf 32}~(7), 823--837.

\bibitem[Soldati \& Marchioli(2009)]{soldati2009}
{\sc Soldati, Alfredo \& Marchioli, Cristian} 2009 Physics and modelling of
  turbulent particle deposition and entrainment: Review of a systematic study.
  {\em International Journal of Multiphase Flow\/} {\bf 35}~(9), 827--839.

\bibitem[Stickel \& Powell(2005)]{stickel2005}
{\sc Stickel, Jonathan~J \& Powell, Robert~L} 2005 Fluid mechanics and rheology
  of dense suspensions. {\em Annual Review of Fluid Mechanics\/} {\bf 37}, 129--149.

\bibitem[Uhlmann(2005)]{uhlmann2005}
{\sc Uhlmann, Markus} 2005 An immersed boundary method with direct forcing for
  the simulation of particulate flows. {\em Journal of Computational Physics\/}
  {\bf 209}~(2), 448--476.

\bibitem[Vinuesa {\em et~al.\/}(2014)Vinuesa, Noorani, Lozano-Dur{\'a}n,
  Khoury, Schlatter, Fischer \& Nagib]{vinuesa14}
{\sc Vinuesa, Ricardo, Noorani, Azad, Lozano-Dur{\'a}n, Adri{\'a}n, Khoury,
  George K~El, Schlatter, Philipp, Fischer, Paul~F \& Nagib, Hassan~M} 2014
  Aspect ratio effects in turbulent duct flows studied through direct numerical
  simulation. {\em Journal of Turbulence\/} {\bf 15}~(10), 677--706.

\bibitem[Virk(1975)]{virk1975}
{\sc Virk, Preetinder~S} 1975 Drag reduction fundamentals. {\em AIChE
  Journal\/} {\bf 21}~(4), 625--656.

\bibitem[Wagner \& Brady(2009)]{wagner2009}
{\sc Wagner, Norman~J \& Brady, John~F} 2009 Shear thickening in colloidal
  dispersions. {\em Physics Today\/} {\bf 62}~(10), 27--32.

\bibitem[Waleffe(1997)]{waleffe}
{\sc Waleffe, Fabian} 1997 On a self-sustaining process in shear flows. {\em
  Physics of Fluids\/} {\bf 9}~(4), 883--900.

\bibitem[Winkler {\em et~al.\/}(2004)Winkler, Rani \& Vanka]{Winkler04}
{\sc Winkler, CM, Rani, Sarma~L \& Vanka, SP} 2004 Preferential concentration
  of particles in a fully developed turbulent square duct flow. {\em
  International Journal of Multiphase Flow\/} {\bf 30}~(1), 27--50.

\bibitem[Zhao {\em et~al.\/}(2010)Zhao, Andersson \& Gillissen]{zhao2010}
{\sc Zhao, LH, Andersson, Helge~I \& Gillissen, JJJ} 2010 Turbulence modulation
  and drag reduction by spherical particles. {\em Physics of Fluids
  (1994-present)\/} {\bf 22}~(8), 081702.

\end{thebibliography}
%

\end{document}